\documentclass[PAPER, atlasdraft=false,USenglish,cernpreprint, texmf, orcidlogo]{atlasdoc}
\usepackage{atlaspackage}
\usepackage{enumitem}
\usepackage{atlasbiblatex}

\usepackage[jetetmiss=true, snippets=true]{atlasphysics}
\graphicspath{{logos/}{figures/}}

\usepackage{zjets-paper-defs}
\usepackage{listings}
\usepackage{xcolor}
\usepackage{tikz}

\definecolor{codegreen}{rgb}{0,0.6,0}
\definecolor{codegray}{rgb}{0.5,0.5,0.5}
\definecolor{codepurple}{rgb}{0.58,0,0.82}
\definecolor{backcolour}{rgb}{0.95,0.95,0.92}

\lstdefinestyle{mystyle}{
backgroundcolor=\color{backcolour}, commentstyle=\color{codegreen},
keywordstyle=\color{magenta},
numberstyle=\tiny\color{codegray},
stringstyle=\color{codepurple},
basicstyle=\ttfamily\footnotesize,
breakatwhitespace=false,
breaklines=true,
captionpos=b,
keepspaces=true,
numbers=left,
numbersep=5pt,
showspaces=false,
showstringspaces=false,
showtabs=false,
tabsize=2
}

\AtlasTitle{A fully differential charged-particle phase-space measurement of $Z$+jets production in $pp$ collisions at $\sqrt{s}=13$ \TeV with the ATLAS detector}

\AtlasAbstract{%
The production of a $Z$~boson decaying to muons in association with hadronic jets is a benchmark process in proton--proton collisions at the Large Hadron Collider. It provides a clean experimental signature with high selection efficiency and purity while probing a wide range of Standard Model dynamics. Measurements of this process are typically reported as binned differential cross sections in a limited set of predefined observables, which reduces the information retained and hinders reinterpretation.
In this paper, the \omnifold\ method is used to produce an unbinned and highly differential measurement of the $Z$+jets cross section using \SI{140}{\per\femto\barn} of $\sqrt{s} = \SI{13}{\TeV}$ proton--proton collision data collected by the ATLAS detector.
This constitutes the first published unbinned, fully differential measurement of a collider cross section,
describing the kinematics of individual charged particles within a well-defined region of phase space.
This unbinned event representation is used to construct measurements in four example applications that illustrate the broad range of observables constrained with a single measurement, including several that would be difficult or impossible to access using traditional unfolding methods.
The measurement is published as particle-level Monte Carlo events together with unfolded event weights, allowing arbitrary observables to be constructed after publication and enabling a broad range of future applications and reinterpretations.
}

\AtlasRefCode{STDM-2024-02}

\PreprintIdNumber{CERN-EP-2026-288}

\AtlasJournal{PRX Intelligence}
\AtlasCoverEgroupAnalysisTeam{atlas-stdm-2024-02-analysis-team@cern.ch}

\hypersetup{pdftitle={ATLAS document},pdfauthor={The ATLAS Collaboration}}

\begin{document}

\maketitle

%

%
%

%

%
\section{Introduction}

Precision differential cross-section measurements of Standard Model (SM) processes at the Large Hadron Collider (LHC) play a central role in contemporary particle physics.
Such measurements provide stringent tests of theoretical predictions, constrain free parameters of the SM, and can reveal deviations associated with physics beyond the Standard Model (BSM).
Measurements of the production of a $Z$ boson decaying to two muons are widely used for these purposes~\cite{STDM-2012-04,STDM-2015-14,STDM-2018-49,STDM-2020-17,CMS-EWK-11-021,CMS-SMP-12-004,CMS-SMP-12-017,CMS-SMP-12-019,CMS-SMP-14-013,CMS-SMP-16-015,CMS-SMP-19-010,CMS-SMP-20-009,CMS-SMP-19-009,CMS-SMP-20-010,CMS-SMP-21-003,LHCb:2019qoc,LHCb:2022rky}.
In events where the $Z$ boson is produced with large transverse momentum (\pt), the dimuon system recoils against a collimated spray of hadrons, referred to as a jet.
By using the two muons for both triggering and event selection,
the $Z$ boson can be used to select a clean and largely unbiased sample of high-\pt{} jets.
Measurements of these jets provide powerful probes of the dynamics of the strong interaction at high energies~\cite{Larkoski:2017jix,Kogler:2018hem}.
They also provide important inputs to precision quantum chromodynamics (QCD) calculations and to the tuning of Monte Carlo (MC) event generators~\cite{ATL-PHYS-PUB-2011-014}.

An essential component of collider measurements is correcting for the effects of finite detector resolution and acceptance on the measured distributions.
In most current differential cross-section measurements, this procedure, known as \textit{unfolding}, is performed by binning the data into histograms using a list of predefined observables of interest and correcting the resulting histogram counts using the simulated detector response~\cite{Cowan:2002in,1974AJ.....79..745L,Richardson:72,DAgostini:1994fjx,Hocker:1995kb,Schmitt:2012kp}.
The resulting measurement is therefore reported as a collection of differential cross sections in fixed histogram bins.
While highly successful, this approach performs an irreversible compression of the data before detector effects are corrected for statistically.
Binned unfolding methods scale poorly with dimensionality, making simultaneous measurements of many observables computationally and statistically challenging.
Moreover, the choice of observables and binning scheme are fixed, so most analyses focus on a small set of observables chosen before the measurement is performed.
Information not captured by those choices is discarded, limiting reinterpretation and preventing the construction of new observables after publication.
There are also classes of observables that are difficult or impossible to access within the traditional framework, including statistical moments of differential cross sections and observables for which no practical response matrix can be constructed.

To overcome these limitations, a new class of unfolding methods based on artificial intelligence (AI) techniques was developed~\cite{Andreassen:2019cjw,Gagunashvili:2010zw,Glazov:2017vni,Datta:2018mwd,Bellagente:2019uyp,Bellagente:2020piv,Vandegar:2020yvw,Howard:2021pos,Backes:2022sph,Diefenbacher:2023wec,Shmakov:2023kjj,Chan:2023tbf,Shmakov:2024gkd,Desai:2024yft,Butter:2024vbx,Zhu:2024drd,Pazos:2024nfe,Favaro:2025psi,Butter:2025via,Butter:2025mek,Petitjean:2025tgk,Qureshi:2026oiv,Ore:2026qgp,Huetsch:2024quz}.
One such method, \omnifold{}~\cite{Andreassen:2019cjw, Andreassen:2021zzk}, is used in several measurements by experimental collaborations~\cite{H1:2021wkz,H1:2023fzk,H1:2024mox,LHCb:2022rky,Song:2023sxb,Pani:2024mgy,Huang:2025ziq,STDM-2020-17,STDM-2022-05,CMS-SMP-23-008}.
\omnifold{} produces an unbinned representation of the unfolded cross section by assigning weights to particle-level simulated events, allowing observables to be constructed after unfolding without requiring a separate unfolding procedure for each observable.
In most existing applications, however, the unfolded event sample is subsequently projected onto a limited set of observables and binned for visualization and interpretation; see Ref.~\cite{Canelli:2025ybb} for a review and discussion of unbinned measurements.
A recent ATLAS measurement of $Z$+jets events, for example, used \omnifold{} to perform a simultaneous differential cross-section measurement of 24 observables~\cite{STDM-2020-17}. This will hereafter be referred to as the \multifold{} analysis.

A particularly powerful application of AI-based unfolding methods is the ability to operate directly on the kinematics of final-state particles, rather than on a predetermined set of summary observables.
Because the number of particles produced in each collision varies event-by-event, averaging about 50 charged particles and sometimes exceeding 250 in this measurement, the resulting phase space is both high-dimensional and variable-dimensional.
Unlike traditional response-matrix techniques, AI-based unfolding methods can learn detector-response corrections directly in high-dimensional variable-length event representations.
Rather than publishing a finite collection of differential cross sections, the primary result is an unfolded representation of the event-level phase space from which differential cross sections and other observables can subsequently be derived by summing the appropriate event weights. In the case of \omnifold{}, the published measurement consists of particle-level simulated events together with a nominal unfolded event weight and associated alternative weight variations that encode the measurement uncertainties. The nominal event weights are normalized to the measured fiducial cross section and therefore have units of femtobarns (fb). Consequently, the measured cross section corresponding to any particle-level selection can be obtained simply by summing the event weights of the selected events.

The execution of this measurement requires extending \omnifold{} to operate on variable-dimensional event representations using transformer neural networks, together with a large-scale pretraining procedure on simulation to achieve sufficient precision in the vastly enlarged phase space.
To enable direct validation, the simulated event samples, event selection, and target phase space are identical to those used in the \multifold{} analysis, allowing comparisons with the 24 observables measured previously while simultaneously extending the measurement to substantially higher dimensionality.
By measuring the kinematics of the individual charged particles produced in each collision, a large number of observables are implicitly measured by a single cross-section measurement~\cite{Altheimer:2013yza}.
The measurement therefore shifts the experimental focus from publishing observables to publishing an event-level representation from which observables can be constructed.
This includes observables whose measurement would be prohibitively complex or impossible to access within conventional binned unfolding frameworks.
For example, traditional measurements of jet substructure observables require careful calibration of jets built with a particular clustering algorithm~\cite{JETM-2022-01}.
With fully differential charged-particle phase-space measurements, any jet clustering algorithm can be applied to the particle-level event representation, and valid measurements of jet observables can be constructed from the resulting jets without the need for additional calibration.
This provides access to measurements of a much broader range of jet observables than are currently practically accessible to particle physics experiments.
Because the measurement remains unbinned, observables can be rebinned to suit downstream analyses and statistical moments can be computed without introducing binning effects.
Several examples of these applications are presented in this paper.
Specifically, Sections~\ref{sec:jet_shape_observables}, \ref{sec:ecc}, and \ref{sec:intrinsic_dimensionality_observables} present measurements of jet-shape observables defined with multiple jet clustering algorithms, energy correlators, and the intrinsic dimensionality of jet data, respectively, all derived from the same unfolded charged-particle event sample.
A limitation of the present measurement is that it constrains only the charged-particle component of the final state, since robust systematic uncertainties for neutral-particle measurements are not currently available~\cite{STDM-2017-04,JETM-2023-06}.

Given the breadth of possible applications, this paper focuses on four representative use cases introduced in Section~\ref{sec:result}.
To encourage further study and reinterpretation, the published measurement is released publicly on the CERN open data portal~\cite{cern_open_data_portal}.
This release enables the measurement of any observable that is a function of the measured phase space.
All physics results presented in this paper are derived from this single underlying measurement.
Additionally, a user guide with detailed examples and usage recommendations is provided to accompany the data~\cite{gitlab}.


%
\section{Experimental setup}
\label{sec:data_simulation_event}

This measurement is performed using the full Run~2 proton--proton collision dataset collected by the ATLAS detector~\cite{PERF-2007-01} at $\sqrt{s} = \SI{13}{\TeV}$, corresponding to an integrated luminosity of \SI{140}{\per\femto\barn}~\cite{DAPR-2021-01}.
The ATLAS experiment is a multipurpose particle detector with a forward--backward symmetric cylindrical geometry and a near \(4\pi\) coverage in solid angle.\footnote{\AtlasCoordFootnote}
It consists of an inner tracking detector surrounded by a thin superconducting solenoid providing a \qty{2}{\tesla} axial magnetic field, electromagnetic and hadronic calorimeters, and a muon spectrometer.

The ATLAS inner tracking detector covers the pseudorapidity range \(|\eta| < 2.5\).
It consists of silicon pixel, silicon microstrip, and transition radiation tracking detectors.
Lead/liquid-argon (LAr) sampling calorimeters provide electromagnetic (EM) energy measurements with high granularity within the region \(|\eta|< 3.2\).
A steel/scintillator-tile hadronic calorimeter covers the central pseudorapidity range (\(|\eta| < 1.7\)).
The endcap and forward regions are instrumented with LAr calorimeters for EM and hadronic energy measurements up to \(|\eta| = 4.9\).
The muon spectrometer surrounds the calorimeters and is based on three large superconducting air-core toroidal magnets with eight coils each.
The muon spectrometer includes a system of precision tracking chambers up to \(|\eta| = 2.7\) and fast detectors for triggering up to \(|\eta| = 2.4\).
The luminosity is measured mainly by the LUCID--2~\cite{LUCID2} detector which is located close to the beampipe.
A two-level trigger system was used to select events~\cite{TRIG-2016-01}.
The first-level trigger is implemented in hardware and uses a subset of the detector information to accept events at a rate close to \qty{100}{\kHz}.
This is followed by a software-based trigger that reduces the accepted rate of complete events to \qty{2}{\kHz} on average depending on the data-taking conditions.
A software suite~\cite{SOFT-2022-02} is used in data simulation, in the reconstruction and analysis of real and simulated data, in detector operations, and in the trigger and data acquisition systems of the experiment.

\subsection{Particle-level phase-space definition}
\label{sec:fiducial}
The signal is defined as events containing a $Z\to\mu\mu$ decay.
The particle-level phase space, which defines the cross section being measured, consists of final-state charged particles with a mean lifetime satisfying $c\,\tau_0 > \SI{10}{\milli\meter}$.
The fiducial region is defined by the following requirements.
Events are required to contain at least two muons that do not originate from hadron decays and must each satisfy transverse momentum $\pt > 25$ \GeV and pseudorapidity $|\eta| < 2.4$.
The two highest-\pt{} muons are selected and required to have opposite charge, dimuon invariant mass $m_{\mu\mu} \in (81, 101)$\,\GeV, and dimuon transverse momentum $\ptmm > 200$\,\GeV.
The requirement $\ptmm>200$ \GeV selects the boosted regime, where the recoiling hadronic system contains energetic jets without imposing any explicit selection on the jets themselves.
This provides an unbiased sample for studies of charged-particle jet substructure while maintaining a high signal purity and sufficient statistical precision for the measurement. All such charged particles with $\pt > 0.5$ \GeV and $|\eta| < 2.5$ are included in the particle-level event representation.

Each particle-level event is represented by the set of charged particles satisfying these requirements, where each particle is described by its kinematic properties $(\pt,\eta,\phi,m)$.
At particle level, the mass assigned to each charged particle corresponds to its particle species.
This information is not constrained by the unfolding procedure and is simply taken from the MC simulated samples used to construct the measurement (see Section~\ref{sec:limitations}).
Because the number of particles varies from event to event, the resulting event representation is variable in length.

\subsection{Data and simulated event samples}
Data events are collected using single-muon triggers~\cite{TRIG-2018-01}.
Muons are reconstructed by matching charged-particle tracks in the inner tracking detector and muon spectrometer, accounting for energy loss in the calorimeters~\cite{MUON-2018-03}.
They are required to fulfill \textit{Medium} identification
criteria and \textit{PflowLoose} isolation~\cite{MUON-2018-03},
and must satisfy $|d_{0}|/\sigma_{d_{0}} < 3$ and
$|z_{0}\sin(\theta)| < \SI{0.5}{\milli\meter}$.
The transverse impact parameter $d_0$ is the signed transverse distance from the beamline to the point of closest approach of the muon track, while the longitudinal impact parameter $z_0$ is the difference between the longitudinal coordinate of this
point and that of the reconstructed primary vertex.
These requirements ensure that the selected muons originate from the interaction vertex, i.e.\ the primary vertex with the highest sum of associated track $p_\mathrm{T}^2$~\cite{ATL-PHYS-PUB-2015-026}.
Charged-particle tracks used for jet building are required to satisfy the \textit{Loose} track-quality criteria, the \textit{Tight} track-to-vertex association criterion to suppress tracks originating from pileup interactions~\cite{ATL-PHYS-PUB-2015-051}, and must not be associated with the selected muons.
Identical kinematic requirements are applied to the reconstructed quantities as those used to define the particle-level fiducial volume.

MC simulated event samples are used to provide theoretical predictions at both the reconstructed and particle level, and to perform the unfolding.
Most $Z$+jets events that satisfy the selection originate from the Drell--Yan process.
As in the \multifold{} analysis, the nominal Monte Carlo prediction is provided by~\MGNLO{2.6.5}~\cite{Alwall:2014hca,Frederix:2012ps,PMGR-2021-01} interfaced to \PYTHIA[8.240]~\cite{Sjostrand:2014zea} (denoted \MADGRAPHPY{} in the following), and \SHERPA[2.2.11]~\cite{Bothmann:2019yzt,PMGR-2021-01} provides an alternative.
Both the nominal and alternative Monte Carlo predictions use the \NNPDF[3.0nnlo] parton distribution function (PDF) set~\cite{Ball:2014uwa}.
Electroweak $Z$+jets events are simulated with \herwig~7.2.0~\cite{Bahr:2008pv,Bellm:2015jjp}, interfaced with VBFNLO v3.0.0~\cite{Baglio:2011juf} using the MMHT2014LO PDF set~\cite{Harland-Lang:2014zoa}. Diboson production is generated at next-to-leading-order accuracy with Sherpa 2.2.1 or 2.2.2~\cite{Bothmann:2019yzt}, employing the \NNPDF[3.0nnlo] PDF set~\cite{Ball:2014uwa}. Background contributions from top processes ($t\bar{t}$, $tV$, single top) are modeled by \POWHEG[v2]~\cite{Alioli:2010xd} interfaced to \PYTHIA[8.230], and an alternative description of the \ttbar background is provided by \SHERPA[2.2.1]~\cite{Bothmann:2019yzt}.
Both the nominal and alternative descriptions of the \ttbar background utilize the \NNPDF[3.0nnlo] PDF set~\cite{Ball:2014uwa}.
The detector response is simulated using a dedicated \GEANT{}-based model~\cite{Agostinelli:2002hh} of the ATLAS detector~\cite{SOFT-2010-01}.
Simulated inelastic minimum-bias events are overlaid to model additional $pp$ collisions in the same and neighboring bunch crossings (``pileup'')~\cite{SOFT-2010-01}.
Simulated events are reconstructed using the same procedure as for data.

The analysis measures inclusive $Z$+jets production and makes no attempt to separate the Drell--Yan, diboson, and electroweak processes.
Although the measurement does not explicitly separate these processes, the released particle-level simulated samples contain a per-event label that makes it possible to obtain individual predictions of each of the components.
See Appendix~\ref{app:usage} for a discussion of proper procedures for these use cases.
The resulting inclusive $Z$+jets signal sample consists of approximately 95.3\% Drell--Yan, 2.9\% diboson (primarily $ZV\to\mu\mu jj$), and 1.6\% electroweak $Zjj$ production.
These samples are used as the starting point for the top background subtraction and unfolding procedures described below.


%
\section{Analysis methodology}

This section describes the methodology used to construct the unbinned measurement of the charged-particle final-state phase space. The measurement is obtained through a sequence of steps including background subtraction, iterative unfolding with \omnifold{}, and the assignment of unfolded weights to particle-level MC events.
These steps are described below, along with additional components of the methodology including the event representation, neural-network architecture, pretraining procedure, and generation of the pseudodata sample used in the simulation-based validation.

\subsection{Background subtraction}
\label{sec:bkg_sub}

Top-quark processes are expected to account for approximately 0.25\% of the selected event yield and constitute the only non-negligible background in the fiducial region. Prior to unfolding, this small residual background is removed from the data using an unbinned subtraction procedure. In the \multifold{} measurement of the same process, explicit subtraction of the top-quark background was deemed unnecessary given its sufficiently small contribution. Instead, its estimated size was assigned as a systematic uncertainty. In the present analysis, the expected background is directly subtracted as part of the measurement procedure rather than being treated as a systematic uncertainty, thereby reducing the overall measurement uncertainty.

The subtraction is performed using a neural-network-based reweighting method~\cite{Nachman:2020fff}.
In principle, the reweighting could be derived in the full event phase space used for the unfolding.
However, because the top-quark contribution is extremely small,  it was found that a sufficiently accurate reweighting could not be achieved using the full high-dimensional event representation.
Instead, the subtraction is performed in a reduced space consisting of five observables, including the output of a classifier trained on the full charged-particle phase space to distinguish \ttbar events from $Z$+jets events.
Incorporating this highly sensitive discriminator
improves the \ttbar subtraction throughout
the charged-particle phase space, even in observables that were not explicitly used in the subtraction procedure.
The subtraction procedure assigns an event weight to each data event, yielding a statistically background-subtracted data sample that is subsequently used as input to the \omnifold{} procedure.
See Appendix~\ref{app:bkg_sub} for additional details and results of the subtraction procedure.

\begin{figure}[t]
\centering
\subfloat[]{
\includegraphics[width=0.98\textwidth]{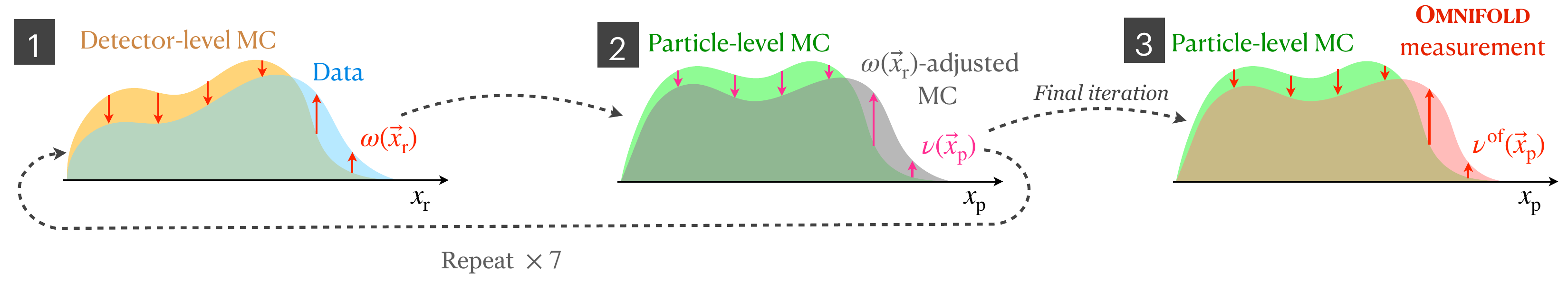}
} \\
\subfloat[]{
\includegraphics[width=0.98\textwidth]{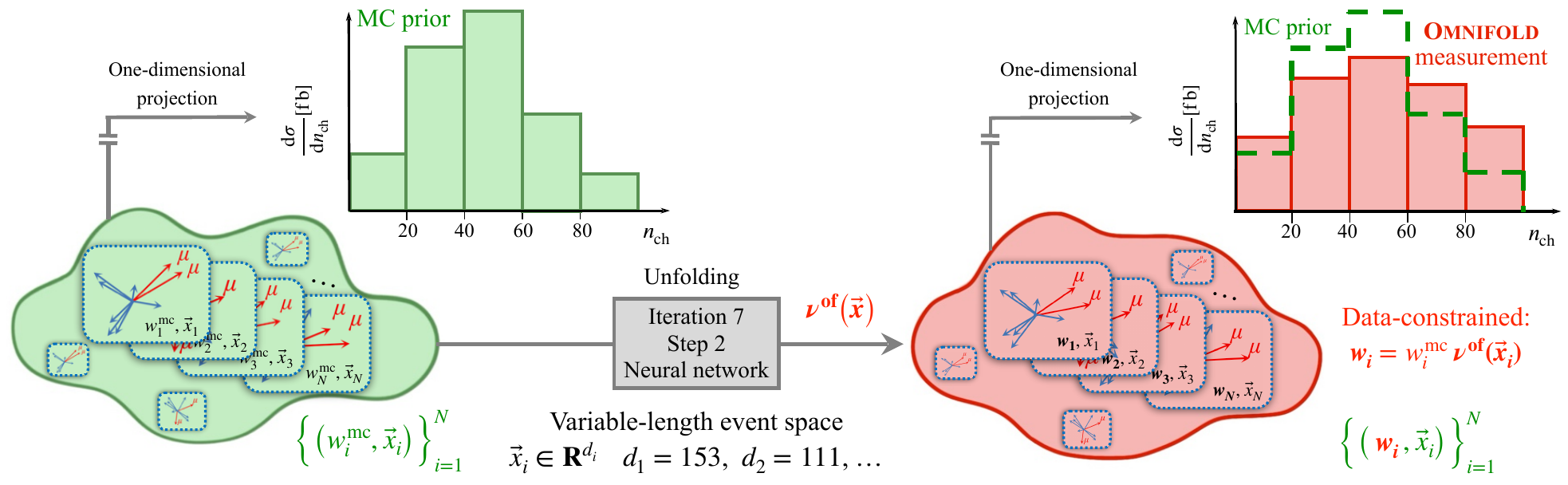}
}
\caption{
Schematic illustrating (a) the \omnifold{} procedure and (b) the variable-dimensional nature of the fully differential charged-particle phase-space measurement.
\omnifold{} starts by deriving a reweighting function \wxr{} of the detector-level MC events \xr\ to match the data distribution (step 1).
This reweighting function is then propagated
from detector level to particle level, and a second reweighting function \wxp\ of the particle-level MC events \xp{} is derived to match the raw particle-level MC distribution to the \wxr{}-adjusted particle-level MC distribution (step 2).
The result of the second step is then propagated
from particle level back to detector level and the process repeats for a total of seven iterations.
The final result of the procedure is the particle-level reweighting function $\nu^\mathrm{ of}(\xp)$ after seven iterations.
This reweighting function is used to produce an event weight for each event in the nominal particle-level MC sample.
These event weights are well defined despite the variable dimensionality because the neural networks are set-processing transformer architectures, allowing observables such as the number of charged hadrons, \nch, to be measured even though they are not explicitly provided to the networks.
The first two events in the published dataset contain 51 and 37 charged particles, corresponding to variable-length event representations with 153 and 111 dimensions, respectively.}
\label{fig:omnifold_schematic}
\end{figure}

\subsection{The \omnifold\ method}

\omnifold{} is an iterative unfolding algorithm that uses machine-learning classifiers to estimate likelihood ratios between the probability distributions describing data and MC simulation.
The final result is a set of weights assigned to particle-level simulated events, which together form an unbinned representation of the unfolded cross section.

Schematics illustrating the \omnifold{} procedure and the variable-length event representation are shown in Figure~\ref{fig:omnifold_schematic}.
One iteration of the procedure estimates the detector-level and particle-level reweighting functions \wxr{} and \wxp{}, which depend on the detector-level and particle-level event representation, \xr{} and \xp{}, respectively. The detector-level function \wxr\ reweights simulated events to reproduce the observed data distribution. These weights are then propagated through the detector simulation to the corresponding particle-level events, which are used to train the particle-level function \wxp. The resulting particle-level weights define the particle-level MC prior for the next iteration.
The procedure is repeated over multiple iterations, yielding the final \omnifold{} weights $\nu^\mathrm{of}(\xp)$.
As in the \multifold{} analysis, the unfolded event weights are subsequently normalized such that their sum reproduces the measured fiducial cross section obtained from the data event count, luminosity, efficiency, and fiducial acceptance. Consequently, the published event weights have units of femtobarns (fb), and the measured cross section corresponding to any particle-level selection is obtained by summing the weights of the selected events. The overall unfolding procedure is an example of an expectation-maximization algorithm and is formally the zero bin-width limit of the commonly used Iterative Bayesian Unfolding (IBU) method~\cite{DAgostini:1994fjx}.

The unfolding is performed using seven iterations.
This choice was determined using the pseudodata validation procedure described in Section~\ref{sec:pseudodata}. The agreement between the unfolded pseudodata measurement and the known particle-level pseudodata target is quantified using a distance metric defined as the sum of one-dimensional Wasserstein distances~\cite{Ramdas:2015} computed over many event-level observables. No significant improvement was observed beyond seven iterations. The unfolding is derived in a slightly enlarged fiducial region ($\ptll>\SI{190}{\GeV}$) to reduce sensitivity to migrations across the fiducial boundary. Once derived, the unfolded weights associated with each particle-level event are fixed and can be reused to construct arbitrary observables without retraining or rerunning the unfolding procedure.

The published measurement is constructed from a statistically independent test subset of the nominal MC prediction.
The effective event yield of this test set is equal to approximately 2.5 times that of the data.
This test subset is used to define the published measurement and does not participate in the neural-network training or in the determination of the final reweighting function $\nu^\mathrm{of}(\xp)$.

\subsection{Variable-length event representation and neural-network architecture}
\label{sec:nn_arch}

Unlike the \multifold{} analysis, where the unfolded weights define cross sections only in a fixed set of 24 observables, the present measurement assigns weights directly to particle-level event configurations and therefore provides a measurement of the full charged-particle final-state phase space.
This is achieved by replacing the multi-layer perceptron architectures used by the \multifold{} analysis with Particle Transformer (ParT) networks~\cite{Qu:2022mxj}, which can operate directly on variable-length particle clouds.
Each event is represented as a variable-length particle cloud consisting of the four-vectors
($\pt, \eta, \phi, m$) of all charged particles.
At both detector and particle level, the mass of the charged particles is set to the pion mass, except for the muons produced in the  $Z$ boson decay, which are assigned the muon mass.
This representation differs from that used in defining the fiducial volume in Section~\ref{sec:fiducial} and used in subsequent analyses of the measurement in Section~\ref{sec:result}, where all charged particles are assigned their particle-level masses.
This design choice was made to prevent the reweighting produced by \omnifold{} from altering the fractions of particle-level hadron identities, which are not constrained at detector level.
Instead, the measurement assumes the particle-level hadron fractions predicted by the \MADGRAPHPY{} generator.

In addition to the particle four-vectors, each particle is assigned the index of the jet to which it belongs.
These jet indices are calculated by clustering all particles with the \akt algorithm with a radius parameter of $R=0.4$.
Particles associated with jets beyond the fifth-highest-\pt{} jet are grouped into a single additional category.
These six classes of particles are one-hot encoded and provided as additional features to the network.
These auxiliary features are fully determined by the three-vectors ($\pt, \eta, \phi$) of all particles, but were found to improve the accuracy of the likelihood ratios predicted by the neural networks.
Events passing the fiducial selection contain, on average, about 50 charged particles, implying that the unfolding is performed simultaneously in roughly $3\times 50 = 150$ kinematic coordinates.
Charged-particle multiplicities can exceed 250, corresponding to more than 750 simultaneous kinematic coordinates.

\subsection{Pretraining and optimization}
\label{sec:training}
About \num{247000} data events pass the detector-level selection; as a result, the first-stage classification problems within \omnifold{} are strongly statistics limited. To address this limitation, a large-scale pretraining procedure is employed in which neural networks are trained from scratch on a large sample of simulated collision events, then fine-tuned on the density ratio estimation tasks within the \omnifold{} procedure.
Specifically, networks are pretrained to distinguish Drell--Yan events generated with \MADGRAPHPY{} from those generated with \SHERPA.
Dedicated subsets of the simulated event samples from both generators are used for pretraining and not used in the following \omnifold{} procedure.
In total, approximately 14 million events are used in the pretraining.
Ten pretrained networks are trained independently from scratch for both the detector-level and particle-level phase spaces.\footnote{Ten independent pretrained networks are used to reduce correlations arising from repeatedly initializing the training from the same pretrained network. Empirically, ten independent pretrainings were found to be sufficient for this purpose.} These pretrained networks are subsequently used as warm-start checkpoints for the neural-network training within \omnifold{}, with one of the ten checkpoints corresponding to the appropriate phase space selected at random for each training.
The pretraining procedure was found to substantially reduce both the method bias observed in pseudodata studies and the variance between independent unfolding runs.
It was an essential addition to the \omnifold{} procedure that enabled an accurate unfolding in an $\mathcal{O}(100)$-dimensional phase space.
This finding motivates additional development of large-scale pretrained \textit{foundation models} for particle physics data~\cite{Mikuni:2024qsr,Bhimji:2025isp,Bardhan:2025icr,Leigh:2024ked,Golling:2024abg,Harris:2024sra,Hallin:2025ywf,Birk:2024knn,Hsu:2026sww,Birk:2025fbs}, which could be valuable tools for additional AI-based unfolding and other simulation-based inference applications~\cite{SOFT-2023-01,HIGP-2024-14} in the future.

The ParT networks are implemented in PyTorch~\cite{paszke2019pytorch} and trained on the Perlmutter supercomputer at NERSC using between 16 and 64 NVIDIA A100 GPUs.
A ParT model with six attention blocks, one class attention block, and a dropout probability of 0.02 applied to all attention weights provided the best performance in the validation studies.
These networks contain approximately 1.6 million trainable parameters and were trained using the Lion optimizer~\cite{chen2023symbolic} with a weight decay parameter of 0.01 and a cosine-annealing learning-rate schedule.
The network architecture and training hyperparameters were determined through an extensive random hyperparameter search. Details of the hyperparameter optimization procedure and the final hyperparameter choices are provided in Appendix~\ref{app:hpo}.

\subsection{Construction of pseudodata event samples}
\label{sec:pseudodata}

The primary validation of the measurement is performed using \textit{pseudodata} generated from statistically independent simulated event samples. The dominant Drell--Yan contribution, comprising approximately 95\% of the selected events, is modeled using \SHERPA{} rather than the nominal \MADGRAPHPY{} prediction, thereby providing an independent description of the hard scattering, parton shower, and hadronization. The smaller electroweak $Zjj$ and diboson contributions are generated using the same event generators as in the nominal prediction, but with statistically independent simulated event samples.

To avoid potential circularity in the validation, the pseudodata are not constructed using neural-network-based reweighting methods. Instead, the detector-level simulated events are reweighted to resemble the observed data using an independent sequence of Gaussian-kernel reweighting functions. These smooth reweighting functions are derived sequentially from one-dimensional projections of event-level observables, including the 24 observables measured by the \multifold{} analysis, together with additional observables that constrain soft hadronic activity, such as charged-hadron multiplicities in different regions of transverse momentum and pseudorapidity, and the scalar sum of the transverse momenta of all charged particles in each event. The reweighting functions are derived and applied iteratively until good agreement with the data is achieved across this broad set of observables.

The reweighted detector-level events are statistically downsampled to match the observed event count in data and used as pseudodata. The corresponding particle-level events, which retain their full statistical precision and have an effective event yield approximately 12 times that of the data, provide the particle-level target against which the unfolded measurement is compared.
These pseudodata samples form the basis of the validation studies described in Section~\ref{sec:validation}.

\subsection{Considerations for using the measurement}
\label{sec:limitations}

Several considerations should be kept in mind when using the unbinned measurement. First, a measurement of a given observable should be validated using the pseudodata measurement before drawing any conclusion based on the measurement with real data.
This is particularly important because it is possible to define observables using the particle-level phase space that are not constrained by the measurement, for example those sensitive to the particle-level hadron fractions predicted by the \MADGRAPHPY{} generator.
Second, the unfolding process assumes that a sufficient number of events fall in each bin of the particle-level histogram, so a binning should be chosen that ensures at least \num{5000} effective events are in each bin.
Finally, all events are guaranteed to contain two muons given the event selection, but for a small fraction of events, most charged-hadron tracks fall outside the $|\eta| < 2.5$ acceptance, resulting in minimal hadronic activity in the fiducial volume.
When studying jet observables it is advisable to remove these events with a small minimum charge-hadron-jet \pt\ requirement.
These considerations are discussed in more detail in the documentation accompanying the public codebase and in Appendix~\ref{app:usage}, which provide practical recommendations for using the measurement.


%
\section{Uncertainties and validation}
\label{sec:uncerts}

The unbinned measurement is defined by a nominal set of particle-level events and their associated unfolded weights.
Uncertainties are propagated through the full analysis chain and represented by alternative realizations of the measurement, corresponding to alternative sets of unfolded particle-level event weights.
These alternative weight sets can be projected onto arbitrary observables and combined to construct covariance matrices in the same manner as for conventional differential cross-section measurements.

In total, the measurement is accompanied by 162 weight variations that quantify its uncertainties. Twenty-seven weight variations encode independent systematic uncertainty sources. The remaining 135 weight variations encode stochastic uncertainties arising from the finite statistical precision of the data and simulated event samples. These are evaluated using bootstrap replicas~\cite{ATL-PHYS-PUB-2021-011}.

\subsection{Experimental and modeling uncertainties}
\label{sec:experimental_modeling_uncertainties}

Experimental sources of uncertainty include systematic bias due to the muon efficiency and calibration~\cite{MUON-2018-03}, track reconstruction~\cite{ATL-PHYS-PUB-2015-051}, pileup modeling, and the luminosity measurement~\cite{DAPR-2021-01}.
Theoretical uncertainties in the Drell--Yan component of the MC are evaluated for variations of PDF and $\alpha_s$ choices~\cite{PMGR-2021-01}, QCD scales~\cite{PMGR-2021-01}, and the generator tune~\cite{STDM-2012-23}.
To approximate the expected impact of theoretical uncertainties on the small contributions from diboson and electroweak $Zjj$ processes, flat yield variations of 30\% and 20\% are applied to the predicted yield of each source, respectively,
corresponding to conservative estimates of the level of agreement to data as observed in Refs.~\cite{ATL-PHYS-PUB-2016-002} and~\cite{STDM-2013-02}.
The uncertainty associated with the top-background subtraction is evaluated by repeating the subtraction procedure using the alternative \SHERPA{} \ttbar prediction in place of the nominal model.
The resulting alternative subtraction weights are propagated through the full \omnifold{} procedure, and the difference relative to the nominal measurement is assigned as an uncertainty.

The relative importance of the individual uncertainty sources depends on the observable under consideration. For observables primarily sensitive to the muon kinematics, the dominant experimental uncertainty arises from the muon isolation and reconstruction efficiencies. In contrast, for observables that depend strongly on charged-particle kinematics, such as the jet kinematic observables, the charged-hadron multiplicity ($n_\mathrm{ch}$), and the scalar sum of charged-hadron transverse momenta ($H_\mathrm{T}$), the dominant experimental uncertainty is typically associated with the track reconstruction efficiency. Across most measured observables, the experimental uncertainties exceed the theoretical modeling uncertainties, with the latter generally contributing at a level comparable to or smaller than the luminosity uncertainty~\cite{DAPR-2021-01}.
The remaining uncertainty sources are associated with the unfolding procedure itself and are discussed in the following subsection.

\subsection{Unfolding uncertainties}

In addition to the experimental and theoretical uncertainties described above, several uncertainty sources associated with the unfolding procedure itself are evaluated: training stochasticity, data-driven finite-iteration uncertainty, hadron-composition dependence, and hidden-variable dependence.

Neural-network optimization is inherently stochastic and yields slightly different unfolded weights for different training realizations. This uncertainty is estimated by repeating the \omnifold{} procedure with identical inputs but different random seeds governing the training and validation splits, as well as minibatch construction. In contrast to Ref.~\cite{STDM-2020-17}, ten repetitions are used instead of one hundred because the pretraining procedure described in Section~\ref{sec:nn_arch} substantially reduces the variance between independent training realizations while maintaining the accuracy of the unfolded results. The resulting spread in the unfolded weights is assigned as the stochastic training uncertainty.

Like IBU, \omnifold{} returns a biased estimator of the true particle-level distribution. The residual bias arising from the finite number of iterations is evaluated using a data-driven procedure.
The \MADGRAPHPY{} sample used for the nominal measurement is reweighted at particle level such that the corresponding detector-level distributions reproduce the observed data.
To ensure independence from the neural-network-based reweighting used by \omnifold{}, the same Gaussian-kernel-based reweighting procedure described in Section~\ref{sec:pseudodata} is used.
The resulting reweighted sample is then treated as pseudodata and unfolded using the nominal procedure.
The difference between the unfolded result and the known particle-level target distribution is assigned as the data-driven finite-iteration uncertainty.

The unfolded result can also be biased by latent features in the simulation of the detector response that are not included as input variables.
The most significant of these \textit{hidden variables} are the particle-level hadron fractions predicted by the MC sample used to derive the unfolding.
Measurements of the kinematic properties of charged hadrons have different biases and resolutions depending on the hadron type, so changing the charged hadron composition will change the detector response.
To assess potential biases in the unfolded results arising from differences in hadron composition, the hadron content of the nominal \MADGRAPHPY{} sample is reweighted at particle level to match that of the \SHERPA{} sample.
Four hadron categories are defined for this procedure: pions, kaons, protons, and an `other' category that includes any charged hadron that is not identified as a pion, kaon, or proton.
The majority of hadrons in this category are strange baryons, predominantly $\Omega$, $\Xi$, and $\Sigma$ baryons.
The reweighting is performed to match several key distributions within each category to the \SHERPA{} sample.
These include the distribution of the hadron multiplicity fraction, the distribution of the fraction of jet momentum carried by each hadron category in the leading and subleading jets, and the distribution of the fraction of the total event transverse momentum carried by each hadron category when excluding tracks inside the leading and subleading jets.
The unfolding is then repeated with the reweighted \MADGRAPHPY{} sample, and the difference from the nominal result is taken as the hadron composition uncertainty.

Even a highly differential unfolding may remain sensitive to latent detector-response variables that are not explicitly included in the event representation.
To estimate the impact of such effects, the unfolding is repeated with the alternative \SHERPA{} sample used as the MC sample, after it is reweighted using the same Gaussian-kernel-based methodology described in Section~\ref{sec:pseudodata} to match the \MADGRAPHPY{} sample at particle level.
The difference between the resulting unfolded spectra and the nominal results derived with the \MADGRAPHPY{} sample is taken as the hidden variable uncertainty.\footnote{This uncertainty is assessed using independent simulated events from those used for the nominal results. This emphasizes that although the measurement is unbinned, it is not defined on an event-by-event basis since there is no way to assess the hidden variable uncertainty on a single event from the \MADGRAPHPY{} sample.}

The \omnifold{} measurement is considerably more differential than conventional binned unfolding methods and is therefore, in principle, sensitive to fewer hidden detector-response variables. Nevertheless, important latent variables remain. In particular, the particle-level hadron composition is not constrained by the detector-level event representation and therefore cannot be inferred directly from the observed data. Consequently, even a measurement of the fully differential charged-particle phase space retains a non-negligible hidden-variable uncertainty.

Compared with IBU, the hidden-variable uncertainty can be slightly larger for \omnifold{}, particularly in low-statistics regions, owing both to these remaining latent variables and to the finite accuracy of the neural-network likelihood-ratio estimates. Comparisons of the total uncertainty and hidden-variable uncertainty for the IBU, \multifold{}, and \omnifold{} measurements are provided in Appendices~\ref{app:uncert_comp} and~\ref{app:hv_uncert_comp}, respectively. For the 24 observables measured in the \multifold{} analysis, the unfolding uncertainty of the \omnifold{} measurement is generally comparable in magnitude to the leading experimental uncertainties across much of the fiducial phase space, although it can become significantly larger in low-statistics regions. It therefore constitutes one of the dominant contributions to the overall uncertainty budget.

\subsection{Validation}
\label{sec:validation}

The measurement is validated in multiple ways, with the resulting recommendations for downstream use of the published measurement summarized in Appendix~\ref{app:usage}.
The primary validation is based on an auxiliary measurement in which the pseudodata described in Section~\ref{sec:pseudodata} are treated as data and unfolded.
The resulting pseudomeasurement is compared with the known particle-level pseudodata target using binned $\chi^2$ goodness-of-fit tests across a broad range of observables, accounting for all sources of uncertainty and their bin-to-bin correlations through the full covariance matrix.

\begin{figure}[b]
\centering
\subfloat[]{
\includegraphics[width=0.45\textwidth]{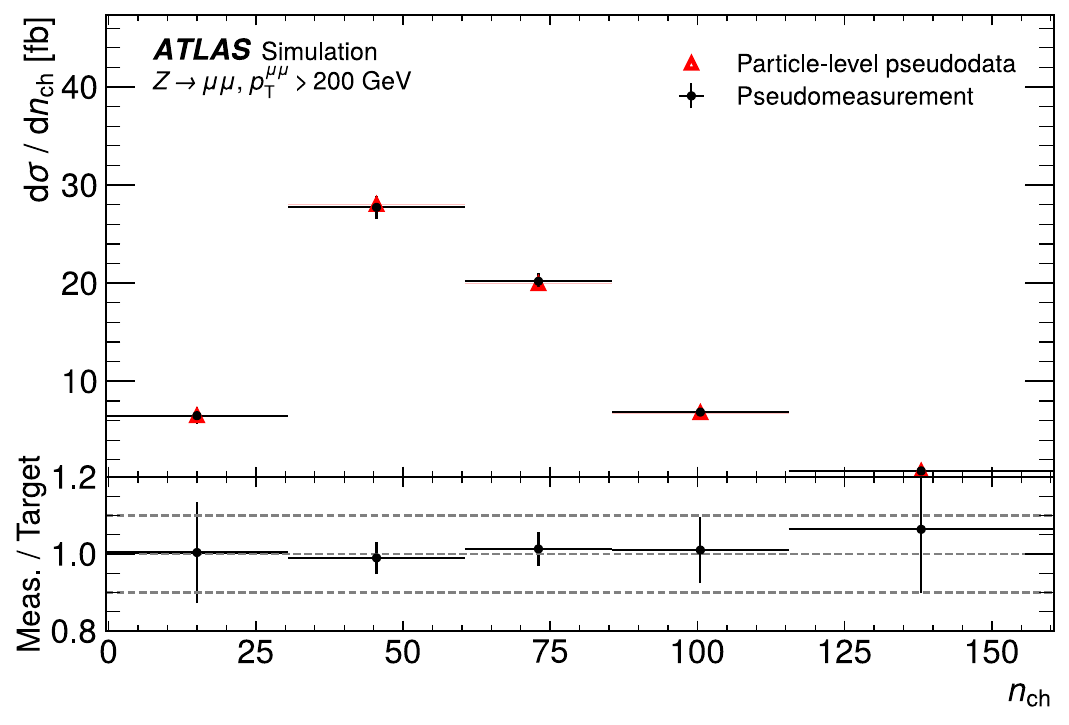}
\label{fig:Ntracks_pseudodata}
}
\subfloat[]{
\raisebox{-0.5mm}{\includegraphics[width=0.44\textwidth, height=0.223\textheight, keepaspectratio]{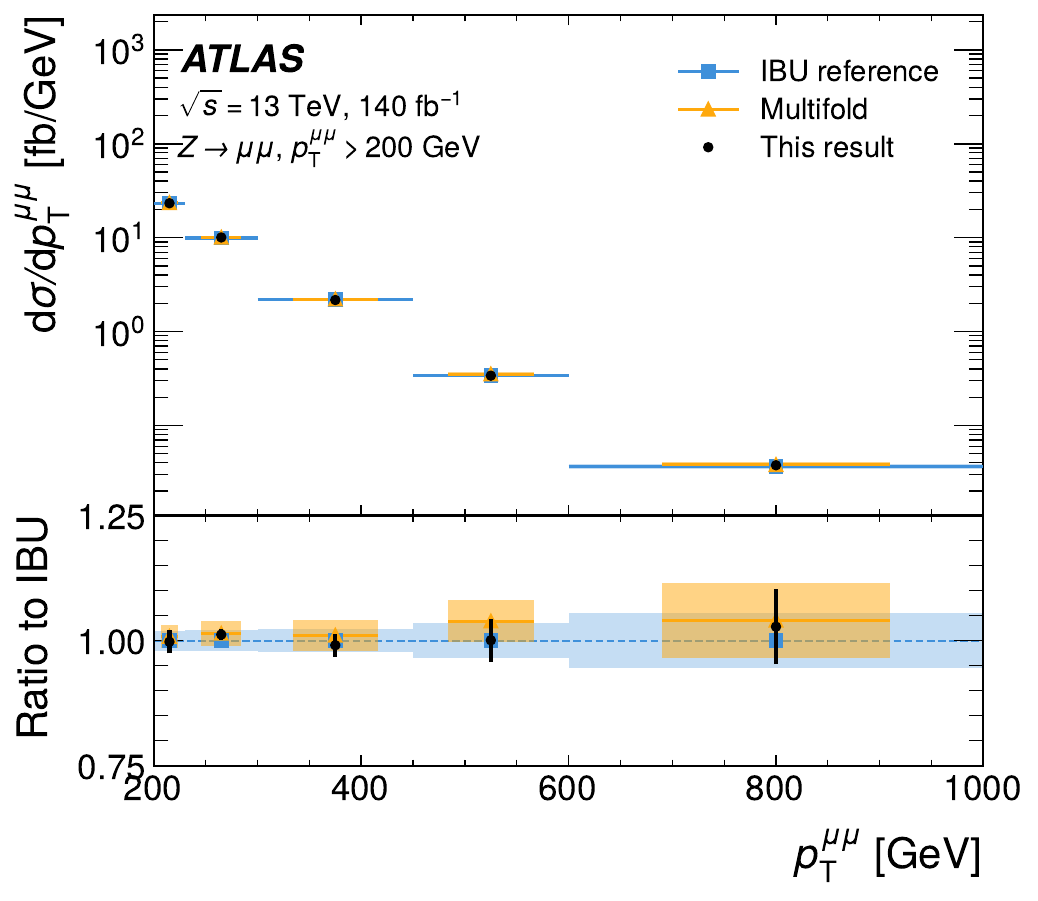}
}
\label{fig:cross_section_comparison_pt_ll}
}
\caption{
Representative validation results. (a) Comparison of the unfolded pseudomeasurement (points) with the known particle-level target (triangles) for the charged-hadron multiplicity, \nch{}. The corresponding goodness-of-fit test yields $\chi^2=2.41$ with a $p$-value of 0.790. (b) Comparison of the differential cross section,
$\mathrm{d}\sigma/\mathrm{d}\ptmm$,
measured using IBU, \multifold{} from Ref.~\cite{STDM-2020-17}, and \omnifold{} (denoted ``This result'').
}
\label{fig:validation}
\end{figure}

Figure~\ref{fig:Ntracks_pseudodata} shows a representative example of this validation for \nch{}, one of the new observables constructed from the variable-length event representation and presented in Section~\ref{sec:event_level_observables}.
Good agreement with the known particle-level pseudodata target is observed across the distribution.
Appendix~\ref{app:pseudodata} provides further details of the statistical methodology and summarizes the resulting $\chi^2$ values, numbers of degrees of freedom, and p-values for the 24 observables measured in Ref.~\cite{STDM-2020-17}, together with two additional observables, \nch{} and $H_\text{T}$, presented in Section~\ref{sec:event_level_observables}.
The resulting p-values span a broad range and show no systematic indication of disagreement between the pseudomeasurement and the particle-level pseudodata target.
The appendix also presents the full pseudodata results, including uncertainty breakdowns and correlation matrices, for these two additional observables.
The performance of the method was further studied using pseudodata for a broader set of projections, including two-dimensional distributions and measurements in restricted kinematic regions, such as diboson- and vector-boson-fusion-enhanced regions.

An additional pseudodata validation targets regions in which the nominal \MADGRAPHPY{} sample predicts a much lower cross section than the \SHERPA{} sample used to construct the pseudodata. A ParT-based classifier is trained to distinguish particle-level Drell--Yan events generated by \MADGRAPHPY{} and \SHERPA{} using the event representation described in Section~\ref{sec:nn_arch}. Since the classifier is optimized to identify differences between the two generator distributions, its output provides a sensitive observable for probing regions where the particle-level densities predicted by the two generators differ substantially.

The high-\SHERPA{}-probability tail of the classifier output is binned following the statistical requirements of Appendix~\ref{app:usage}. In this region, the \MADGRAPHPY{}-based nominal event density is approximately an order of magnitude smaller than that of the \SHERPA{}-based pseudodata target. Nevertheless, the unfolded pseudomeasurement is consistent with the known particle-level pseudodata target, with a $\chi^2$ test yielding a p-value of 0.863. Further details are provided in Appendix~\ref{app:pseudodata}. The unfolded event-weight distribution provides a complementary indication of the magnitude of the reweighting required in the measurement: only 0.01\% of events have dimensionless weights $\nu^\mathrm{of}$ greater than 50.

Additional validation is provided through comparisons with corresponding IBU-based measurements and the published \multifold{} measurement~\cite{STDM-2020-17}.
Figure~\ref{fig:cross_section_comparison_pt_ll} shows a representative comparison of the differential cross section measured using IBU, \multifold{}, and \omnifold{} for the transverse momentum of the dimuon system, \ptll.
The three measurements are consistent within their uncertainties across the full kinematic range.
More extensive comparisons across the 24 observables measured in Ref.~\cite{STDM-2020-17}, together with the two additional observables presented in this paper, are provided in Appendix~\ref{app:comparison}.
Together, these studies motivate the usage recommendations in Appendix~\ref{app:usage}, which are followed for the results presented in the next section.


%
\section{Results}
\label{sec:result}

The published measurement~\cite{cern_open_data_portal,gitlab} constitutes the primary result of this work.
The trained networks produced by the \omnifold{} procedure are also released publicly, allowing appropriate unfolded event weights to be calculated for arbitrary samples generated with \MADGRAPHPY{} and \SHERPA{}.

Because the published measurement is unbinned, high-dimensional, and supports a broad range of downstream applications, it is neither practical nor desirable to present an exhaustive set of projected observables in this paper.
Instead, four representative applications are presented to illustrate how the measurement can be used to construct observables spanning a wide range of physics topics, including event-level kinematics, jet substructure, energy-flow observables, and machine-learning-inspired characterizations of jet structure.
The examples are arranged from conventional observables commonly measured in collider experiments to observables that are difficult or impossible to access using traditional unfolding techniques.

In all figures presented in this section, the data are shown with the full measurement uncertainties described in Section~\ref{sec:uncerts}. The measurements are compared with predictions for the sum of the Drell--Yan, electroweak $Zjj$ and diboson contributions. The dominant Drell--Yan component is modeled using either the nominal \MADGRAPHPY{} prediction or the alternative \SHERPA{} prediction, while the electroweak $Zjj$ and diboson contributions are modeled using \HERWIG[7] and \SHERPA{}, respectively. The uncertainties on the predictions include the experimental and modeling uncertainties described in Section~\ref{sec:experimental_modeling_uncertainties}, with the dominant theoretical contributions arising from variations of the QCD scales and the generator tune.


\subsection{Event-level observables}
\label{sec:event_level_observables}

The most direct application of the full charged-particle phase-space measurement is the construction of conventional differential cross-section measurements by projecting the event-level representation onto observables of interest.
Though such measurements can in principle be performed using traditional binned unfolding techniques, the event-level representation provides considerably greater flexibility for downstream analyses.
Because the measurement is unbinned, additional event selections can be imposed after publication to define measurements in restricted regions of the fiducial phase space, provided sufficient statistical precision is maintained.
Similarly, the binning of any observable can be modified on-the-fly to suit downstream applications such as Monte Carlo generator tuning.
More generally, any observable that can be constructed from the unfolded muon and charged-particle kinematics is implicitly measured.
This includes observables that were not considered during the analysis design and were not used as inputs to the unfolding procedure.
The reinterpretation potential of the measurement therefore extends the long-term physics value of the ATLAS Run~2 dataset by enabling future measurements that do not require access to the detector simulation.

\begin{figure}[t]
\centering
\subfloat[]{
\includegraphics[width=0.48\textwidth]{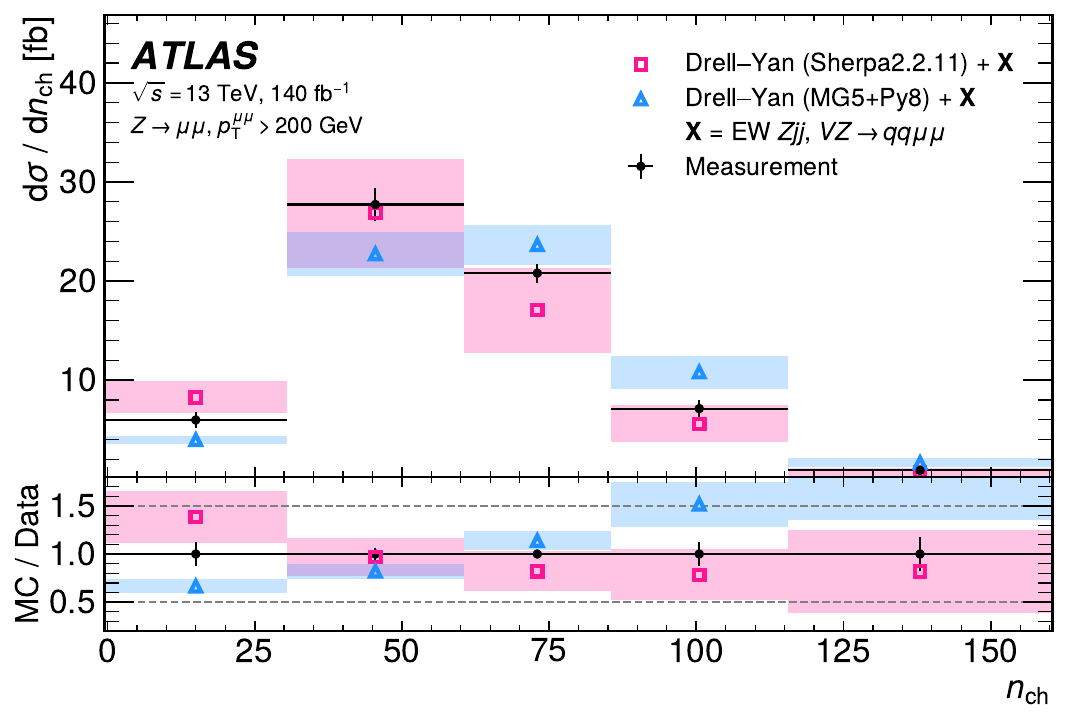}
}
\subfloat[]{
\includegraphics[width=0.49\textwidth]{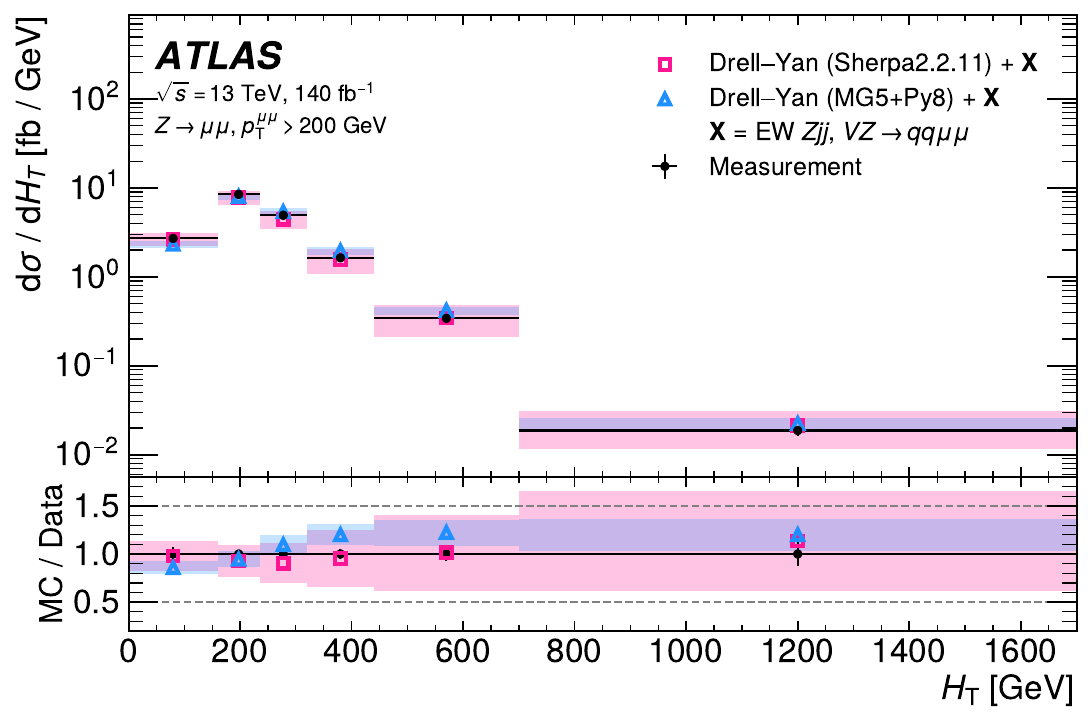}
}
\caption{Measurements of (a) the number of charged hadrons \nch{} and (b) the scalar sum of the transverse momenta of all charged hadrons, \HT{}. The measurement is shown as points, and particle-level predictions employing \MADGRAPHPY{} or \SHERPA{} for the Drell--Yan contribution are shown as triangles and squares, respectively. Uncertainties on the measurements are shown as error bars and uncertainties on the predictions are shown as shaded bands.}
\label{fig:event_level_observables}
\end{figure}

As representative examples, measurements of \nch and \HT are shown in Figure~\ref{fig:event_level_observables}.
The muons from the $Z$ boson decay are not included in either observable.
Neither \nch{} nor \HT{} enters the unfolding procedure directly; both are obtained entirely as projections of the published event-level representation, providing an independent demonstration of the predictive power of the published measurement.
Validation studies using pseudodata are presented in Appendix~\ref{app:pseudodata}, including full uncertainty breakdowns and correlation matrices.

The uncertainty bands associated with generator predictions reflect the modeling uncertainties described in Section~\ref{sec:experimental_modeling_uncertainties} and are dominated by the contribution arising from QCD scale variations in the Drell--Yan components. Both generator predictions exhibit significant discrepancies with the measured distributions. The largest discrepancies occur in the \nch{} distribution, where the simulation employing \MADGRAPHPY{} (\SHERPA{}) for the Drell--Yan contribution predicts systematically too many (few) charged particles, with deviations reaching approximately 50\% in some regions of phase space.
The \HT{} distribution is generally better described, although noticeable shape differences remain, particularly for the \MADGRAPHPY{}-based prediction.
These results demonstrate the ability of the measurement to constrain poorly modeled aspects of hadronic final states and provide valuable input for future tuning of Monte Carlo event generators.

\subsection{Jet-shape observables}
\label{sec:jet_shape_observables}

Jet-substructure observables provide sensitive probes of QCD radiation patterns and are widely used to study the internal structure of quark and gluon jets.
In conventional measurements~\cite{STDM-2010-10,CMS-TOP-17-013,TOPQ-2019-22}, jet observables are unfolded for a fixed jet definition, making systematic comparisons between clustering algorithms or radius parameters impractical. As the \omnifold{} measurement is differential in the kinematics of individual charged particles, jets can be built directly from the published event-level unfolded data using arbitrary clustering algorithms and parameters.
This enables measurements of jet observables for many jet definitions without repeating the unfolding procedure.

\begin{figure}
\centering
\subfloat[]{
\includegraphics[width=0.48\textwidth]{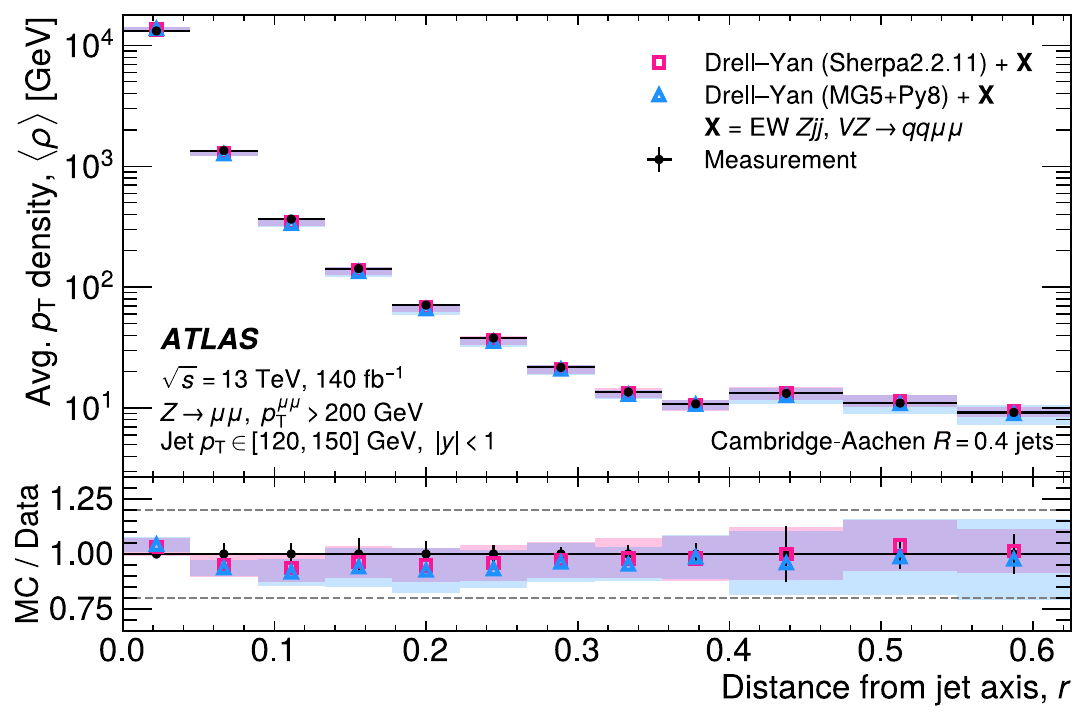}
\label{fig:jet_shape_gen}
}
\subfloat[]{
\includegraphics[width=0.48\textwidth]{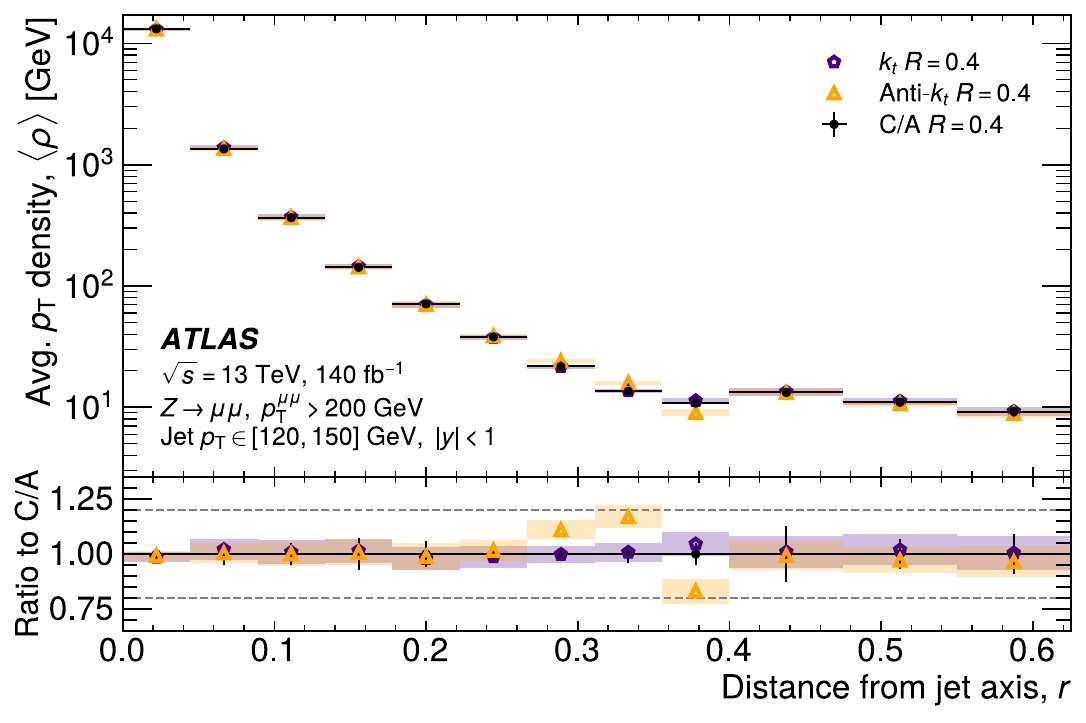}
\label{fig:jet_shape_alg}
}
\caption{
Measurements of the average transverse-momentum density $\langle \rho(r) \rangle$ as a function of distance from the jet axis.
(a): Comparison of Cambridge--Aachen $R=0.4$ jets between the unfolded measurement (points) and particle-level predictions with the Drell--Yan contribution modeled using \MADGRAPHPY{} or \SHERPA{} (triangles and squares, respectively). Uncertainties on the measurement are shown as error bars and uncertainties on the generator predictions are shown as shaded bands.
(b): Comparison of $\langle \rho(r) \rangle$ measured using jets clustered from the unfolded charged hadrons with three jet algorithms, all with a radius parameter of $R=0.4$: \kt (pentagons), \akt (triangles), and \ca (points). Uncertainties on the \ca measurements are shown as error bars while uncertainties on \kt and \akt measurements are shown as shaded bands.}
\label{fig:jet_shape}
\end{figure}

As an example, Figure~\ref{fig:jet_shape} shows a measurement of the average transverse-momentum density of charged hadrons as a function of distance from the jet axis, $\langle\rho(r)\rangle$. Following Refs.~\cite{Cacciari:2007fd,Cacciari:2008gn}, the transverse-momentum density is defined by
\begin{equation}
\langle\rho(r_i)\rangle =
\left\langle
\frac{p_{\mathrm T,i}}{A_i}
\right\rangle_{\mathrm{jets}},
\label{eq:jet_shape}
\end{equation}
where $p_{\mathrm T,i}$ is the magnitude of the vector sum of the transverse momenta of all charged hadrons contained within a region in $(y,\phi)$-space with area $A_i$.
The quantity $r$ describes the angular distance to the jet axis in $\Delta y\times\Delta \phi$ space, and each $i$ denotes an annulus region spanning a radial interval $\Delta r$ around $r_i$. A given bin shown in Figure~\ref{fig:jet_shape} covers the range $r \in [r_i-\Delta r,r_i+\Delta r]$ with an annulus area of $A_i=\pi(r_i+\Delta r)^2-\pi(r_i-\Delta r)^2$. %
The angle brackets in Equation~\ref{eq:jet_shape} denote the weighted average over the selected jet sample. Because the observable is defined as a weighted average over jets, it is insensitive to the overall normalization of the measured cross section. Only jets with $\pt \in [120,150]$~\GeV and $|y|<1$ are considered, as this interval provides a reasonably narrow selection of central jets located in the falling region of the leading jet $\pt$ distribution while retaining a large number of events.

Figure~\ref{fig:jet_shape_gen} illustrates a measurement of the average \pt density of \ca $R=0.4$ jets compared with theory predictions from the \MADGRAPHPY{} and \SHERPA{} samples, shown as the triangular and square data points, respectively.
Both generators tend to predict a higher transverse-momentum density close to the jet axis ($r<0.04$) and a lower density at intermediate radii ($0.04<r<0.3$), suggesting that the simulated jets are somewhat narrower than those observed in data.
Figure~\ref{fig:jet_shape_alg} shows measurements of the same observable but using jets clustered by three different jet algorithms using the FastJet package~\cite{Fastjet}.
While the \ca and \kt jet algorithms produce similar results, the \akt algorithm produces a large increase and subsequent decrease in \pt density relative to the other two algorithms near the jet boundary: $r\approx R$.
Although only results for $R=0.4$ are shown here, the same qualitative behavior is observed for other jet-radius parameters that were studied.
This behavior can be understood as a consequence of the circular and greedy nature of the \akt{} algorithm, which prioritizes high-\pt constituents during clustering.
In the presence of a relatively uniform soft background, upward
fluctuations near the jet boundary can influence the jet axis position,
leading to an enhancement of the measured density just inside the nominal
jet radius and a corresponding depletion near the boundary.
This effect is less pronounced in jets built with the \ca and \kt algorithms as the former relies on purely angular ordering, while the latter constructs jets beginning at low transverse momenta and so is not greedy.
The corresponding enhancement just outside the nominal jet radius is a consequence of the same effect and is present for all three clustering algorithms in both the unfolded measurement and the simulations.
Importantly, all three measurements shown in Figure~\ref{fig:jet_shape_alg} are derived from the same published event-level representation and require neither additional detector calibration nor repeated unfolding.

Historically, jet-shape observables measured in $Z+$jets events have attracted particular interest because the selected jets are predominantly quark initiated~\cite{CMS2022QuarkGluon}.
The ability to reconstruct jets directly from the unfolded particle-level event representation provides new opportunities for studying quark-jet properties at high \pt.
More generally, additional selections based on observables constructed from the unfolded event representation can be applied a posteriori, enabling measurements in specialized phase-space regions without repeating the unfolding procedure. Examples include measurements in quark- and gluon-enriched jet samples, studies of alternative jet definitions, and investigations of novel jet-substructure observables that were not envisioned when the measurement was performed.
%

%
%
%
%
%


\subsection{Energy correlator observables}
\label{sec:ecc}

Energy correlators are a class of observables that quantify energy-weighted correlations in particle collisions~\cite{Moult:2025nhu}. They can be calculated to high precision in perturbative QCD~\cite{Jaarsma:2025tck}, are sensitive to the strong coupling constant $\alpha_s$~\cite{Chen:2023zlx,CMS-SMP-22-015}, and encode fundamental aspects of QCD radiation~\cite{Lee:2022uwt}. The simplest and most widely studied example is the two-point energy correlator (EEC), which measures the correlation between pairs of particles as a function of their opening angle.
It is defined as
\begin{equation}
\mathrm{EEC}(z) = \frac{1}{\sigma} \sum_{i \neq j} \int \mathrm{d}\sigma \frac{E_{\mathrm{T},i}\, E_{\mathrm{T},j}}{Q_\mathrm{T}^2}\, \delta\left(z - \frac{1-\cos\theta_{ij}}{2}\right).
\label{eqn:eec}
\end{equation}
Here $E_{\mathrm{T},i}$ and $E_{\mathrm{T},j}$ are the transverse energies of the two objects, $z$ is the angular separation observable $z = (1 - \cos \theta_{ij}) / 2$ with $\theta_{ij}$ the opening angle between the momentum vectors of the two objects, $Q_\mathrm{T}$ is the sum of the transverse energies of all objects included in the sum, and $\sigma$ is the integrated cross section used for normalization, with $\mathrm{d}\sigma$ denoting the corresponding differential cross-section element.

Many measurements of this observable were performed in which the input objects are hadronic jets produced in electron-positron~\cite{SLD:1994idb,DELPHI:1990sof,L3:1991qlf,ALEPH:1990vew,OPAL:1990reb} and proton--proton collisions~\cite{STDM-2014-03,STDM-2016-10,STDM-2018-51}.
Recently, measurements using the charged constituents within hadronic jets drove interest in these observables by accessing a highly collinear regime which is sensitive to the transition between perturbative and non-perturbative QCD~\cite{Komiske:2022enw}.
Measurements of the two-point energy correlator using the charged particles within jets were performed in electron-positron~\cite{Electron-PositronAlliance:2019cpi,Electron-PositronAlliance:2025fhk}, proton--proton~\cite{CMS-SMP-22-015,STAR:2025jut}, and Pb-Pb collisions~\cite{CMS:2024ovv,ALICE:2024dfl}.

The EEC corresponds to an energy-weighted statistical moment of the pair distribution as a function of the opening angle. Unfolding the EEC with conventional binned techniques requires significant approximations because the observable is constructed from pairwise combinations of particles after detector reconstruction. In contrast, the published event-level representation allows the EEC to be constructed directly from the unfolded charged-particle kinematics, avoiding such approximations.
This paper utilizes the fully differential charged-particle phase space measurement to measure the EEC with all final state charged particles without such approximations.
A further advantage of using all charged particles in the event, rather than only those assigned to reconstructed jets, is that the measurement extends over the full angular range of $z$ without truncation at the jet radius.
This is the first measurement of the EEC using all charged particles in the final state in proton--proton collisions.

\begin{figure}[t]
\centering
\includegraphics[width=0.6\textwidth]{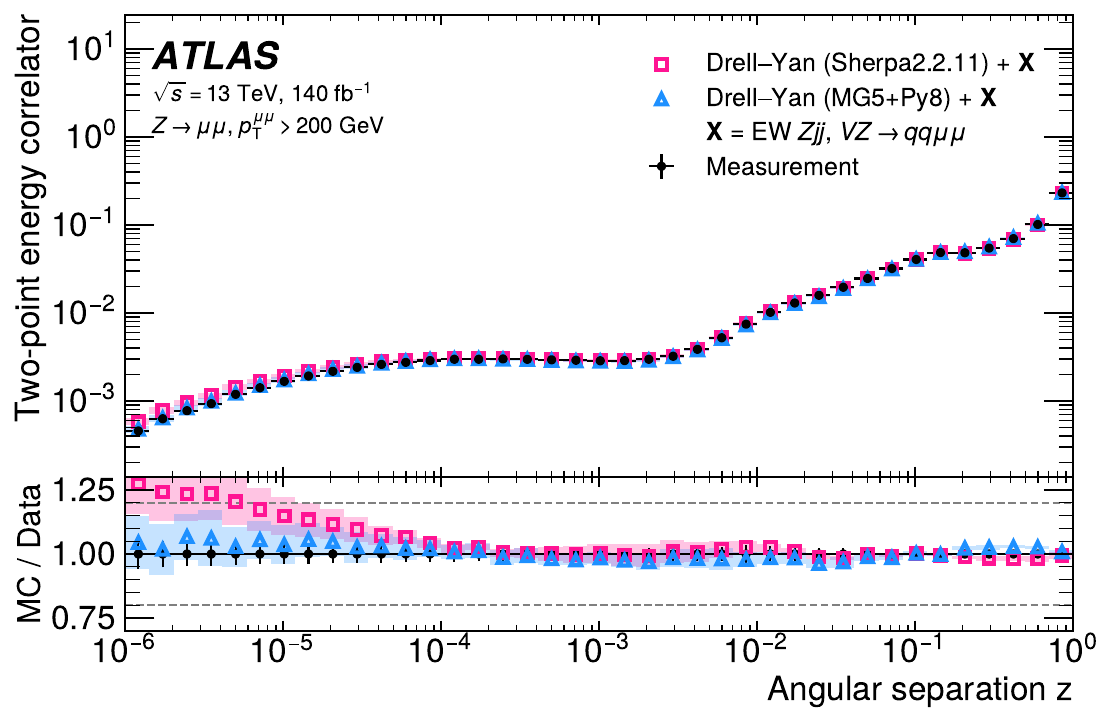}
\caption{Measurement of the two-point energy correlator (EEC), defined in Eq.~(\ref{eqn:eec}), as a function of the angular-separation variable $z$. Uncertainties on the measurement are shown as error bars and uncertainties on the generator predictions are shown as shaded bands.}
\label{fig:eec}
\end{figure}

The measurement of the EEC observable as a function of the angular separation observable $z$ is shown in Figure~\ref{fig:eec}.
The measurement achieves a precision of approximately 5\% or better over the full range of $z$.
The fractional uncertainties and covariance matrix associated with this measurement can be reproduced using the published event representations and analysis code~\cite{cern_open_data_portal, gitlab}.

Three distinct scaling regimes are evident in the behavior of the EEC as a function of $z$.
At large angles ($z > 10^{-2}$), a linear evolution of the EEC with $z$ is observed, corresponding to non-interacting particles. Radiation at these wide angles is dominated by underlying event processes which produce no coherent energy flow.
At intermediate angles, between $z \sim 10^{-3}$ and $z \sim 10^{-4}$, the EEC is constant with $z$. This deviation from the linear scaling behavior expected from a uniform radiation pattern is produced by the QCD dynamics of interacting quarks and gluons.
At small angles ($z < 10^{-4}$), the linear scaling behavior is restored.
This is evidence of the confinement of quarks and gluons into hadrons which do not interact and produce no coherent energy flow.
Significant discrepancies between the data and both generator predictions are observed in the smallest-$z$ region, most notably for the \SHERPA{} prediction.

\subsection{Intrinsic dimensionality of jets}
\label{sec:intrinsic_dimensionality_observables}

Intrinsic dimensionality (ID) is a property of a dataset that quantifies the dimensionality of the low-dimensional manifold on which the data approximately lie within a high-dimensional feature space~\cite{CAMASTRA201626}.
The manifold hypothesis suggests that this dimensionality can be substantially lower than the dimensionality of the feature space naively used to describe data.
Recent theoretical work has suggested using ID estimators on jet-substructure data as useful observables for studying QCD and for testing BSM theories~\cite{Komiske:2019fks,DAgnolo:2025qqr}.
However, no ID estimator has ever been measured in particle physics data, given they are properties of a data set rather than a single jet or event, making it impossible to construct a response matrix.
Unbinned unfolding methods circumvent this problem, since they do not require a response matrix.
This, combined with the full charged-particle phase space nature of this measurement, makes a measurement of a physically meaningful ID estimator possible for the first time.
This paper focuses on the nearest neighbor intrinsic dimension (NNID) estimator~\cite{NIPS2004_74934548}, which was explored in Ref.~\cite{DAgnolo:2025qqr} for tests of BSM theories, to study the ID of a selection of jets taken from the full charged-particle phase space measurement.

Any ID estimator requires a suitable distance metric, a function for calculating the distance between two points that satisfies the triangle inequality.
Reference~\cite{Komiske:2019fks} introduced the energy movers distance (EMD) as a suitable distance metric for jet data.
This metric quantifies the minimal ``cost'' required to rearrange the radiation pattern of one jet into that of another jet.
The per-constituent energy weighting used in the calculation of the EMD ensures that the metric is infrared and collinear safe, meaning it is insensitive to the addition of low momentum or collinear jet constituents.
The first step in measuring the NNID is to calculate the EMD between each pair of jets in the sample.
Jets, clustered with the \akt algorithm~\cite{Cacciari:2008gp} with a radius parameter of $R=1.0$ and required to satisfy $|\eta| < 2.5$ and $\pt \in [330, 370]$~\GeV, are selected from the reweighted particle-level simulated events used to build the fully differential charged-particle phase space measurement.
The narrow transverse momentum interval allows the intrinsic dimensionality to be studied at an approximately fixed jet energy scale.
Only the leading jet is taken from each event and is assigned the weight of the event.
Because the computational cost of evaluating the EMD for all jet pairs scales quadratically with the number of jets, the sample is randomly reduced to \num{50000} jets for this study.

\begin{figure}[t]
\centering
\subfloat[]{
\includegraphics[width=0.5\textwidth]{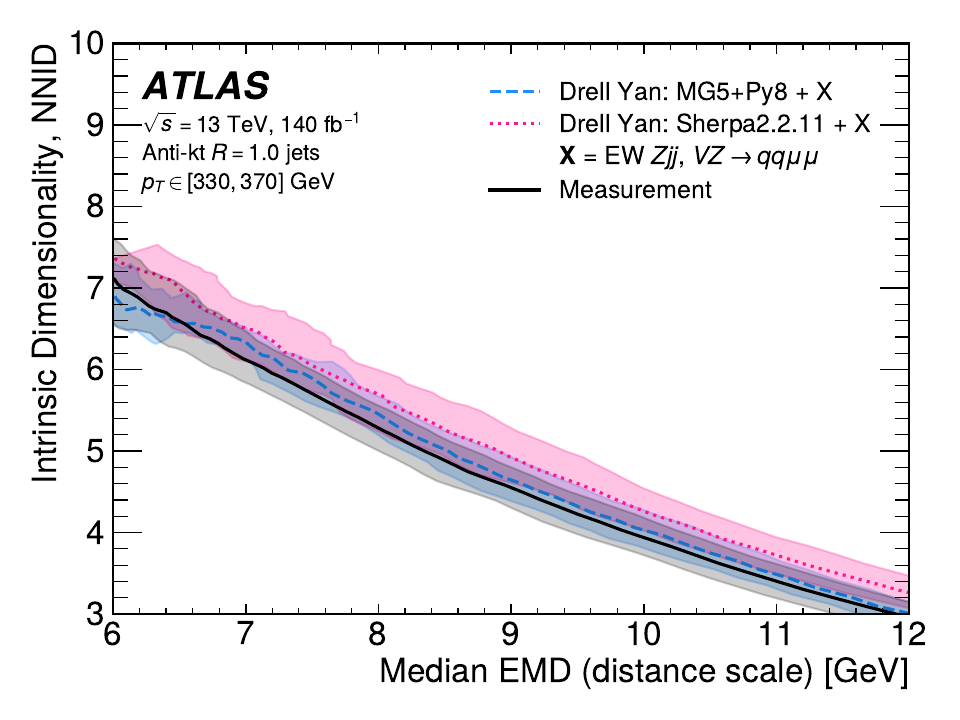}
\label{fig:nnid_low}
}
\subfloat[]{
\includegraphics[width=0.5\textwidth]{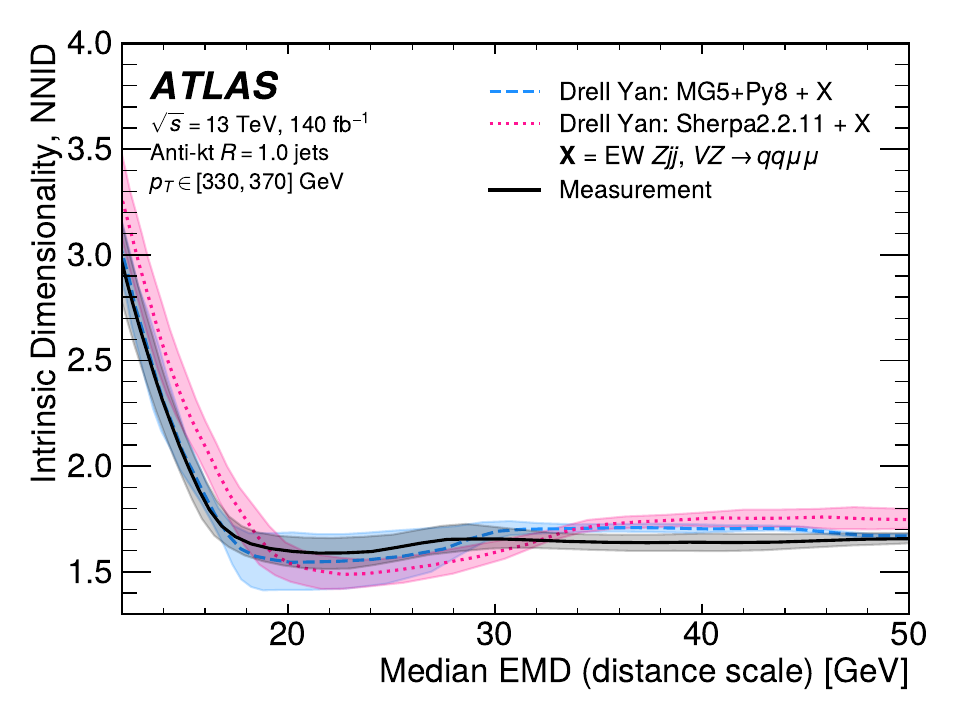}
\label{fig:nnid_high}
}
\caption{Measurement of the nearest neighbor intrinsic dimension (NNID) as a function of the median EMD between jets used to estimate the NNID\@.
The measurement uses anti-$k_t$ jets with $R=1.0$, restricted to $p_\mathrm{T}^\mathrm{jet} \in [330, 370]$~\GeV.
Shaded bands correspond to the total uncertainty on the NNID measurement and generator predictions. Figure~\ref{fig:nnid_low} shows the NNID at low values of the median EMD, and Figure~\ref{fig:nnid_high} shows the NNID at high values.}
\label{fig:nnid}
\end{figure}

Calculation of the NNID proceeds by calculating the ratio of the EMDs between each jet and its $i$th and $j$th nearest neighbor, with nearest neighbors determined by the EMD\@.
Then a minimization of a likelihood function defined over these ratios is performed to extract the NNID\@.
This study follows the recommendations of Ref.~\cite{DAgnolo:2025qqr} and fixes $j = 2i$, making the only free parameter the choice of the algorithmic parameter $i$.
The NNID is calculated for logarithmically spaced choices of $i$.
Each of these choices represents a different observable, while the jets used to calculate the observable remain fixed.
To plot the NNID as a function of a physically meaningful quantity, the median EMD between a jet and its $i$th nearest neighbor is also calculated for each choice of $i$.
This procedure is repeated for each weight variation to produce uncertainties on the NNID and the median EMD\@.
The NNID is then plotted against the median EMD in Figure~\ref{fig:nnid}.

For both the measurement and theory predictions, the NNID decreases as a function of the median EMD\@.
This is an expected property of intrinsic dimension estimators.
When considering jets that have similar substructure (and thus low median EMD), there are many dimensions required to describe how the fine details of each jet's substructure differs from the other.
As the EMD between the jets used to estimate the NNID increases, the variance in the jet data becomes confined to a few directions in phase space and the intrinsic dimensionality decreases.
The NNID decreases approximately linearly with the median EMD at values of the median EMD below 16~\GeV, and becomes approximately constant for values of the median EMD above 30~\GeV.
Between these regimes there is a minimum in the NNID, first decreasing then slightly increasing with larger median EMD\@.
The theoretical predictions provided by \MADGRAPHPY{} and \SHERPA{} generally describe the data within uncertainties, but some departure between the data and generator level predictions is visible, especially for \MADGRAPHPY{} in the region of the minimum.
The decrease and subsequent increase in the NNID at intermediate scales is present in both generators, but is more pronounced than what is observed in data.

Finally, across all scales the intrinsic dimensionality is fewer than ten: much smaller than the many tens of dimensions required to describe a jet in terms of the four-vectors of its constituents.
This observation provides direct experimental evidence that QCD dynamics constrain physically realized jet configurations to occupy only a small subset of the available feature space.
Both generators reproduce this qualitative behavior, although modest differences with the data remain at the largest distance scales.


%
\section{Conclusion}

This paper has presented the first published unbinned collider cross-section measurement performed directly in the full phase space of final-state charged particles. Rather than publishing a finite collection of differential cross sections, the measurement is represented by a set of particle-level events together with associated unfolded event weights, allowing arbitrary observables to be constructed after publication.
Realizing this measurement required extending \omnifold{} to operate on high-dimensional variable-length event representations using transformer neural networks together with large-scale pretraining on simulated data.
The trained neural networks produced in the \omnifold{} procedure are also released publicly.

Four representative applications were presented, spanning event-level observables, jet-shape measurements, energy correlators, and intrinsic-dimensionality estimators. Together these examples demonstrate that a single published event-level representation can reproduce conventional unfolded measurements, support precision studies of observables that would otherwise require dedicated detector calibrations or difficult unfolding procedures, and enable entirely new classes of measurements that are inaccessible to traditional response-matrix techniques.

These four applications illustrate only a small subset of the possible uses of the published measurement. To facilitate future studies, the particle-level event sample and corresponding unfolded event weights are released publicly~\cite{cern_open_data_portal}.
Example analysis code is also provided, reproducing each study presented in this paper and illustrating recommended workflows for downstream analyses~\cite{gitlab}.
Producing this measurement required fine-tuning many thousands of neural networks within the iterative \omnifold{} procedure, representing a significant computational undertaking. Approximately \num{400000} NVIDIA A100 GPU hours were required to produce the final results. Continued advances in high-performance computing resources, together with methodological improvements that reduce the computational cost of uncertainty propagation, will make similar measurements increasingly practical in future analyses.

The techniques presented here provide a path toward treating unfolded event-level datasets as long-lived scientific data products, extending the impact of collider measurements well beyond the specific observables envisioned at the time of publication.
Event-level unfolded measurements of this type are expected to become an increasingly important component of the LHC physics program, complementing traditional differential cross-section measurements while substantially expanding their long-term scientific utility.

%


%
%
%
%
%
\section*{Acknowledgements}
%

%

%
%

%
%

We thank CERN for the very successful operation of the LHC and its injectors, as well as the support staff at
CERN and at our institutions worldwide without whom ATLAS could not be operated efficiently.

The crucial computing support from all WLCG partners is acknowledged gratefully, in particular from CERN, the ATLAS Tier-1 facilities at TRIUMF/SFU (Canada), NDGF (Denmark, Norway, Sweden), CC-IN2P3 (France), KIT/GridKA (Germany), INFN-CNAF (Italy), NL-T1 (Netherlands), PIC (Spain), RAL (UK) and BNL (USA), the Tier-2 facilities worldwide and large non-WLCG resource providers. Major contributors of computing resources are listed in Ref.~\cite{ATL-SOFT-PUB-2026-001}.

We gratefully acknowledge the support of ANPCyT, Argentina; YerPhI, Armenia; ARC, Australia; BMWFW and FWF, Austria; ANAS, Azerbaijan; CNPq and FAPESP, Brazil; NSERC, NRC and CFI, Canada; CERN; ANID, Chile; CAS, MOST, NSFC and FRFCU, China; Minciencias, Colombia; MEYS CR, Czech Republic; DNRF and DNSRC, Denmark; IN2P3-CNRS and CEA-DRF/IRFU, France; SRNSFG, Georgia; BMFTR, HGF and MPG, Germany; GSRI, Greece; RGC and Hong Kong SAR, China; ICHEP and Academy of Sciences and Humanities, Israel; INFN, Italy; MEXT and JSPS, Japan; CNRST, Morocco; NWO, Netherlands; RCN, Norway; MNiSW, Poland; FCT, Portugal; MNE/IFA, Romania; MSTDI, Serbia; MSSR, Slovakia; ARIS and MVZI, Slovenia; DSI/NRF, South Africa; MICIU/AEI, Spain; SRC and Wallenberg Foundation, Sweden; SERI, SNSF and Cantons of Bern and Geneva, Switzerland; NSTC, Taipei; TENMAK, T\"urkiye; STFC/UKRI, United Kingdom; DOE and NSF, United States of America.

Individual groups and members have received support from BCKDF, CANARIE, CRC and DRAC, Canada; CERN-CZ and PRIMUS, Czech Republic; COST, ERC, ERDF, Horizon 2020 and Marie Sk{\l}odowska-Curie Actions, European Union; Investissements d'Avenir Labex, Investissements d'Avenir Idex and ANR, France; DFG and AvH Foundation, Germany; Herakleitos, Thales and Aristeia programmes co-financed by EU-ESF and the Greek NSRF, Greece; BSF-NSF and MINERVA, Israel; NCN and NAWA, Poland; La Caixa Banking Foundation, CERCA and AGAUR programs from Generalitat de Catalunya and PROMETEO and GenT Programmes Generalitat Valenciana, Spain; G\"{o}ran Gustafssons Stiftelse, Sweden; The Royal Society and Leverhulme Trust, United Kingdom; Eric and Wendy Schmidt Fund for Strategic Innovation, United States of America.

In addition, individual members wish to acknowledge support from Chile: Agencia Nacional de Investigaci\'on y Desarrollo (FONDECYT 1230812, FONDECYT 1230987, FONDECYT 1240721, FONDECYT 1240864, FONDECYT 3240661); China: Fundamental Research Funds for the Central Universities (010-63263105), Chinese Ministry of Science and Technology (MOST-2023YFA1605700, MOST-2023YFA1609300), National Natural Science Foundation of China (NSFC 12275265, NSFC-W2543005); Czech Republic: Czech Science Foundation (GACR - 24-11373S), Ministry of Education Youth and Sports (ERC-CZ-LL2327, FORTE CZ.02.01.01/00/22\_008/0004632); EU: H2020 European Research Council (ERC - 101002463); European Union: European Research Council (BARD No. 101116429, ERC - 101219398, ERC 101089007), European Regional Development Fund (HE COFUND GA No.101081355, ERDF), Marie Sklodowska-Curie Actions (GAP-101168829); France: Agence Nationale de la Recherche (ANR-22-EDIR-0002, ANR-24-CE31-0504-01); Germany: Deutsche Forschungsgemeinschaft (DFG - 469666862); China: Research Grants Council (GRF); Italy: Istituto Nazionale di Fisica Nucleare (LHC-MIUR - 28003/2025), Ministero dell'Università e della Ricerca (NextGenEU I53D23000820006 M4C2.1.1, SOE2024\_0000023); Japan: Japan Society for the Promotion of Science (JSPS KAKENHI  JP25H0063, JSPS KAKENHI JP22H04944, JSPS KAKENHI JP24K23939, JSPS KAKENHI JP24KK0251, JSPS KAKENHI JP25H00650, JSPS KAKENHI JP25H01291, JSPS KAKENHI JP25K01011, JSPS KAKENHI JP25K01023, JSPS KAKENHI JP25KK0047); Poland: Polish National Science Centre (NCN OPUS 2023/51/B/ST2/02507, NCN OPUS nr 2022/47/B/ST2/03059, UMO-2019/34/E/ST2/00393, UMO-2021/42/E/ST2/00350, UMO-2022/47/O/ST2/00148, UMO-2023/49/B/ST2/04085, UMO-2023/51/B/ST2/00920, UMO-2024/53/N/ST2/00869); Spain: Agència de Gestió d'Ajuts Universitaris i de Recerca. (AGAUR - 2023 BP 00141), Ministry of Science and Innovation (RYC2019-028510-I, RYC2020-030254-I, RYC2021-031273-I, RYC2022-038164-I), Ministerio de Ciencia, Innovación y Universidades/Agencia Estatal de Investigaci\'on (EU NextGenerationEU (PRTR-C17.I1), PID2022-142604OB-C22); Sweden: Carl Trygger Foundation (Carl Trygger Foundation CTS 22:2312), Swedish Research Council (Swedish Research Council 2023-04654, VR 2021-03651, VR 2022-03845, VR 2022-04683, VR 2023-03403, VR 2024-05451, VR 2025-05940), Knut and Alice Wallenberg Foundation (KAW 2023.0366); United Kingdom: The Binks Trust, Royal Society (NIF-R1-231091); United States of America: U.S. Department of Energy (ECA DE-AC02-76SF00515), John Templeton Foundation (John Templeton Foundation 63206), Neubauer Family Foundation.

%
%


%
%
%
%
%
%
%
%
%

%
\clearpage
\appendix
\part*{Appendix}
\addcontentsline{toc}{part}{Appendix}

\section{Usage recommendations for the unbinned measurement}
\label{app:usage}

The primary result of this paper is a published event-level representation of the measured particle-level cross section, consisting of particle-level MC events together with their associated unfolded weights.
This data product is released publicly to enable the construction of new observables and analyses beyond those presented in this paper.
The recommendations below summarize the conditions under which the measurement was validated and describe best practices for downstream analyses. The same recommendations are implemented in the accompanying public codebase and are reproduced here for convenience.

%
%
%
%
%
%
%

\begin{description}
\item[1. Validation using the pseudo-measurement]~\\Any analysis should always be checked using the pseudo-measurement against the target, to ensure a reliable result is obtained before drawing any conclusion based on the measurement with real data. For any binned measurement, the binning should be chosen such that each bin contains sufficient statistics and the unfolded pseudomeasurement shows good statistical agreement with the target when tested using pseudodata. Examples of such tests are performed in each of the example notebooks before the results of the data measurement are shown.
The key step of this procedure is to construct the measurement covariance matrix that encodes the correlation between bins.
The $\chi^2$ statistic and associated $p$-value are then obtained with a few lines of code that performs matrix operations.
The outcome of these tests will depend on the assumed nuisance parameter smoothing and correlation model as discussed in the next bullet.

\item[2. Minimum jet transverse momentum]~\\When studying jet observables, jets should satisfy  $\pt>\SI{5}{\GeV}$. For observables involving the leading jet, such as $\Delta R(\ell\ell,j_1)$, this requirement should be applied to the leading jet. For dijet observables such as $m_{jj}$, it should be applied to both jets.

\item[3. Sufficient Monte Carlo statistical precision]~\\Each bin should contain a minimum of \num{5000} effective events defined by $n_{\text{eff}} \equiv (\Sigma_iw_i)^2 / \Sigma _iw_i^2$.
This requirement ensures that the Monte Carlo statistical uncertainty is sufficiently small for stable uncertainty estimates.

\item[4. Sufficient data statistical precision]~\\
The statistical uncertainty from the data should not exceed 15\% in any bin,
ensuring that the measurement is supported by sufficient data statistics.
This criterion is most likely fulfilled when $n_{\text{eff}} > \num{5000}$.

\item[5. Uncertainty smoothing and correlation model]~\\
The default recommendation is to smooth the hidden-variable uncertainties using the Gaussian-kernel procedure provided in \texttt{utils.py} and to treat each systematic uncertainty source as fully correlated across bins.
However, there are cases where adjustments should be made to this model. For example, in the default results presented in \texttt{standard\_obs\_data\_results.ipynb} the hidden variable uncertainties \textbf{are not} smoothed for the $m_{j1}$ and $m_{j2}$ observables. This is because these two-point systematic uncertainties are large and have a strong shape, meaning smoothing them would bias the uncertainty and result in poor $p$-values.

Uncertainties with strong shape may additionally switch in sign in the middle of the distribution, potentially resulting in poor $p$-values due to the assumption that bins are fully correlated. This is one example where other correlation models may be explored.
Since the true correlation model is not known, any alternative correlation model should first be validated using the pseudomeasurement before being applied to the data.
For instance, if four bins are used, the systematic uncertainty may be split into four components (‘nuisance parameters’), one per bin. More sophisticated decorrelation schemes could be explored for which each uncertainty component has a smooth uncertainty amplitude spanning the full range, such that at any point (i.e., for any bin) this uncertainty adds in quadrature to match the total uncertainty of the two-point systematics.

\item[6. Drell--Yan-only theory comparisons]~\\
The public theory prediction files contain a vector named \texttt{sample}, whose integer values identify simulated events originating from the Drell--Yan (0), diboson (1), and electroweak (2) processes described in Section~\ref{sec:data_simulation_event}. These files can therefore be used to extract predictions for each individual component.
This is useful, for example, when comparing the measurement with a novel Drell--Yan prediction while no corresponding predictions are available for the electroweak $Zjj$ or $ZV$ contributions. In this case, the public electroweak and diboson predictions can be added to the user-provided Drell--Yan prediction to construct the full prediction for comparison with the published measurement. The published measurement is agnostic to the underlying production process of the particle-level events which are used to construct the published measurement and therefore does not support selections based on the process labels.
An example illustrating how to extract these components and their associated uncertainties for the \pTll\ distribution is provided in \texttt{Drell\_Yan\_results.ipynb}.
\end{description}

Not every observable that can be constructed from the published particle-level events is necessarily constrained by the measurement. For example, the public event representation includes the particle masses, allowing users to identify the hadron species of individual particles. However, the relative abundances of different hadron species are not directly constrained by the detector-level information used in the unfolding. Similarly, the published measurement includes event-level process labels (Drell--Yan, electroweak $Zjj$, and diboson), which identify the simulated production process from which each event originated. These labels are retained for technical consistency with the public theory prediction files, but are not constrained by the unfolding and should therefore not be used to define measured subsamples or fiducial cross sections. Consequently, observables that depend strongly on the particle-level hadron composition or on the event process labels remain prior dependent and will generally fail to satisfy the validation procedure described above. All downstream analyses should therefore follow the recommendations above, with particular emphasis on validation using the pseudodata measurement before interpreting results obtained from the data.

In almost all cases, a successful validation using pseudodata demonstrates that the unfolding procedure is capable of reproducing the target cross section in the bins that the user has defined. It is, however, possible that the predictions of the \MADGRAPHPY{} and \SHERPA[2.2.11] generators happen to agree for some observable that is not constrained by the unfolding. In this case the pseudodata validation is trivial, and the measurement should not be considered capable of constraining this particular observable. When using the information that is known to be unconstrained by the unfolding, the user should verify that the pseudodata validation is non-trivial before proceeding.

The validation procedure described above assumes that the downstream analysis projects the event-level representation to construct a binned measurement. In this case, an appropriate goodness-of-fit test is the binned $\chi^2$ procedure implemented in the example notebooks. More advanced applications may instead operate directly on the high-dimensional event representation without introducing intermediate histograms. Defining statistically rigorous goodness-of-fit tests for such fully unbinned analyses remains an open area of research. Consequently, no recommended procedure is currently provided in the public codebase. Users pursuing such applications should develop an appropriate validation strategy using the pseudodata measurement before drawing conclusions from the corresponding data analysis.

%
%


\section{Details of pseudodata measurement and statistical tests}
\label{app:pseudodata}

The primary validation of this measurement is based on a pseudodata study, in which simulated events are modified such that their detector-level distributions closely resemble those observed in data while retaining a known particle-level target. Because the particle-level target remains known, the complete measurement chain—including background subtraction, unfolding, uncertainty propagation, and statistical interpretation—can be tested under realistic conditions.
In contrast, this validation cannot be performed using collision data because the corresponding particle-level distribution is unknown.

The pseudodata are constructed by combining independent samples of the \SHERPA{} Drell--Yan, electroweak, diboson, and \POWHEG[v2] \ttbar{} processes. Smooth reweighting functions are derived from detector-level observables and applied at particle level such that, after detector simulation, the resulting detector-level distributions closely reproduce those observed in data. The reweighted sample is subsequently divided into statistically independent detector- and particle-level samples. The detector-level sample is randomly downsampled to match the effective statistical precision of the collision data and serves as the pseudodata input to the full analysis chain. The particle-level sample retains the full Monte Carlo statistics and provides a known target against which the unfolded measurement is compared. This combination yields a realistic validation of the measurement while minimizing the statistical uncertainty on the reference target.

All uncertainties are propagated independently for the pseudodata-based and data-based measurements.
The resulting pseudodata-based measurement is then compared with the known particle-level target to assess the statistical consistency of the measurement.
For most applications of this measurement, this should be done with a binned $\chi^2$ statistical test, performed as follows:

\begin{enumerate}
\item
The nominal measurement and all associated weight variations are projected onto the observable of interest.
The varied histograms should then be used to construct a covariance matrix.
The covariance matrix encodes both the variance of each bin and the correlations between bins arising from shared systematic uncertainties and statistical fluctuations. It therefore represents the complete uncertainty model of the binned measurement. Detailed illustration of how to perform these steps is provided in the public codebase.
\item The particle-level pseudodata sample should be binned in the same observable and bins as the pseudodata measurement.
\item{A $\chi^2$ test statistic is then calculated as
\begin{equation}
\chi^2 = (\mathbf{m}-\mathbf{t})^T\,\Sigma^{-1}\,(\mathbf{m}-\mathbf{t}),
\end{equation}
where $\mathbf{m}$ is the vector of measured bin contents, $\mathbf{t}$ is the corresponding particle-level pseudodata target, and $\Sigma$ is the covariance matrix. The test statistic quantifies the statistical compatibility between the measurement and the known target while accounting for the statistical and systematic uncertainties, including their correlations between bins.
}
\item The resulting $\chi^2$ test statistic should be compared to a $\chi^2$ distribution with the number of degrees of freedom $n_\mathrm{dof}$ equal to the number of bins in the histogram, and a $p$-value should be calculated.
A sufficiently large $p$-value indicates that the measurement is statistically consistent with the known target under the assumed uncertainty model. The target hypothesis is not rejected at the 5\% significance level when the p-value is at least 0.05.
\end{enumerate}

\begin{figure}[t]
\centering
\subfloat[]{
\includegraphics[width=0.45\textwidth]{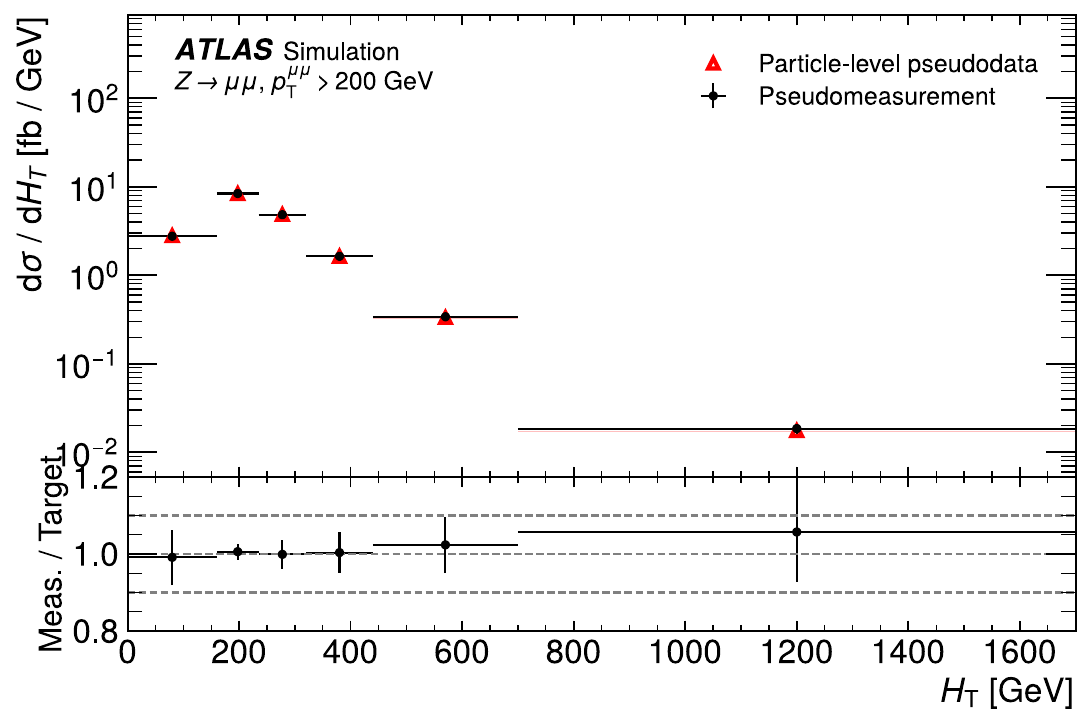}
}\\
\subfloat[]{\raisebox{5.7mm}{
\includegraphics[width=0.45\textwidth]{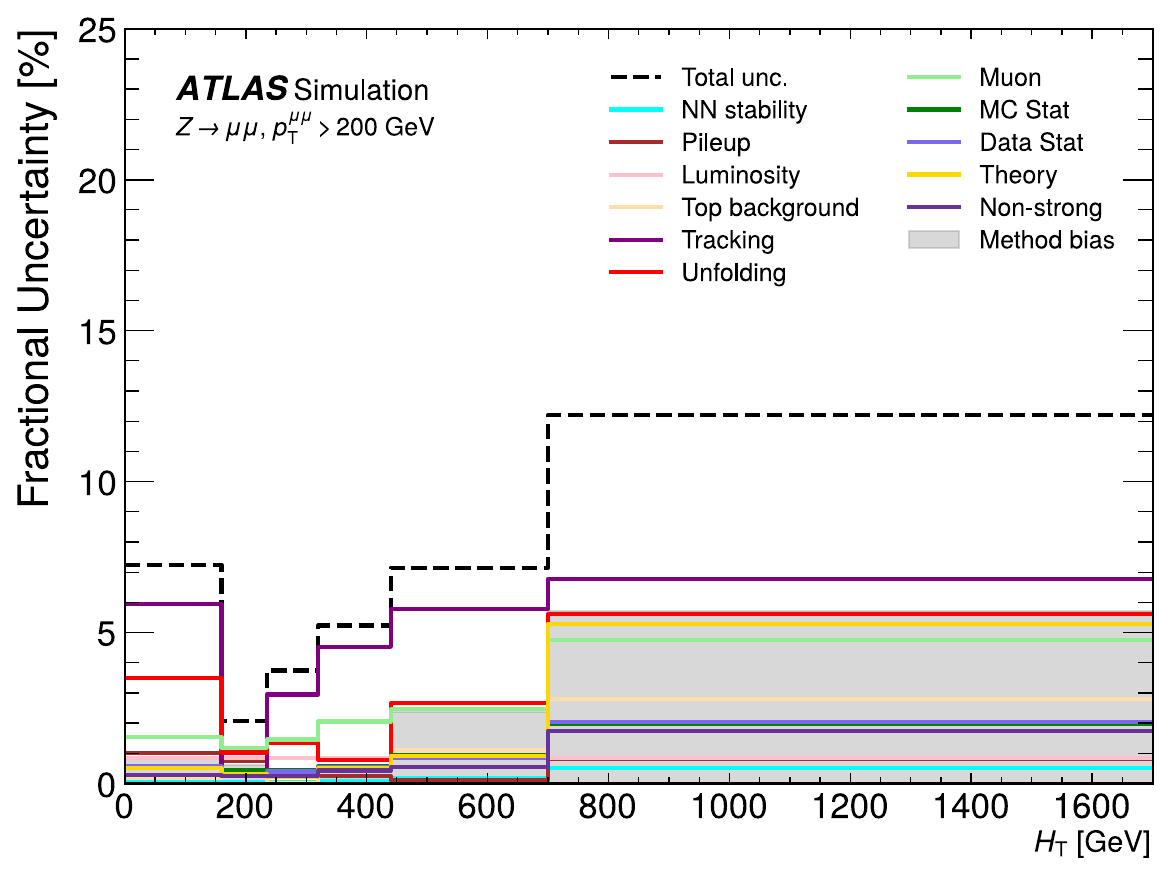}
}}
\subfloat[]{
\includegraphics[width=0.45\textwidth, height=0.28\textheight]{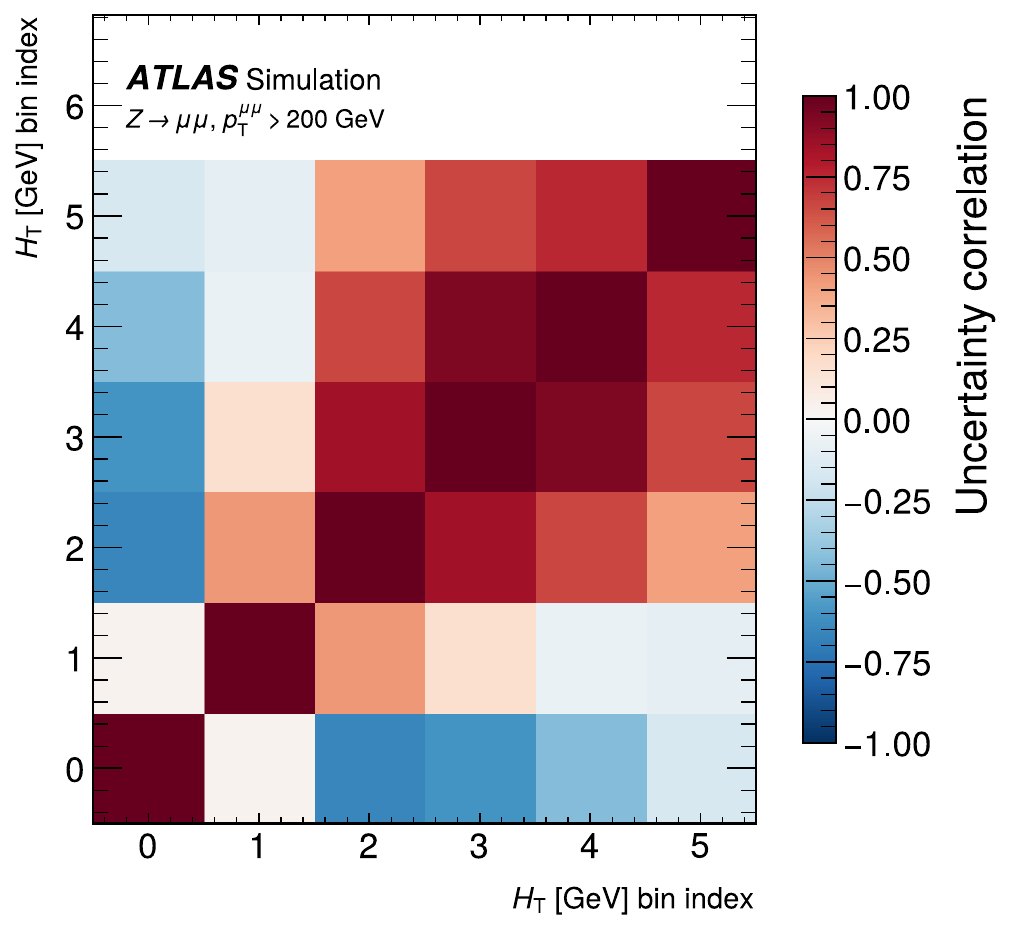}
}\\
\subfloat[]{\raisebox{5.7mm}{
\includegraphics[width=0.45\textwidth]{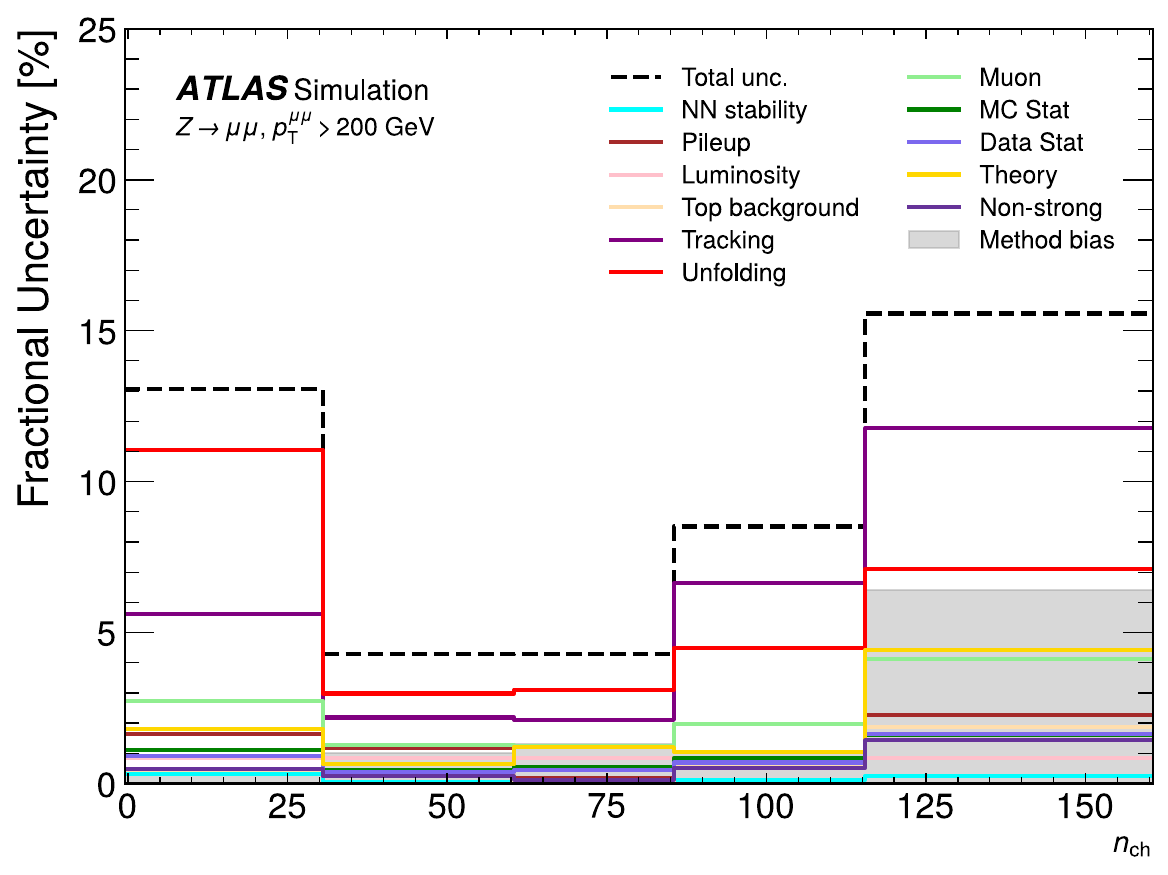}
}}
\subfloat[]{
\includegraphics[width=0.45\textwidth, height=0.28\textheight]{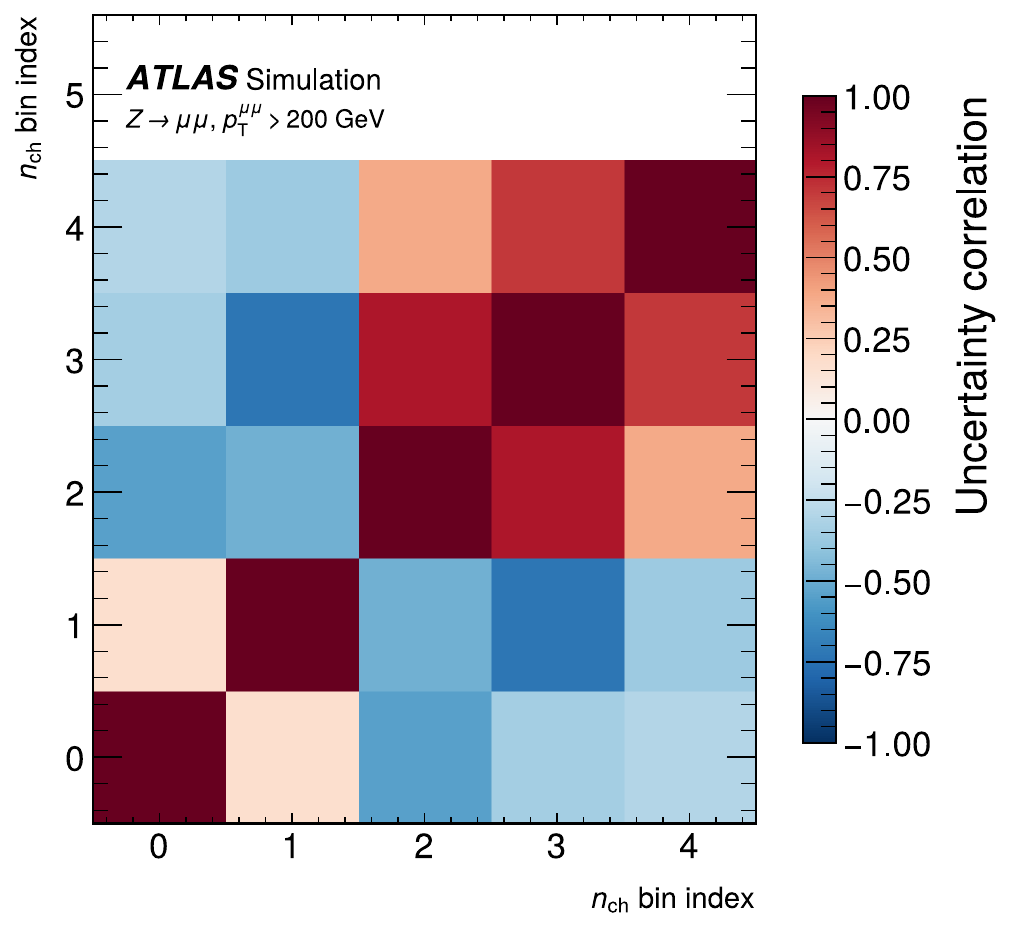}
}
\caption{
Validation of the pseudodata measurement for the \HT{} and \nch{} observables.
(a) Comparison of the pseudodata measurement of the \HT{} observable with the known particle-level pseudodata target. The analogous plot for the \nch{} observable is shown in Figure~\ref{fig:Ntracks_pseudodata}. Breakdowns of the various pseudodata measurement uncertainties are shown for (b) \HT{} and (d) \nch{}. The corresponding uncertainty correlation matrices are shown for (c) \HT{} and (e) \nch{}.}
\label{fig:pseudodata_event_level}
\end{figure}

The results of these statistical tests for the \nch{} and \HT{} observables, for which the real data measurements are presented in Section~\ref{sec:event_level_observables}, are presented in Figure~\ref{fig:pseudodata_event_level}. Here the pseudomeasurement is compared with the known particle-level pseudodata target (the \nch distribution is shown in Figure~\ref{fig:validation}), the pseudodata uncertainty breakdown, and the corresponding correlation matrices. The uncertainties on the pseudomeasurement are generally very similar in magnitude and shape to the uncertainties on the corresponding real measurement.
A numerical summary of the corresponding tests for all 26 standard event-level observables is given in Table~\ref{tab:pseudodata_chi2_omnifold}. The observed $p$-values indicate that the unfolded pseudodata measurements are statistically consistent with their known particle-level targets for all representative observables considered. These tests provide evidence that the assigned uncertainties adequately cover the residual biases of the unfolding procedure for the chosen observables and binnings.

A key advantage of the published event-level representation is that observables can be rebinned after publication without repeating the unfolding. To illustrate this flexibility, Table~\ref{tab:chi2_adaptive_binning} presents the results of statistical comparisons between the unfolded distributions and the pseudodata targets for several binning choices of the \nch{} and \HT{} observables. For \HT{}, good agreement with the particle-level target is observed for all three binnings considered. For \nch{}, the $p$-value decreases as the binning becomes finer, reaching $p=0.041$ for the finest binning. This behavior demonstrates the importance of validating any new binning with the accompanying pseudodata measurement before interpreting results obtained from collision data.

For the finest \nch{} binning, the reduced $p$-value is traced to the current treatment of the two-point unfolding uncertainty. Although the assigned uncertainty provides sufficient overall coverage, the assumed correlation model does not fully capture the pronounced shape of the corresponding method bias. As discussed in the user recommendations in Appendix~\ref{app:usage}, alternative smoothing or decorrelation schemes may be explored when appropriately motivated, but any such choice should first be validated using the pseudodata measurement.

\begin{table}[htb]
\centering
\begin{minipage}{0.48\textwidth}
\centering
\begin{tabular}{c c c c}
\hline \hline
Observable & $n_{\mathrm{dof}}$ & $\chi^2$ & $p$ \\
\hline
\HT{}              & 6 & 6.21 & 0.400 \\
$n_\mathrm{ch}$    & 5 & 2.41 & 0.790 \\
$n_\mathrm{ch,j1}$ & 6 & 10.4 & 0.110 \\
$n_\mathrm{ch,j2}$ & 5 & 4.72 & 0.450 \\
$\eta_{\mu \mathrm{1}}$ & 14 & 11.1 & 0.678 \\
$\eta_{\mu \mathrm{2}}$ & 14 & 4.34 & 0.993 \\
$m_\mathrm{j1}$         & 6 & 15.6 & 0.016 \\
$m_\mathrm{j2}$         & 4 & 9.23 & 0.056 \\
$p_\mathrm{T,\mu1}$     & 6 & 4.57 & 0.600 \\
$p_\mathrm{T,\mu2}$     & 6 & 8.20 & 0.224 \\
$p_\mathrm{T}^{\mu \mu}$  & 5 & 4.43 & 0.490 \\
$p_\mathrm{T,j1}$       & 6 & 11.3 & 0.081 \\
$p_\mathrm{T,j2}$       & 4 & 2.12 & 0.713 \\
\hline \hline
\end{tabular}
\end{minipage}
\hfill
\begin{minipage}{0.48\textwidth}
\centering
\begin{tabular}{c c c c}
\hline \hline
Observable & $n_{\mathrm{dof}}$ & $\chi^2$ & $p$ \\
\hline
$\phi_\mathrm{\mu1}$ & 16 & 13.5 & 0.634 \\
$\phi_\mathrm{\mu2}$ & 16 & 17.3 & 0.364 \\
$\phi_\mathrm{j1}$   & 16 & 7.02 & 0.973 \\
$\phi_\mathrm{j2}$   & 16 & 7.77 & 0.955 \\
$\tau_\mathrm{1,j1}$ &  7 & 3.32 & 0.854 \\
$\tau_\mathrm{1,j2}$ &  5 & 6.97 & 0.223 \\
$\tau_\mathrm{2,j1}$ &  7 & 11.9 & 0.105 \\
$\tau_\mathrm{2,j2}$ &  5 & 5.40 & 0.369 \\
$\tau_\mathrm{3,j1}$ &  4 & 1.87 & 0.760 \\
$\tau_\mathrm{3,j2}$ &  4 & 1.92 & 0.750 \\
$y_{\mu\mu}$         & 14 & 15.7 & 0.334 \\
$y_\mathrm{j1}$      & 18 & 14.0 & 0.727 \\
$y_\mathrm{j2}$      & 10 & 5.69 & 0.840 \\
\hline \hline
\end{tabular}
\end{minipage}
\vspace{8pt}
\caption{$\chi^2$ test statistics and associated $p$-values for the \omnifold{} pseudodata-based measurements. A $p$-value of at least 0.05 indicates that the null hypothesis of agreement between the unfolded measurement and the particle-level pseudodata target is not rejected at the 95\% confidence level. In these tests, Gaussian kernel smoothing is applied to the hidden-variable unfolding uncertainty for all observables except $m_\mathrm{j1}$ and $m_\mathrm{j2}$. A minimum jet $p_\mathrm{T}$ requirement of $p_\mathrm{T,j1}>5$~\GeV for leading-jet observables and $p_\mathrm{T,j2}>5$~\GeV for subleading-jet observables is imposed. No minimum jet requirement is applied to observables with no explicit dependence on hadronic activity.}
\label{tab:pseudodata_chi2_omnifold}
\end{table}

\begin{table}[htb]
\centering
\begin{tabular}{c c c c c}
\hline\hline
Observable & $N_{\mathrm{eff}}$\,/\,bin & $n_{\mathrm{dof}}$ & $\chi^2$ & $p$ \\
\hline
$n_{\mathrm{ch}}$  & 30,000 & 6  & 5.01  & 0.543 \\
$H_{\mathrm{T}}$   & 30,000 & 6  & 3.42  & 0.755 \\
\hline
$n_{\mathrm{ch}}$  & 20,000 & 10 & 13.2 & 0.214 \\
$H_{\mathrm{T}}$   & 20,000 & 10 & 5.96  & 0.819 \\
\hline
$n_{\mathrm{ch}}$  & 10,000 & 20 & 32.2 & 0.041 \\ %
$H_{\mathrm{T}}$   & 10,000 & 20 & 14.1 & 0.826 \\
\hline\hline
\end{tabular}
\vspace{8pt}
\caption{Results of the $\chi^2$ goodness-of-fit tests for several binning choices of the \nch{} and \HT{} observables. The bin edges are chosen such that each bin contains approximately $N_{\mathrm{eff}}$ effective events.
A $p$-value of at least 0.05 indicates that the null hypothesis of agreement between the unfolded measurement and the particle-level pseudodata target is not rejected at the 95\% confidence level. These results illustrate the importance of validating any new binning choice using the accompanying pseudodata measurement. Gaussian kernel smoothing is applied to the hidden-variable uncertainty.}
\label{tab:chi2_adaptive_binning}
\end{table}

As discussed in Section~\ref{sec:validation}, an additional pseudodata validation was performed to probe regions in which the alternative and nominal Drell--Yan generators, \SHERPA{} and \MADGRAPHPY{}, respectively, differ strongly. A ParT classifier, identical to that described in Section~\ref{sec:nn_arch}, is used to distinguish the alternative pseudodata sample, whose Drell--Yan component is generated with \SHERPA{}, from the nominal sample, whose Drell--Yan component is generated with \MADGRAPHPY{}. This task is identical to the pretraining task described in Section~\ref{sec:training}, and one of the networks produced during pretraining is therefore repurposed for this study.

The high-\SHERPA{}-probability tail of the classifier output identifies a region of relatively poor support from the nominal sample. In this region, the density of nominal events is approximately an order of magnitude smaller than that of the \SHERPA{}-based pseudodata target, providing a stringent test of the unfolding procedure. The agreement between the unfolded results and the known particle-level target is evaluated in six bins concentrated in this region of the classifier output in logit space, with each bin containing approximately 5000 effective events in the reweighted nominal sample, following the usage recommendations in Appendix~\ref{app:usage}.

\begin{figure}[htb]
\centering
\includegraphics[width=0.6\textwidth]{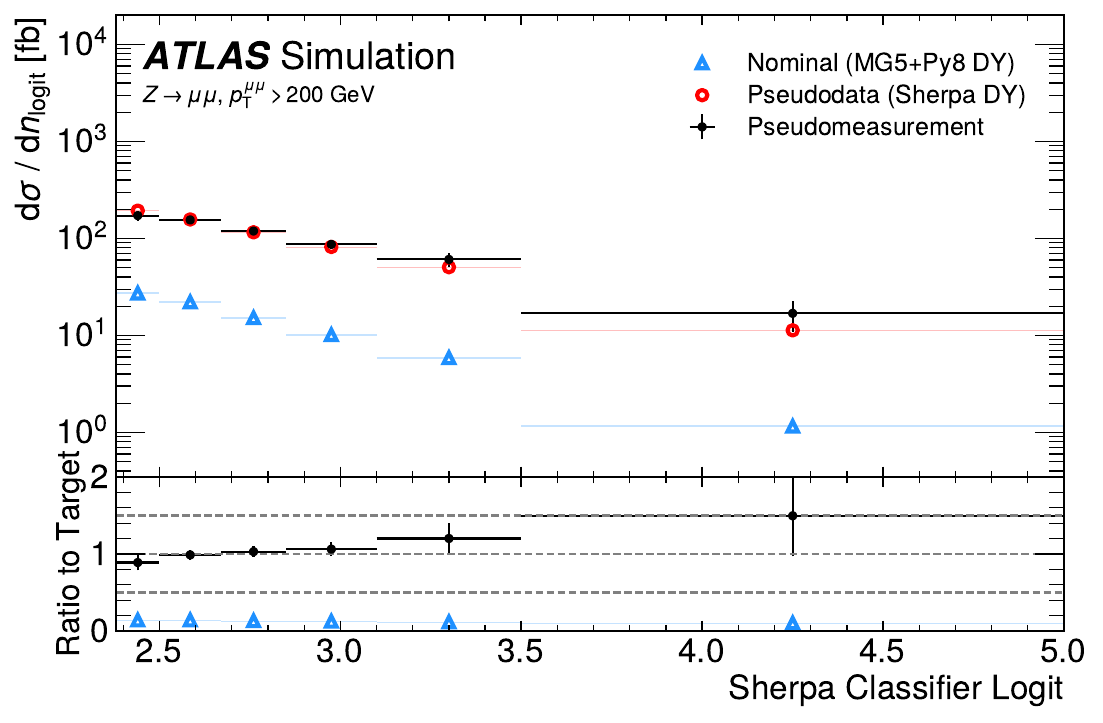}
\caption{Results of the validation test using pseudodata performed in six bins concentrated in the high-\SHERPA{}-probability tail of the classifier output in logit space. The nominal distribution
defining the prior used in the unfolding is shown as triangles, the known particle-level pseudodata target is shown as open circles, and the result of the pseudomeasurement is shown as points. Despite the substantially lower density of nominal events in this region, good agreement is observed, with a p-value of 0.863.}
\label{fig:sherpa_logit}
\end{figure}

The results are shown in Figure~\ref{fig:sherpa_logit}. Despite the substantially lower density of nominal events in this region, the pseudomeasurement is consistent with the known particle-level pseudodata target, with $\chi^2/\mathrm{ndof}=0.424$ and a p-value of 0.863.

\FloatBarrier

\section{Comparison to IBU and \multifold{} measurements}
\label{app:comparison}

This appendix compares the \omnifold{} measurements of a series of standard event-level observables with analogous results obtained using IBU and the previous \multifold{} measurement~\cite{STDM-2020-17}. All three analyses use identical event selections and target the same fiducial phase space, enabling a direct comparison of the measured cross sections.

The uncertainty models are largely identical. Since IBU does not rely on neural-network training, the corresponding measurements do not include a training stochasticity uncertainty. Aside from this difference, the IBU and \omnifold{} measurements propagate the same uncertainty sources. The \multifold{} analysis differs slightly in the treatment of several subdominant uncertainty sources.

\subsection{Measurement comparison}

\begin{figure}[!htbp]
\centering
\subfloat[]{
\includegraphics[width=0.45\textwidth, height=0.27\textheight, keepaspectratio]{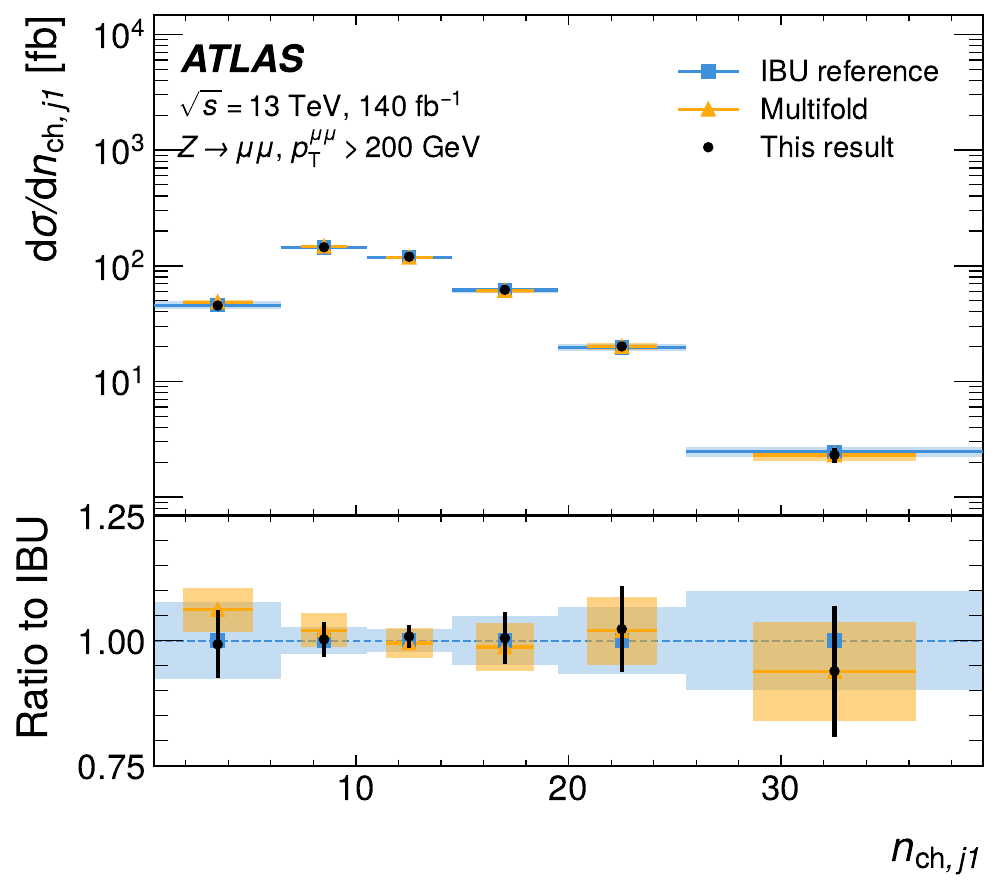}
}
\subfloat[]{
\includegraphics[width=0.45\textwidth, height=0.27\textheight, keepaspectratio]{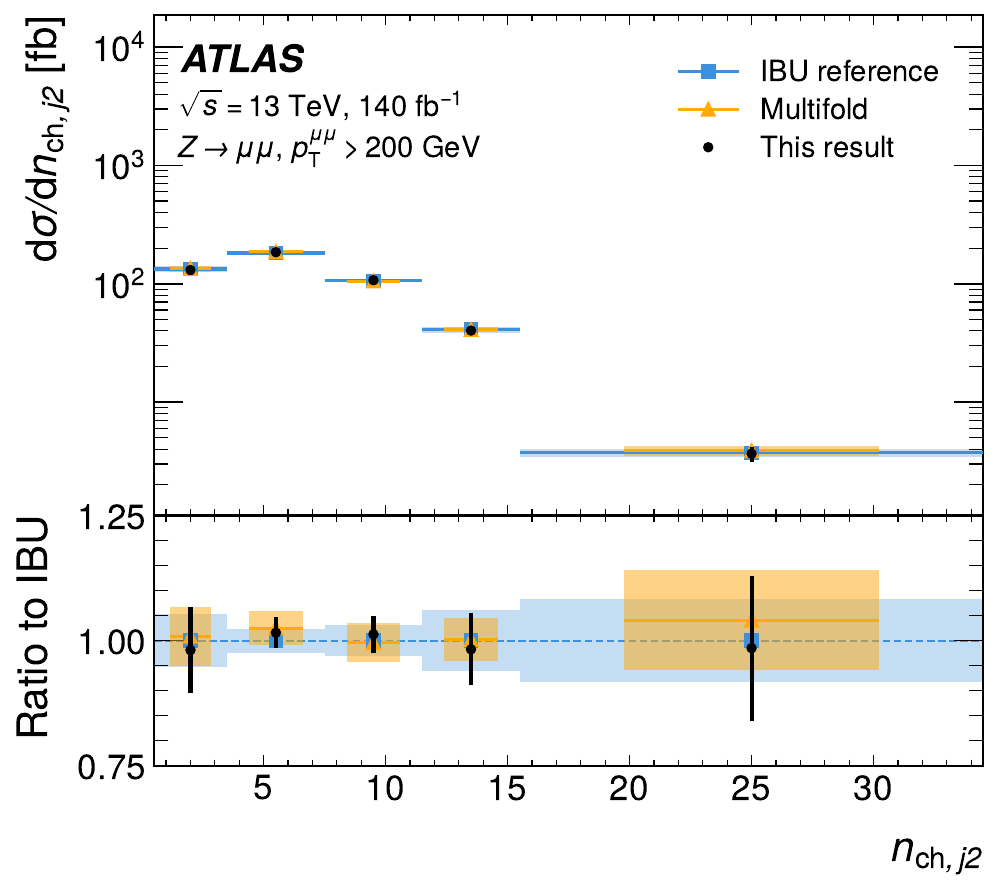}
}\\
\subfloat[]{
\includegraphics[width=0.45\textwidth, height=0.27\textheight, keepaspectratio]{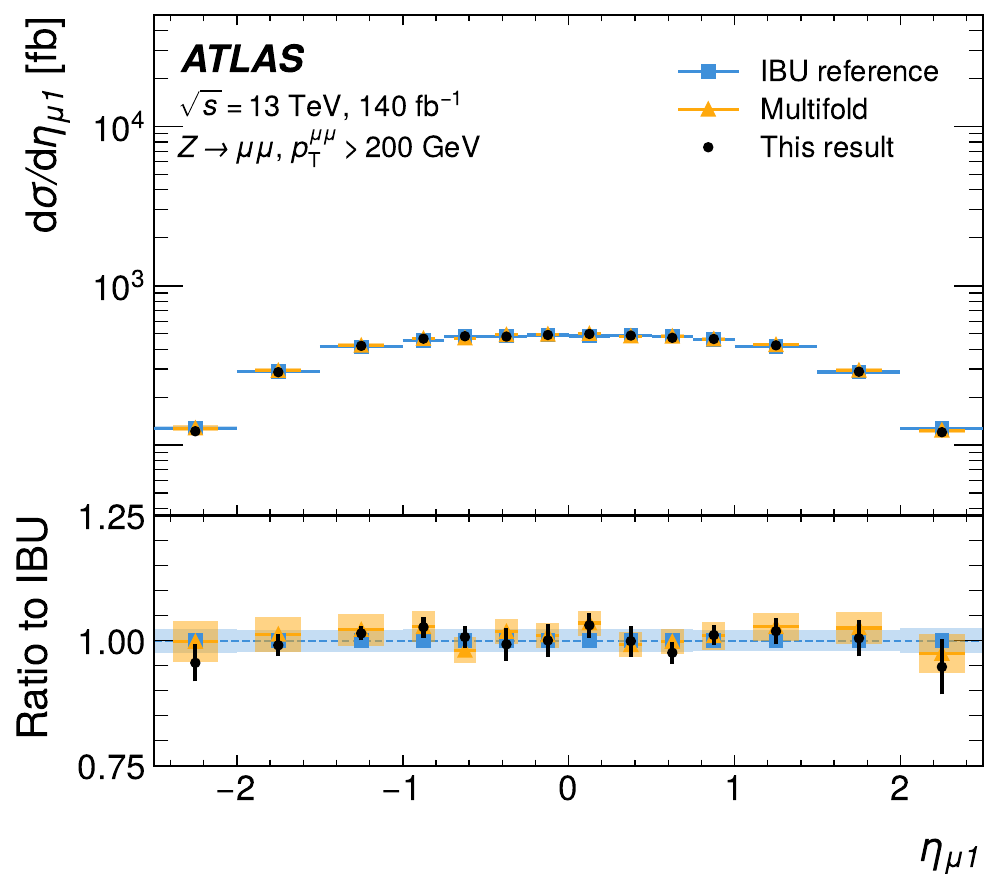}
}
\subfloat[]{
\includegraphics[width=0.45\textwidth, height=0.27\textheight, keepaspectratio]{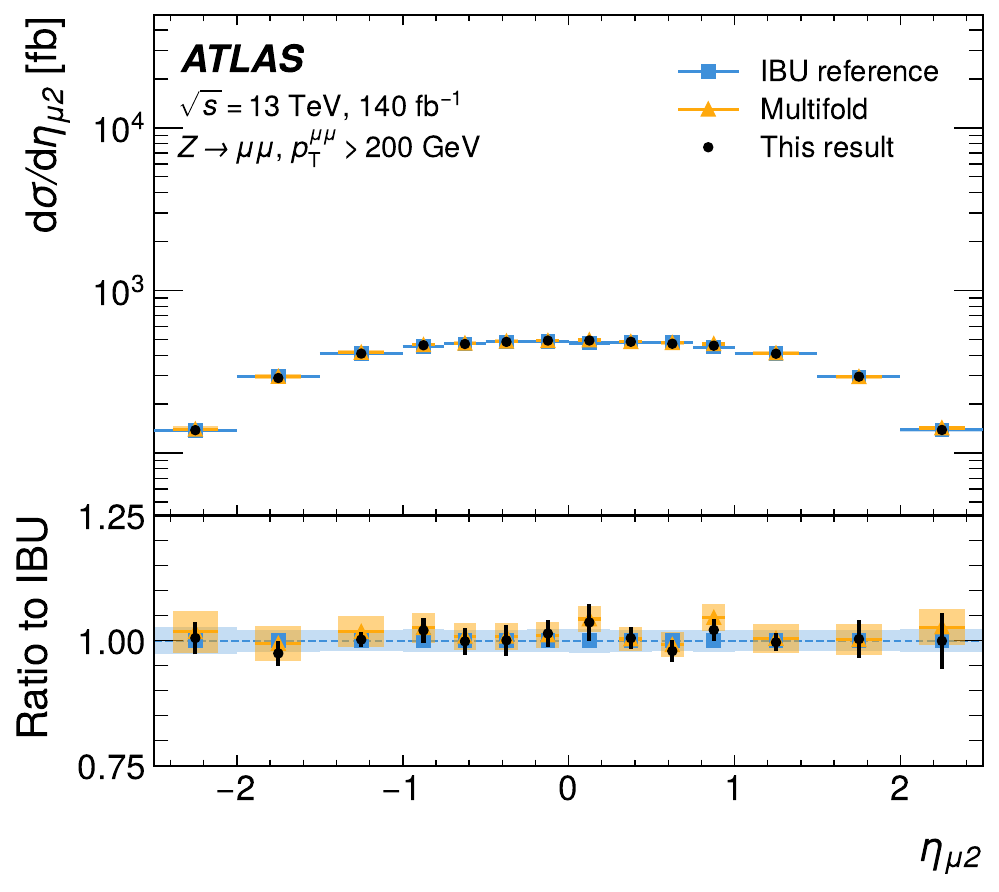}
}\\
\subfloat[]{
\includegraphics[width=0.45\textwidth, height=0.27\textheight, keepaspectratio]{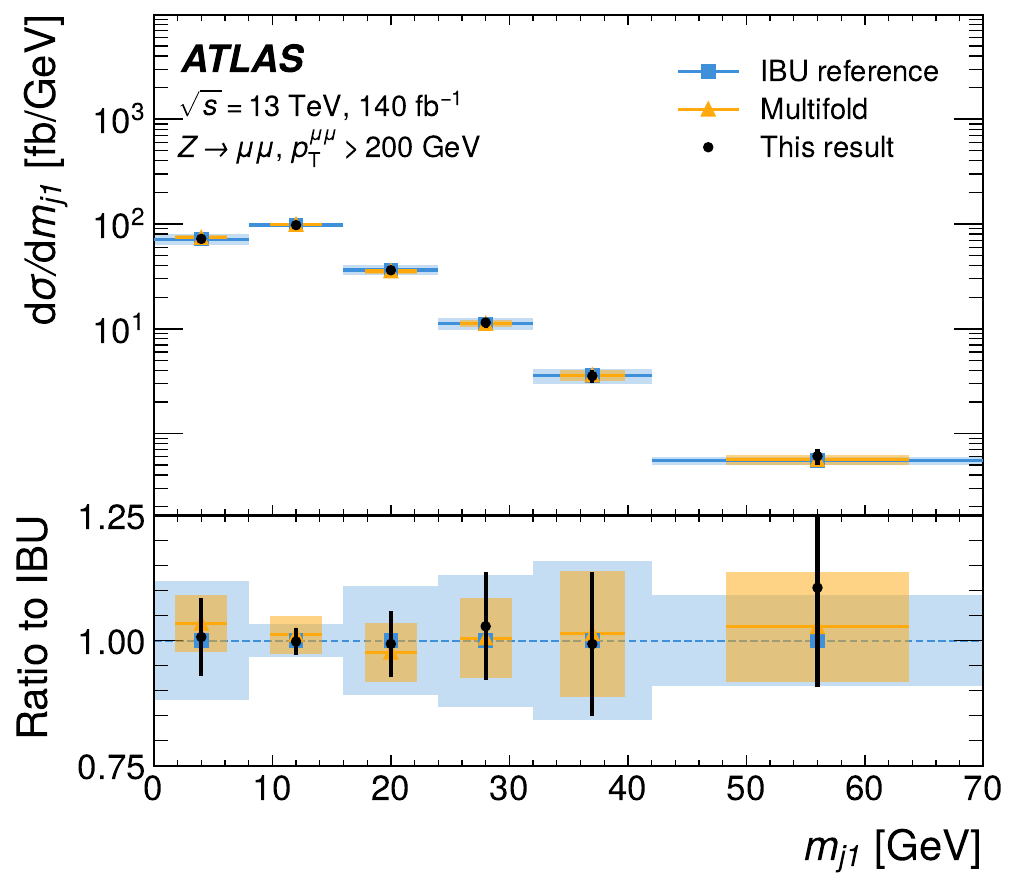}
}
\subfloat[]{
\includegraphics[width=0.45\textwidth, height=0.27\textheight, keepaspectratio]{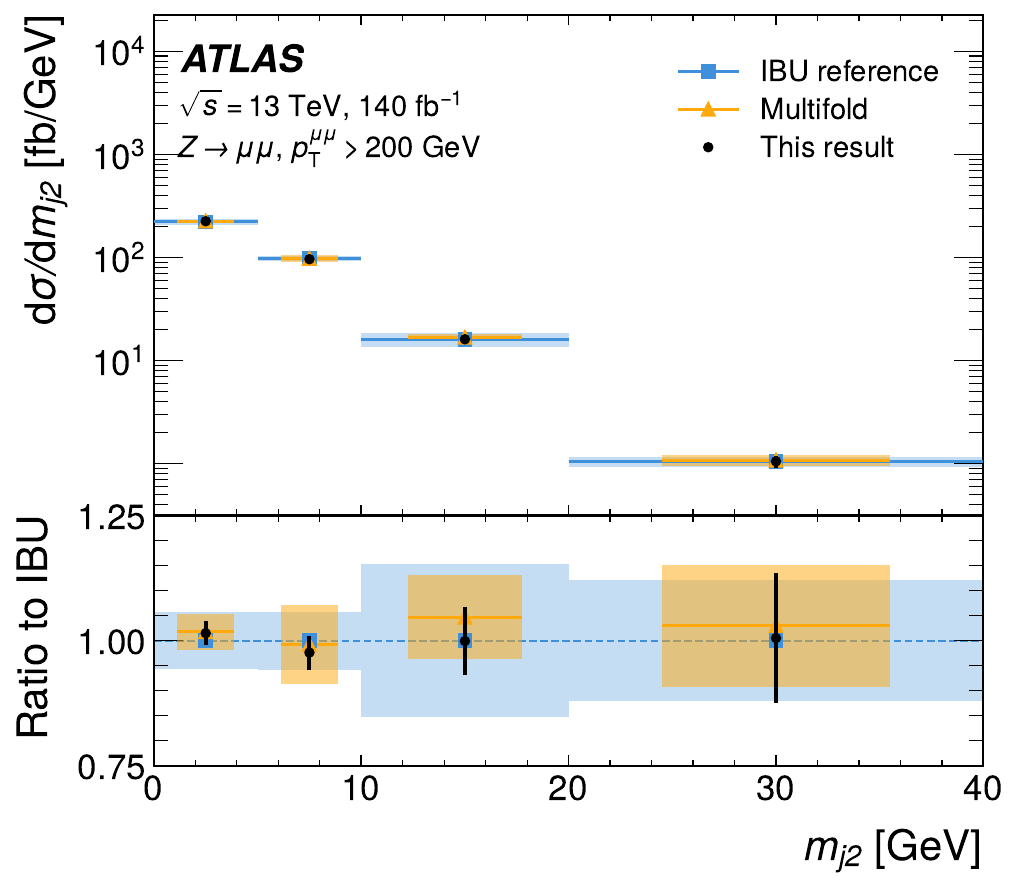}
}
\caption{Comparison of the cross-section measurements obtained with IBU, \multifold{}~\cite{STDM-2020-17} and \omnifold{} (labeled `This result'). The panels show:
the charged-particle multiplicity of the (a) leading and (b) subleading charged-hadron jet; the pseudorapidity of the (c) leading and (d) subleading muon;
the mass of the (e) leading and (f) subleading charged-hadron jet.
Excellent agreement is observed among the three measurements for all observables, despite the substantially different scope of the underlying measurements.}
\label{fig:comparison_cross_section_1}
\end{figure}

\begin{figure}[p]
\centering
\subfloat[]{
\includegraphics[width=0.45\textwidth, height=0.27\textheight, keepaspectratio]{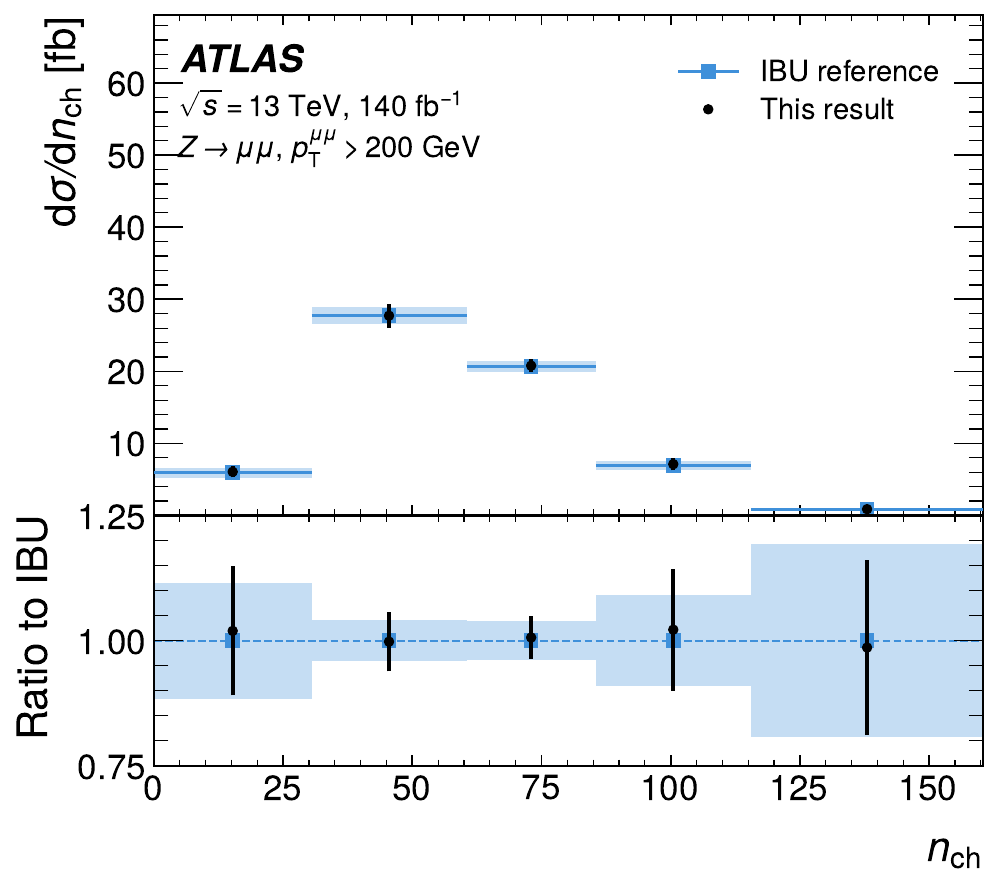}
}
\subfloat[]{
\includegraphics[width=0.45\textwidth, height=0.27\textheight, keepaspectratio]{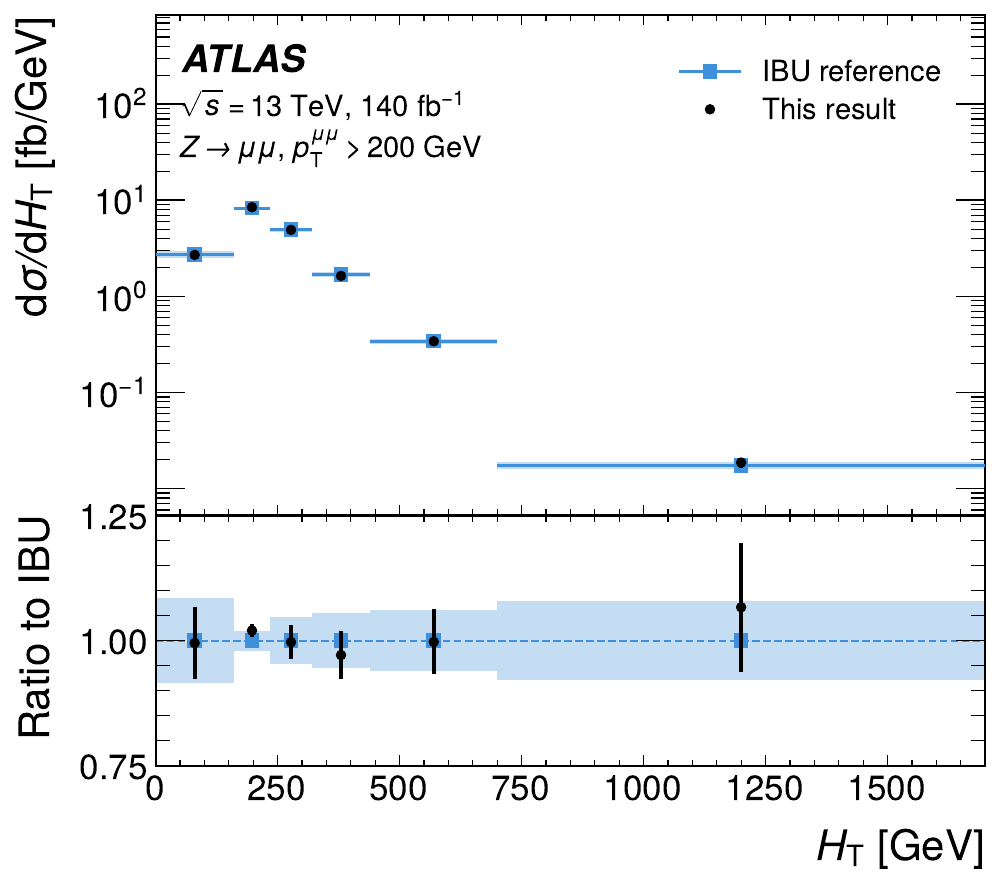}
}
\caption{Comparison of the cross-section measurements obtained with IBU (labeled `IBU') and \omnifold{} (labeled `This result') for the (a) \nch{} and (b) \HT{} observables. These observables were not measured using \multifold{}~\cite{STDM-2020-17}.}
\label{fig:comparison_cross_section_5}
\end{figure}

Figures~\ref{fig:comparison_cross_section_1}--\ref{fig:comparison_cross_section_5} present comparisons of the three measurements for the 24 previously measured observables and the two additional observables considered here. Excellent agreement is observed for all observables considered. Although the projected measurements are similar, the scope of the underlying measurements differs substantially.

Each IBU result corresponds to an independent unfolding performed for a single observable. Consequently, measuring an additional observable requires repeating the complete unfolding procedure. In contrast, \multifold{} simultaneously unfolds 24 observables within a single event-level measurement, although it is not differential in the \nch{} and \HT{} observables and therefore no corresponding \multifold{} results are shown in Figure~\ref{fig:comparison_cross_section_5}. Finally, the \omnifold{} measurement is differential in the full kinematics of the final-state charged particles.
The same published event-level representation is used to construct all 26 observables shown here, and the four example analyses presented in Section~\ref{sec:result}.

\begin{figure}[p]
\centering
\subfloat[]{
\includegraphics[width=0.45\textwidth, height=0.27\textheight, keepaspectratio]{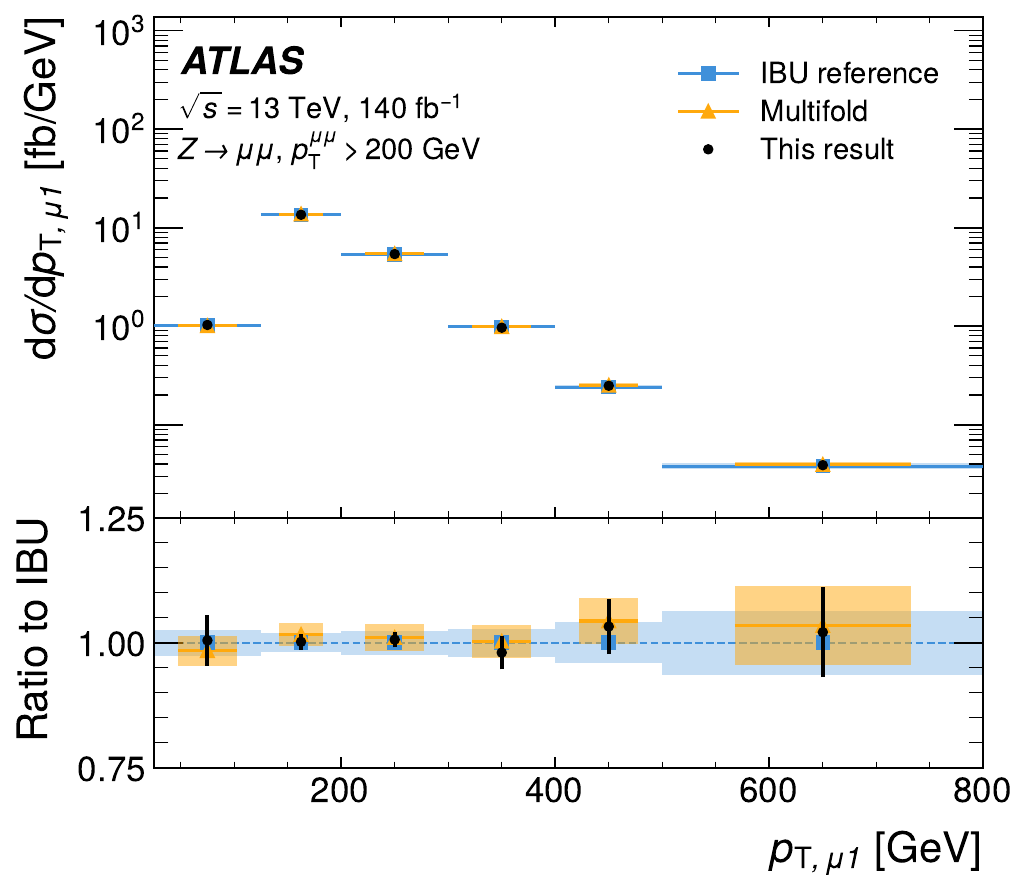}
}
\subfloat[]{
\includegraphics[width=0.45\textwidth, height=0.27\textheight, keepaspectratio]{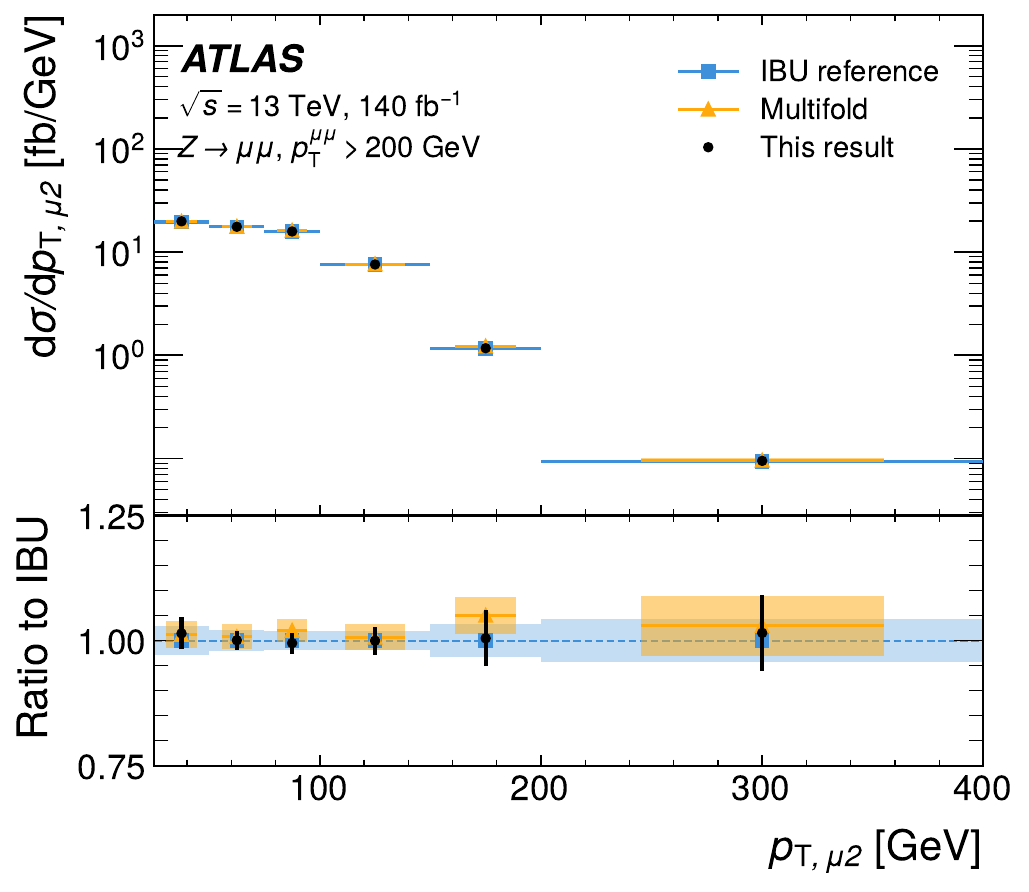}
}\\
\subfloat[]{
\includegraphics[width=0.45\textwidth, height=0.27\textheight, keepaspectratio]{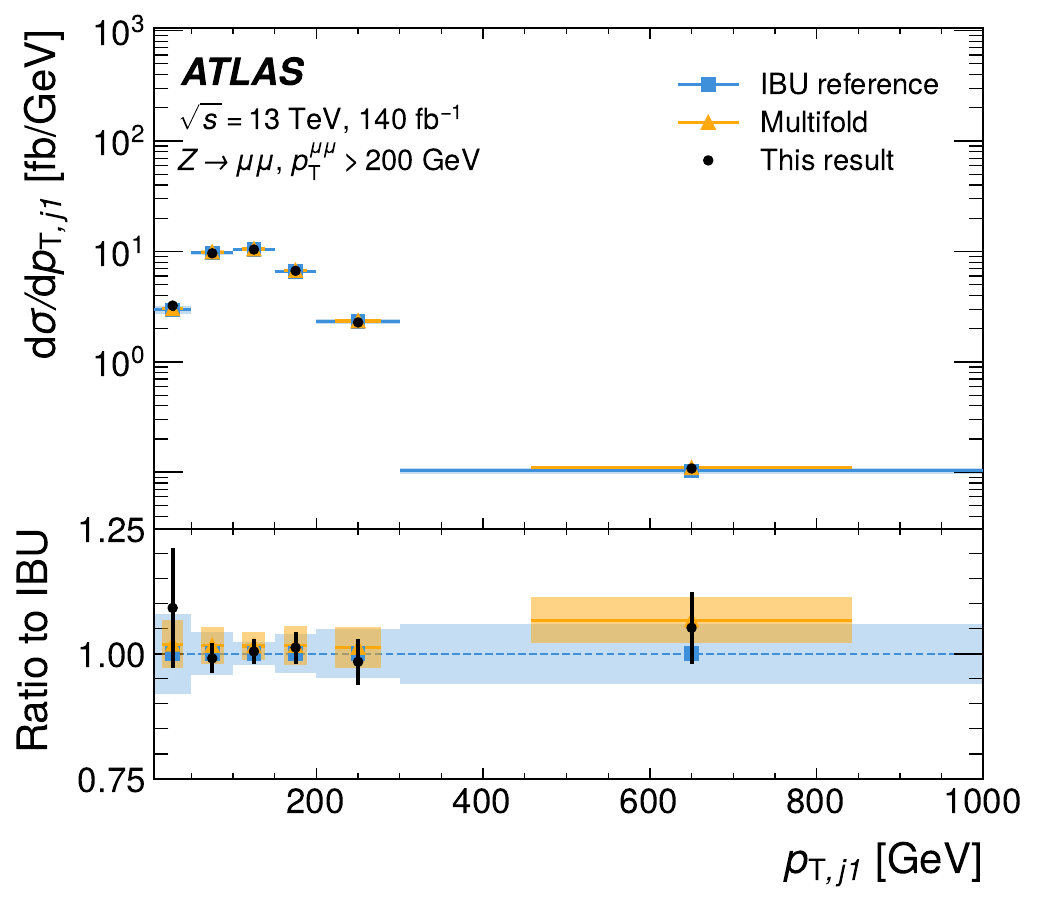}
}
\subfloat[]{
\includegraphics[width=0.45\textwidth, height=0.27\textheight, keepaspectratio]{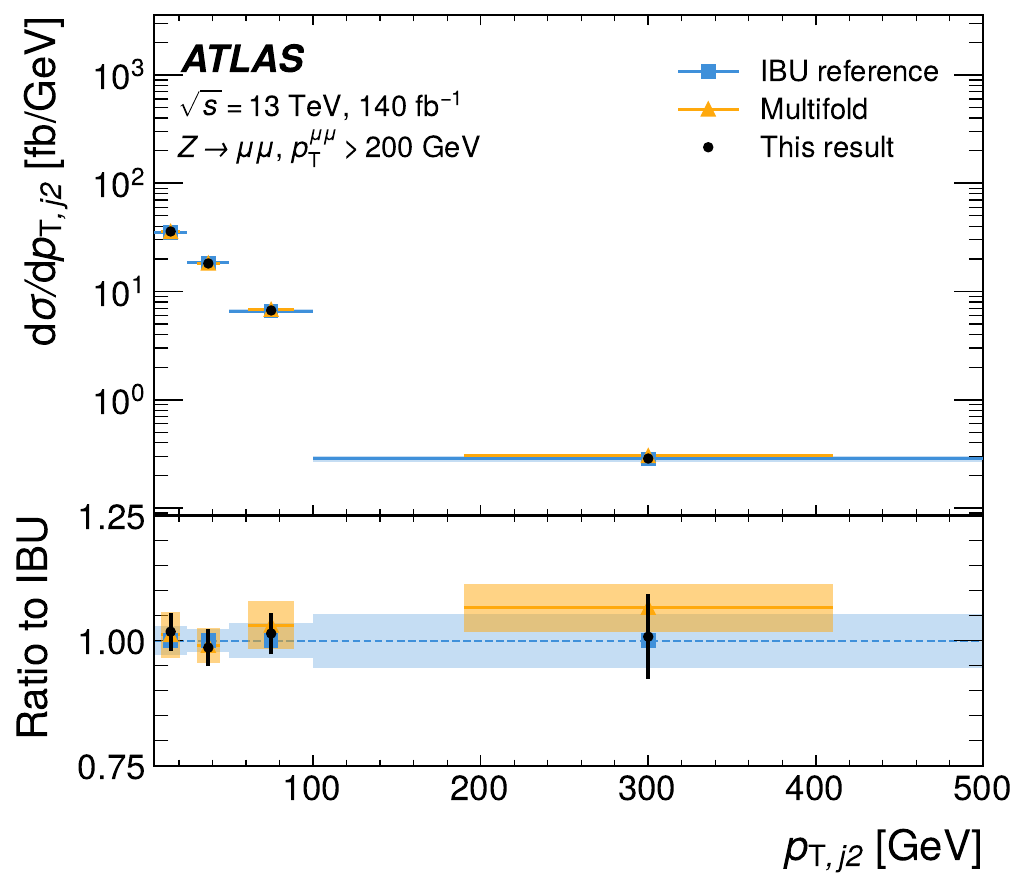}
}\\
\subfloat[]{
\includegraphics[width=0.45\textwidth, height=0.27\textheight, keepaspectratio]{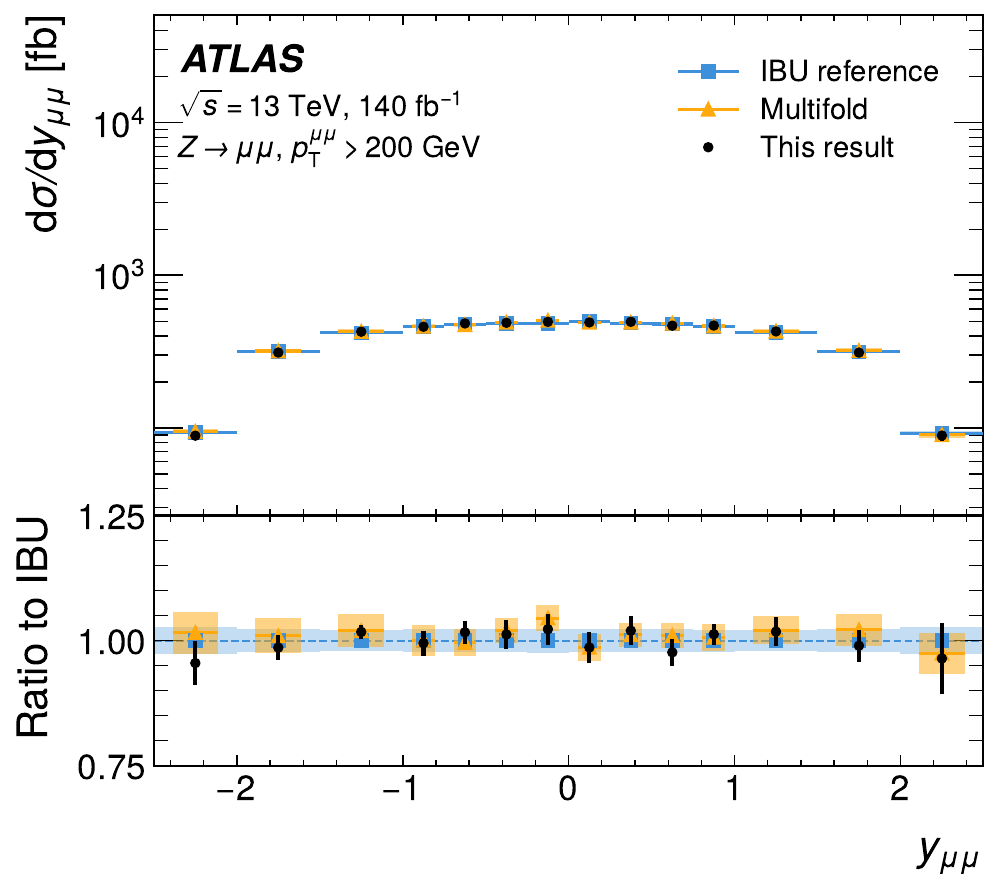}
}\\
\caption{Comparison of the cross-section measurements obtained with IBU, \multifold{}~\cite{STDM-2020-17} and \omnifold{} (labeled `This result') for the transverse momentum of the (a) leading and (b) subleading muon, the transverse momentum of the (c) leading and (d) subleading charged-hadron jets, and (e) the rapidity of the dimuon system.}
\end{figure}

\begin{figure}[p]
\centering
\subfloat[]{
\includegraphics[width=0.45\textwidth, height=0.27\textheight, keepaspectratio]{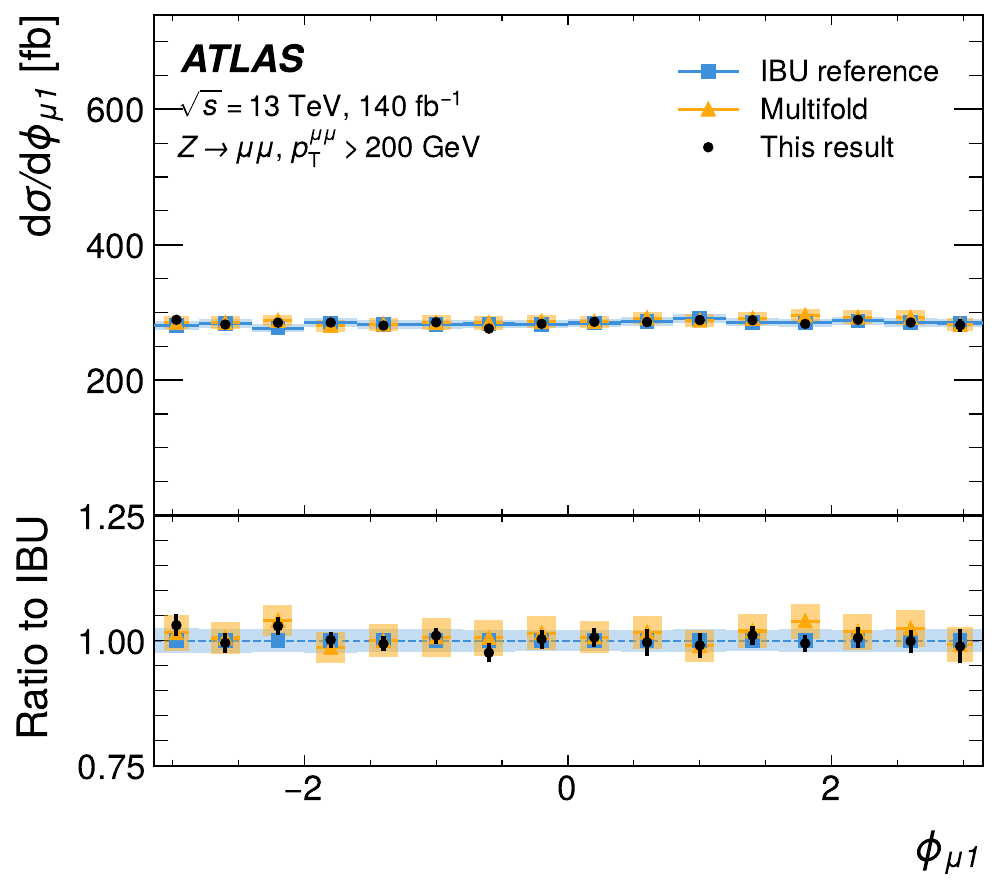}
}
\subfloat[]{
\includegraphics[width=0.45\textwidth, height=0.27\textheight, keepaspectratio]{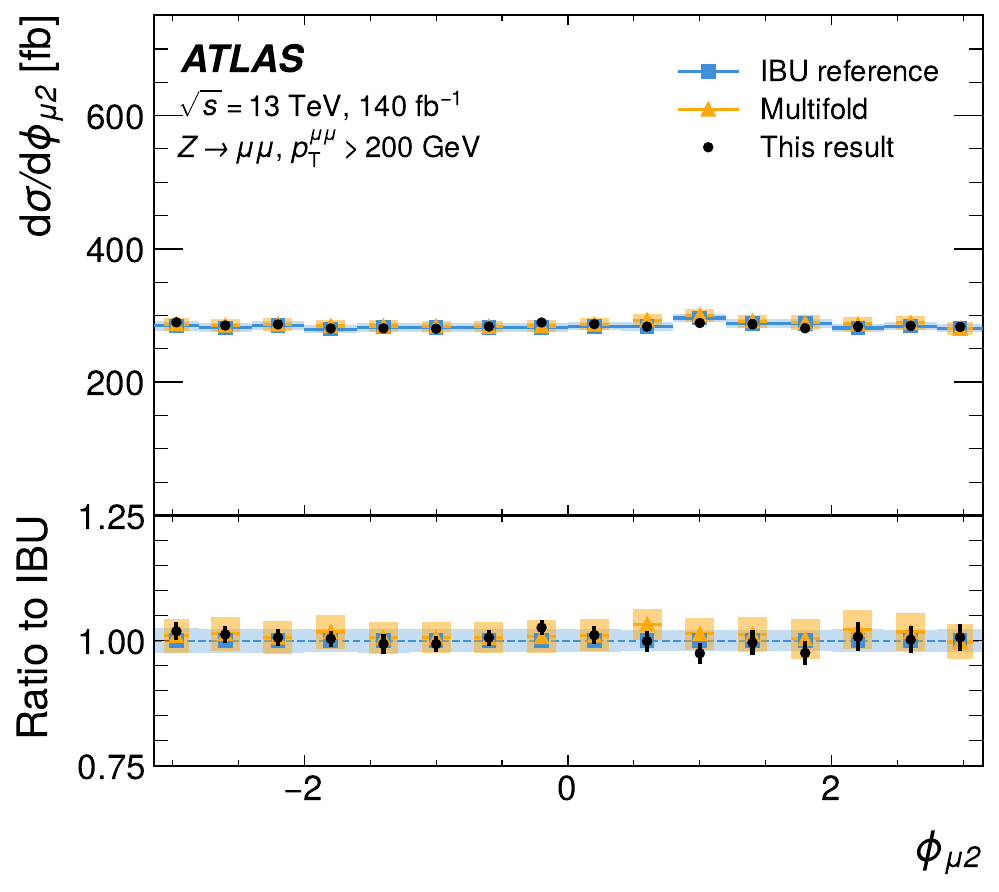}
}\\
\subfloat[]{
\includegraphics[width=0.45\textwidth, height=0.27\textheight, keepaspectratio]{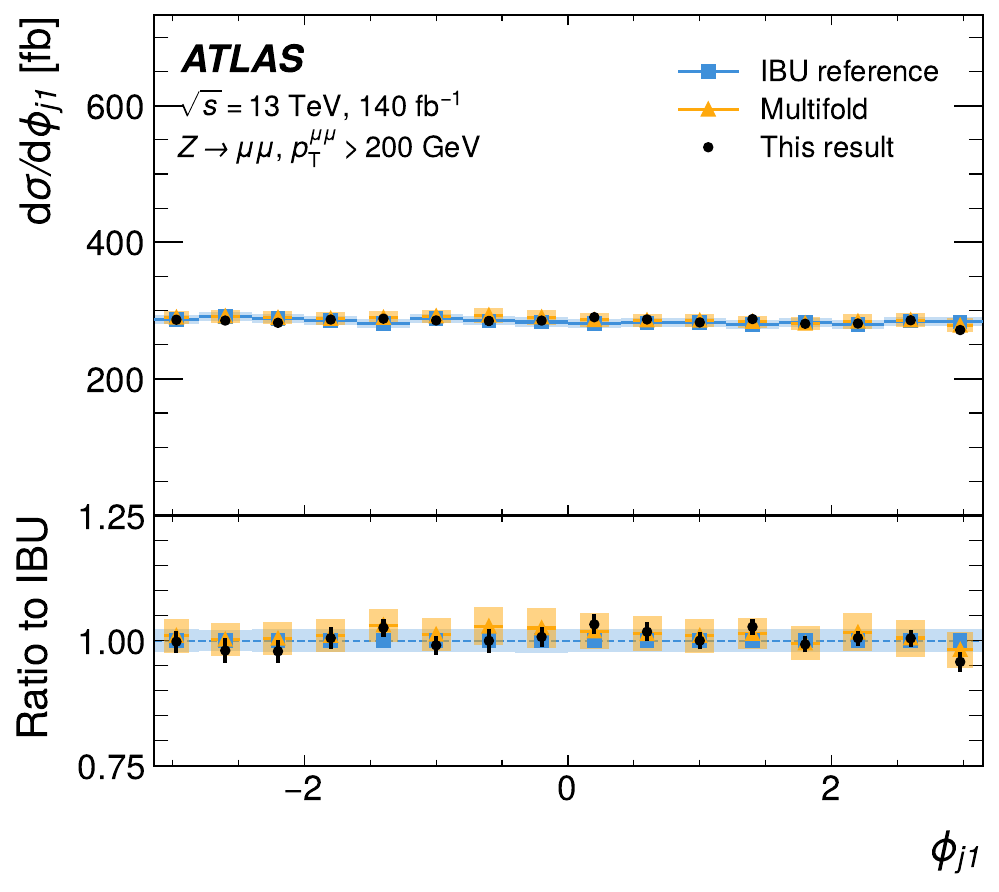}
}
\subfloat[]{
\includegraphics[width=0.45\textwidth, height=0.27\textheight, keepaspectratio]{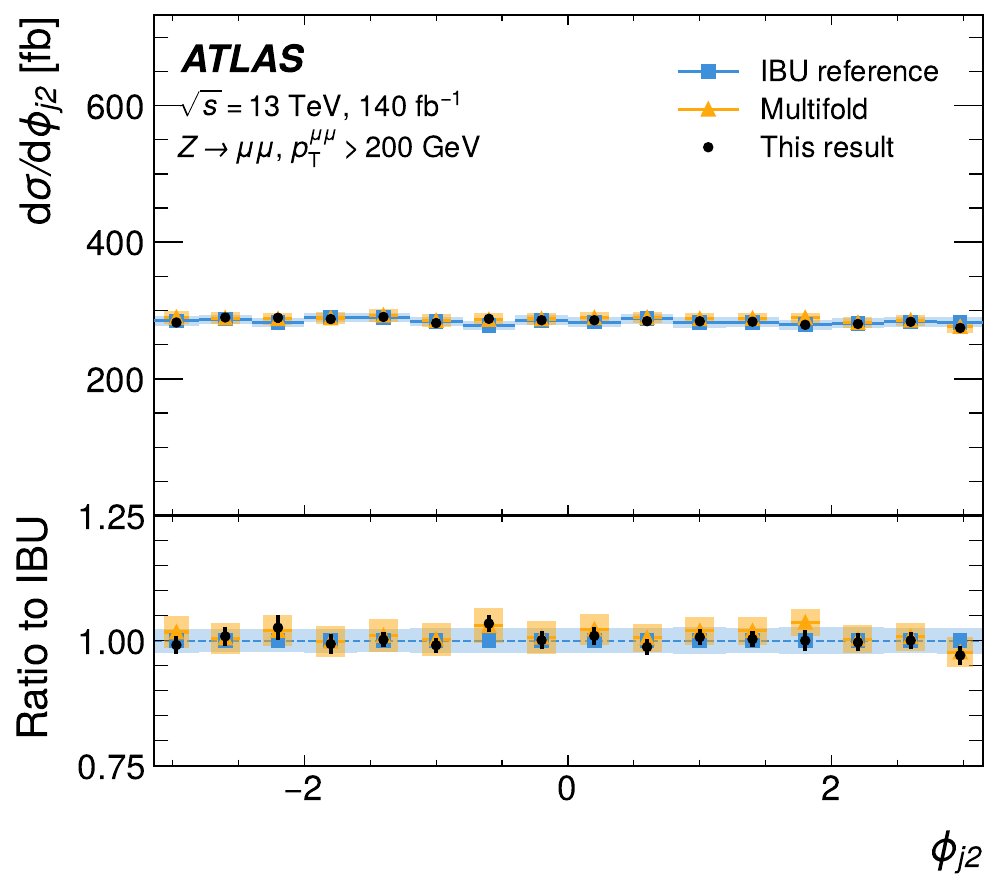}
}\\
\subfloat[]{
\includegraphics[width=0.45\textwidth, height=0.27\textheight, keepaspectratio]{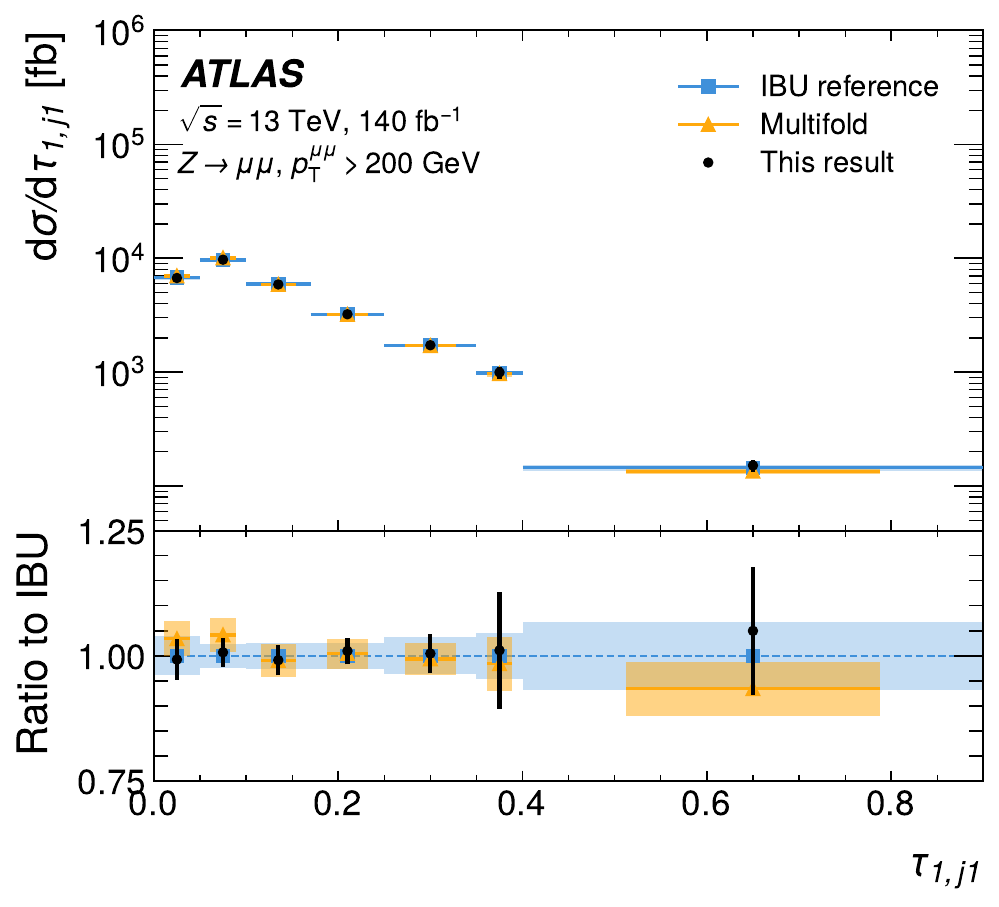}
}
\subfloat[]{
\includegraphics[width=0.45\textwidth, height=0.27\textheight, keepaspectratio]{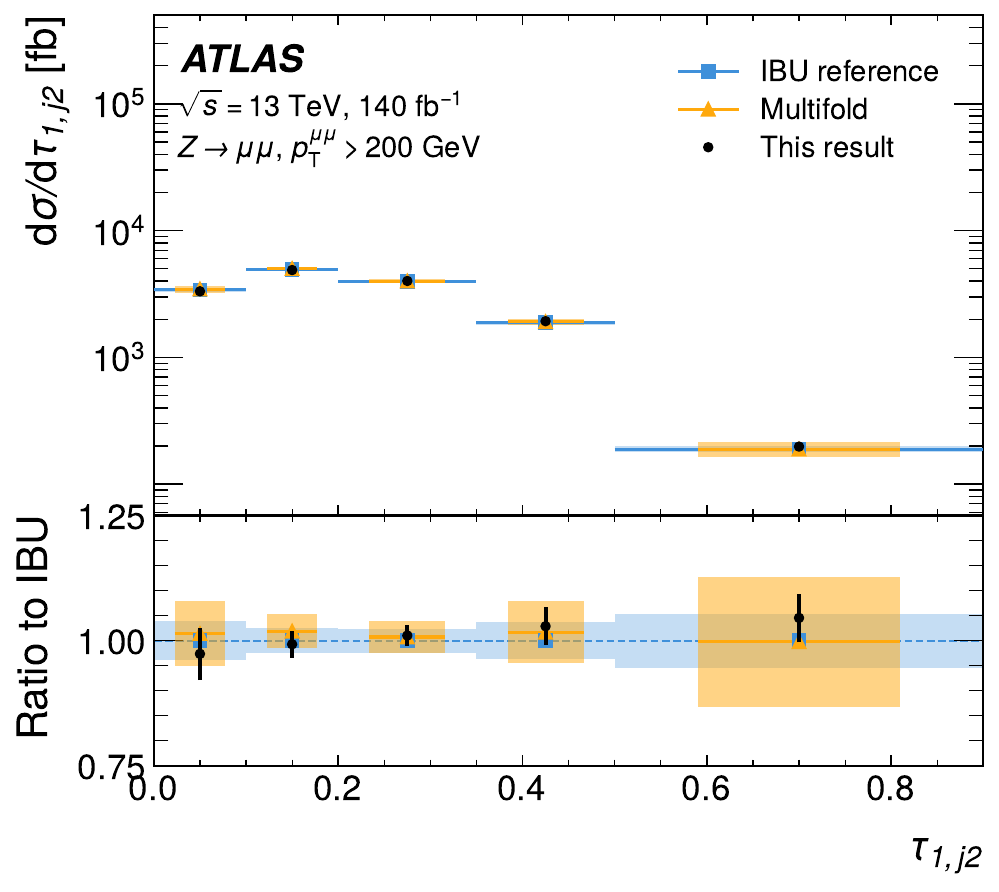}
}
\caption{Comparison of the cross-section measurements obtained with IBU, \multifold{}~\cite{STDM-2020-17} and \omnifold{} (labeled `This result') for the azimuthal angle of the (a) leading and (b) subleading muons, the (c) leading and (d) subleading charged-hadron jets, and the $1$-subjettiness~\cite{Thaler:2010tr} $\tau_1$ of the (e) leading and (f) subleading charged-hadron jets.}
\end{figure}

\begin{figure}[p]
\centering
\subfloat[]{
\includegraphics[width=0.45\textwidth, height=0.27\textheight, keepaspectratio]{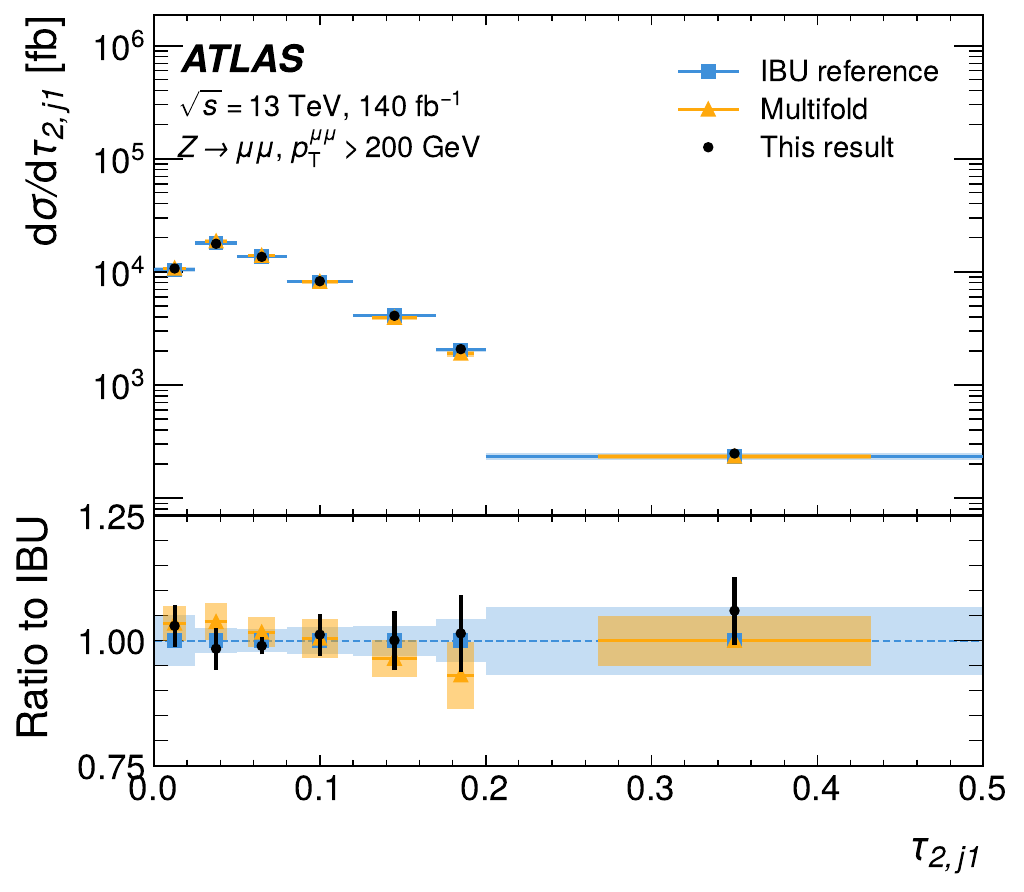}
}
\subfloat[]{
\includegraphics[width=0.45\textwidth, height=0.27\textheight, keepaspectratio]{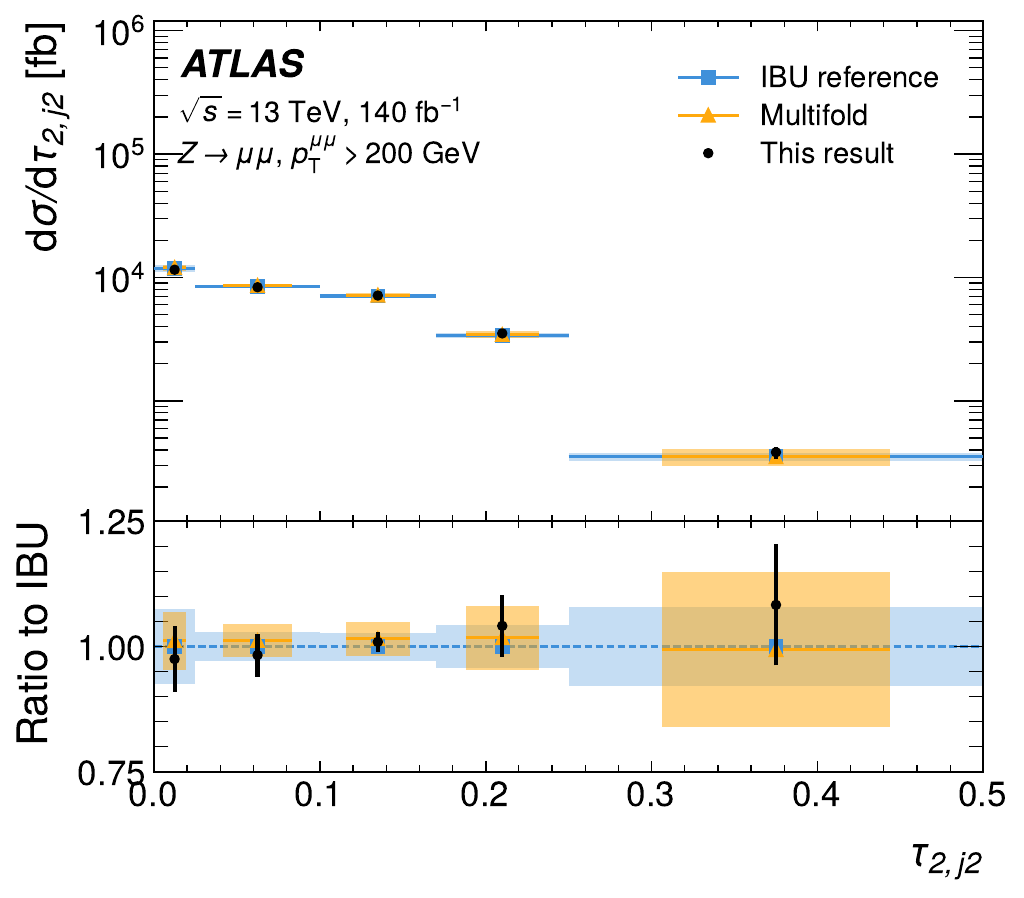}
}\\
\subfloat[]{
\includegraphics[width=0.45\textwidth, height=0.27\textheight, keepaspectratio]{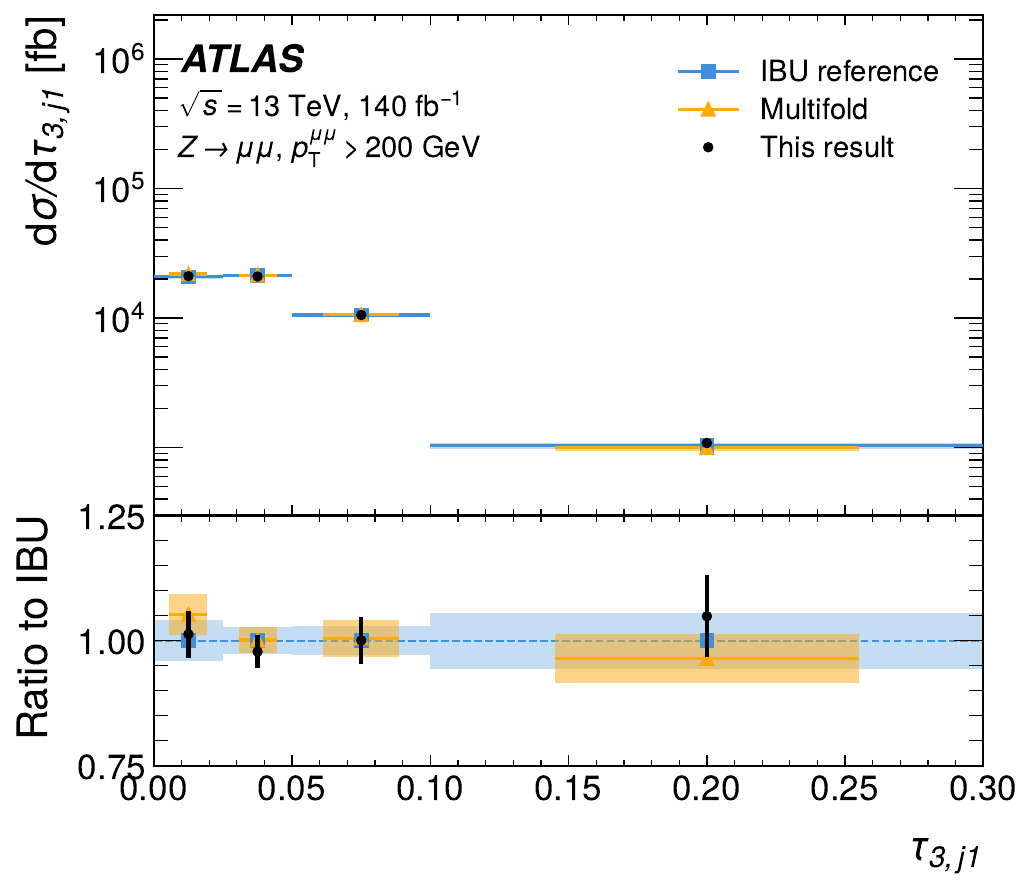}
}
\subfloat[]{
\includegraphics[width=0.45\textwidth, height=0.27\textheight, keepaspectratio]{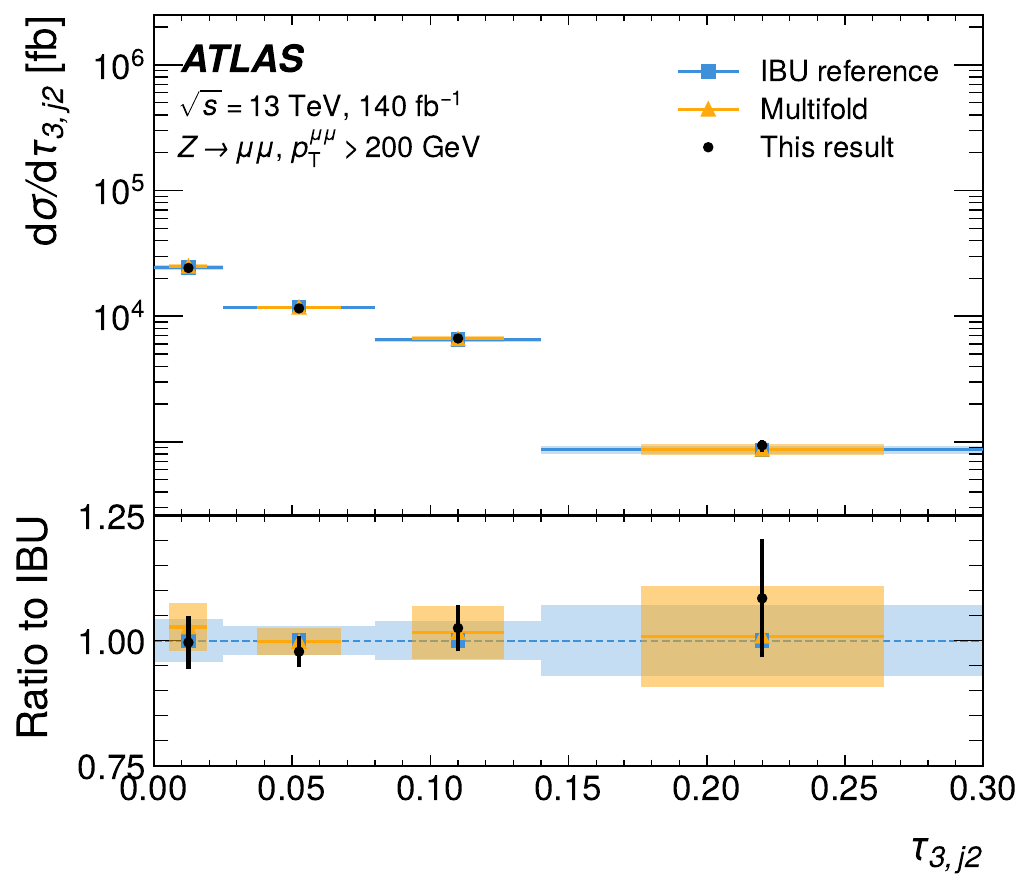}
}\\
\subfloat[]{
\includegraphics[width=0.45\textwidth, height=0.27\textheight, keepaspectratio]{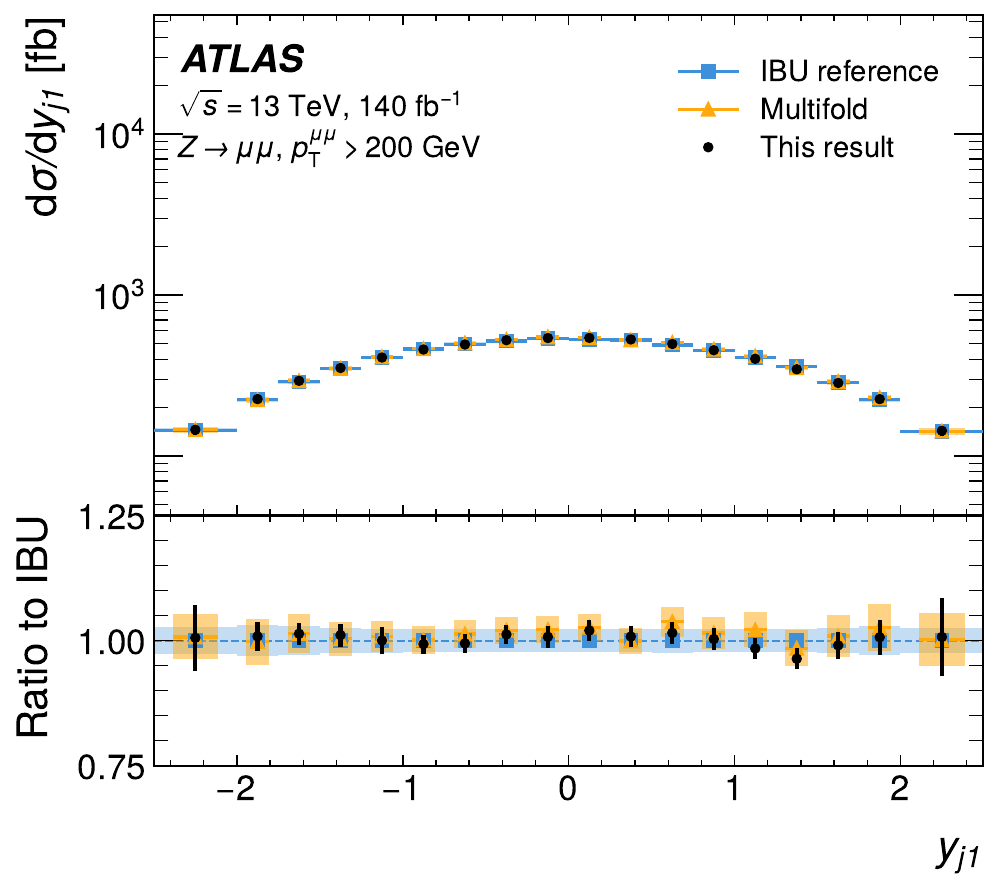}
}
\subfloat[]{
\includegraphics[width=0.45\textwidth, height=0.27\textheight, keepaspectratio]{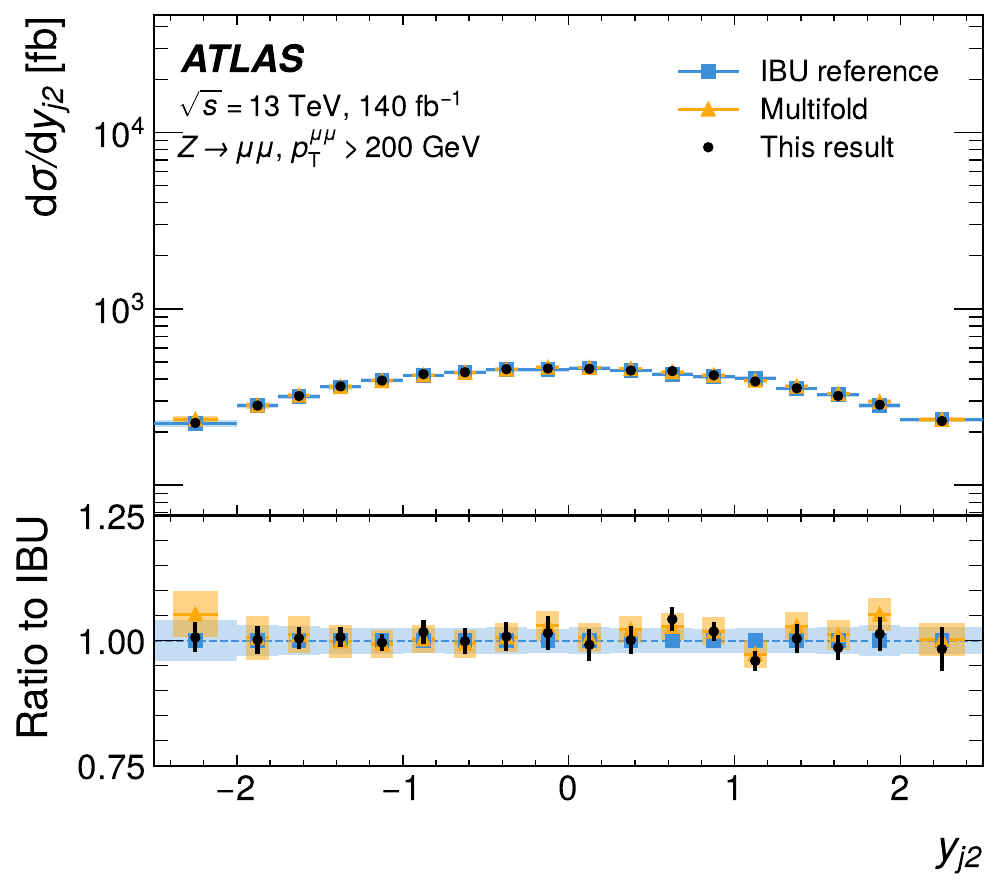}
}
\caption{Comparison of the cross-section measurements obtained with IBU, \multifold{}~\cite{STDM-2020-17} and \omnifold{} (labeled `This result') for the $2$-subjettiness $\tau_2$ of the (a) leading and (b) subleading charged-hadron jets, the $3$-subjettiness $\tau_3$ of the (c) leading and (d) subleading charged-hadron jets, and the rapidity of the (e) leading and (f) subleading charged-hadron jets.}
\end{figure}

\FloatBarrier

\subsection{Total uncertainty comparison}
\label{app:uncert_comp}

\begin{figure}[p]
\centering
\subfloat[]{
\includegraphics[width=0.47\textwidth]{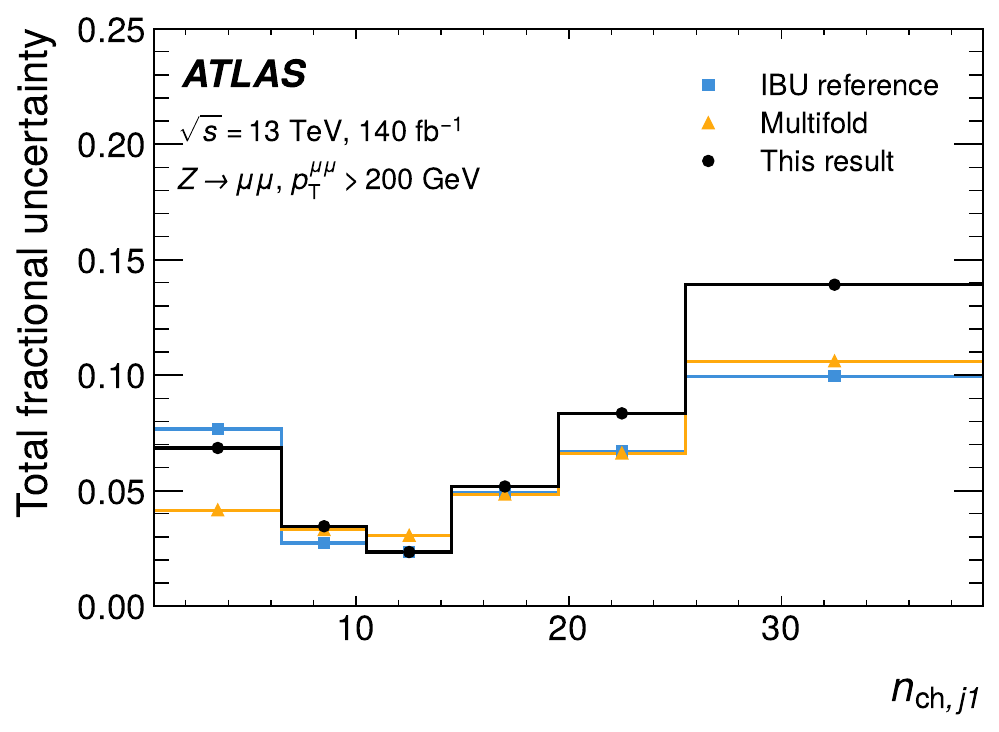}
}
\subfloat[]{
\includegraphics[width=0.47\textwidth]{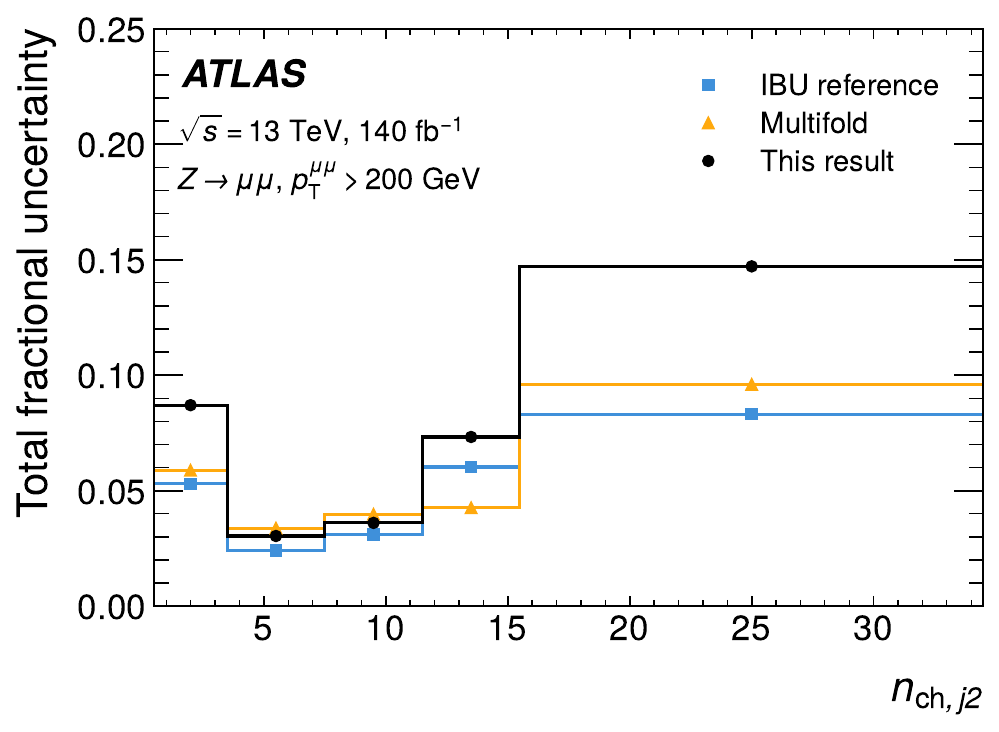}
}\\
\subfloat[]{
\includegraphics[width=0.47\textwidth]{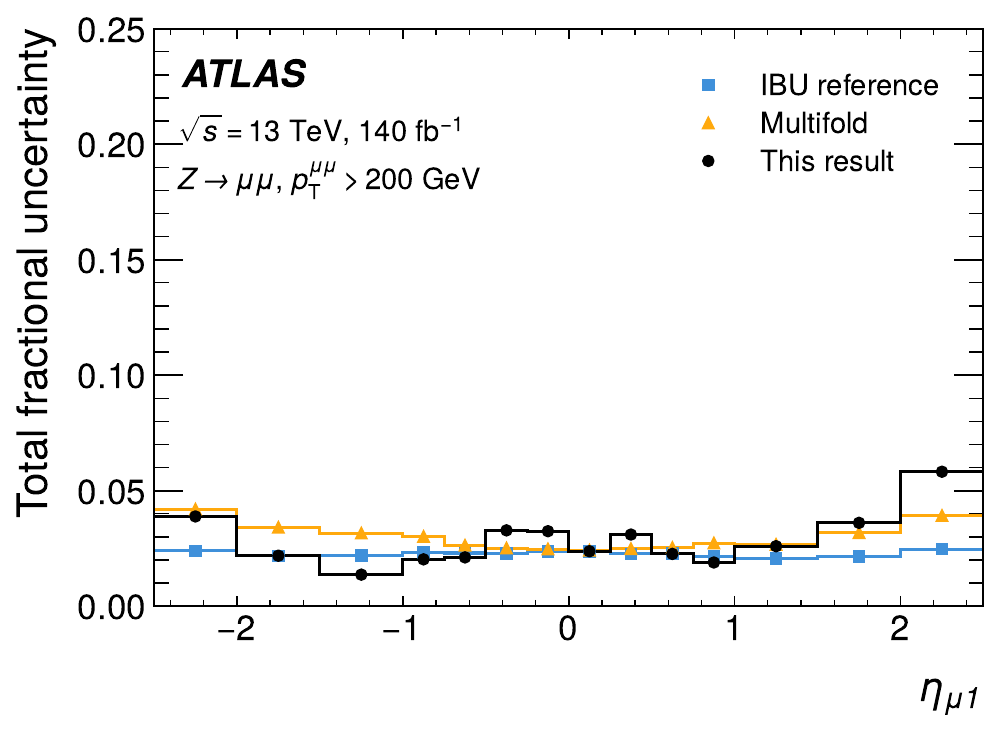}
}
\subfloat[]{
\includegraphics[width=0.47\textwidth]{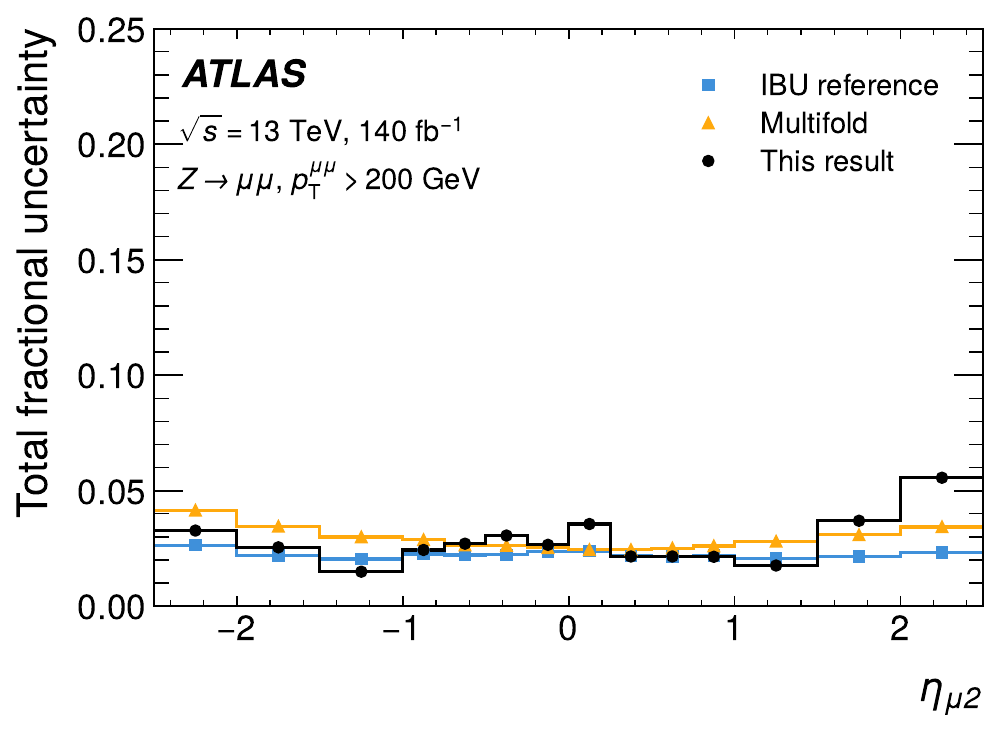}
}\\
\subfloat[]{
\includegraphics[width=0.47\textwidth]{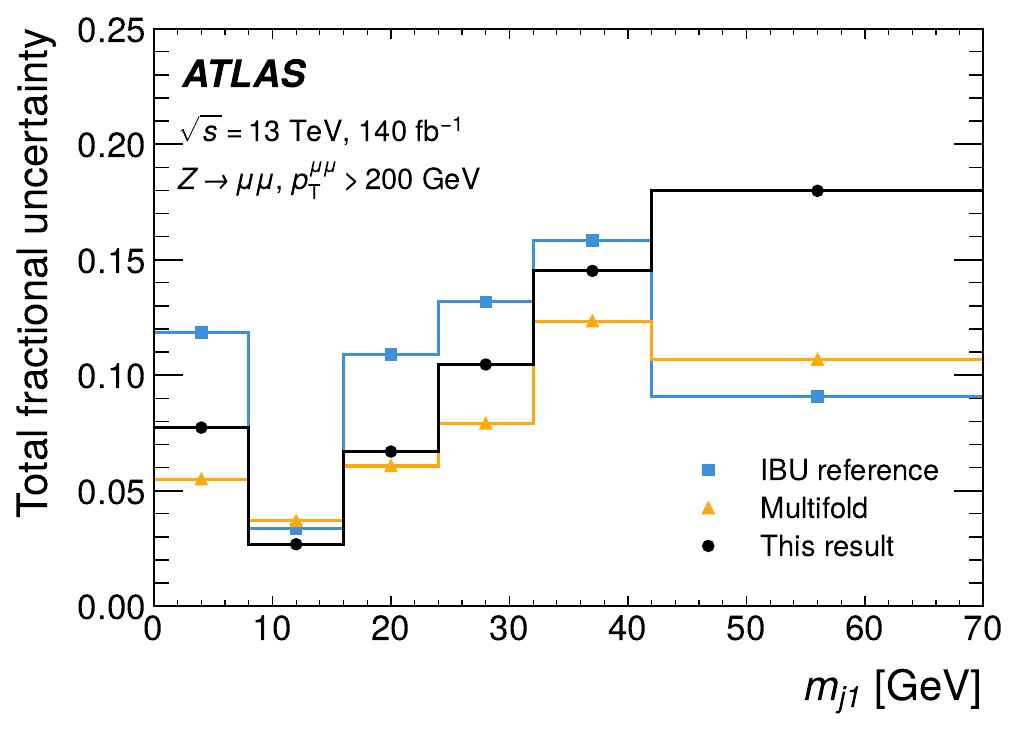}
}
\subfloat[]{
\includegraphics[width=0.47\textwidth]{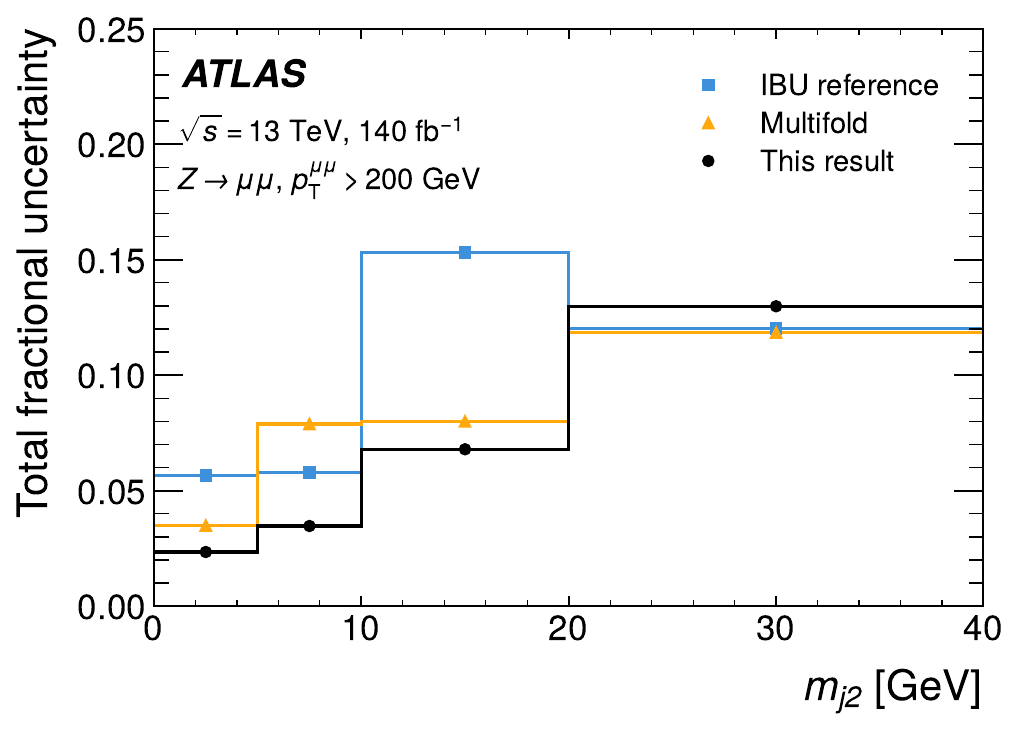}
}
\caption{Comparison of the total uncertainty estimates obtained with IBU, \multifold{}~\cite{STDM-2020-17} and \omnifold{} (labeled `This result').
The panels show the charged-particle multiplicity of the (a) leading and (b) subleading charged-hadron jet, (c) the leading-muon pseudorapidity, (d) the subleading-muon pseudorapidity, the mass of the (e) leading and (f) subleading charged-hadron jet.}
\label{fig:comparison_uncertainty_1}
\end{figure}

\begin{figure}[p]
\centering
\subfloat[]{
\includegraphics[width=0.47\textwidth]{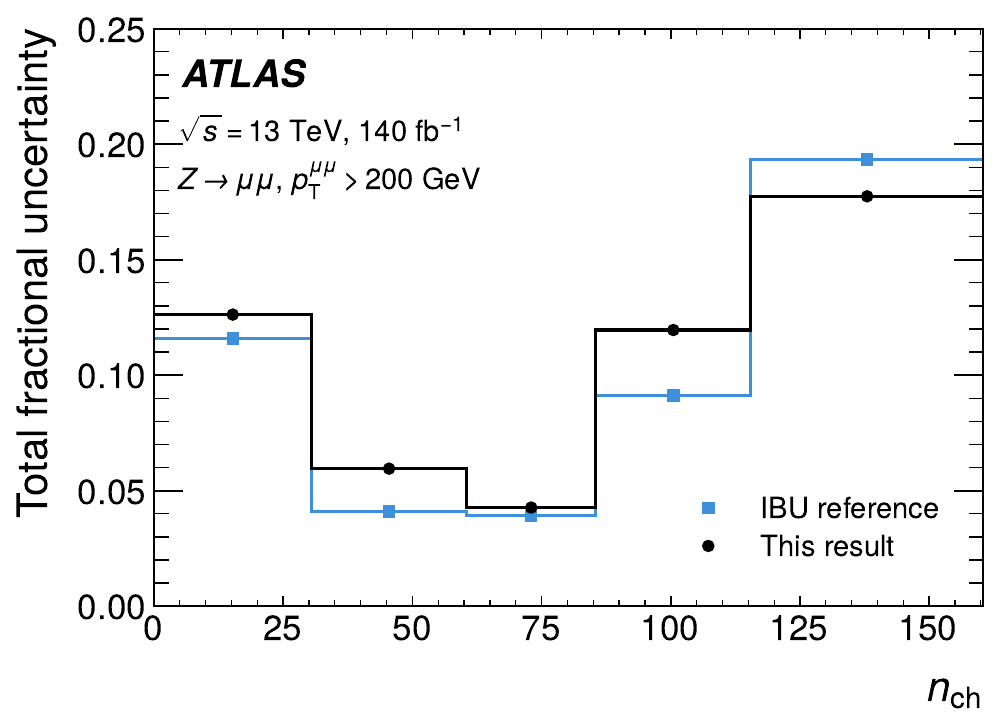}
}
\subfloat[]{
\includegraphics[width=0.47\textwidth]{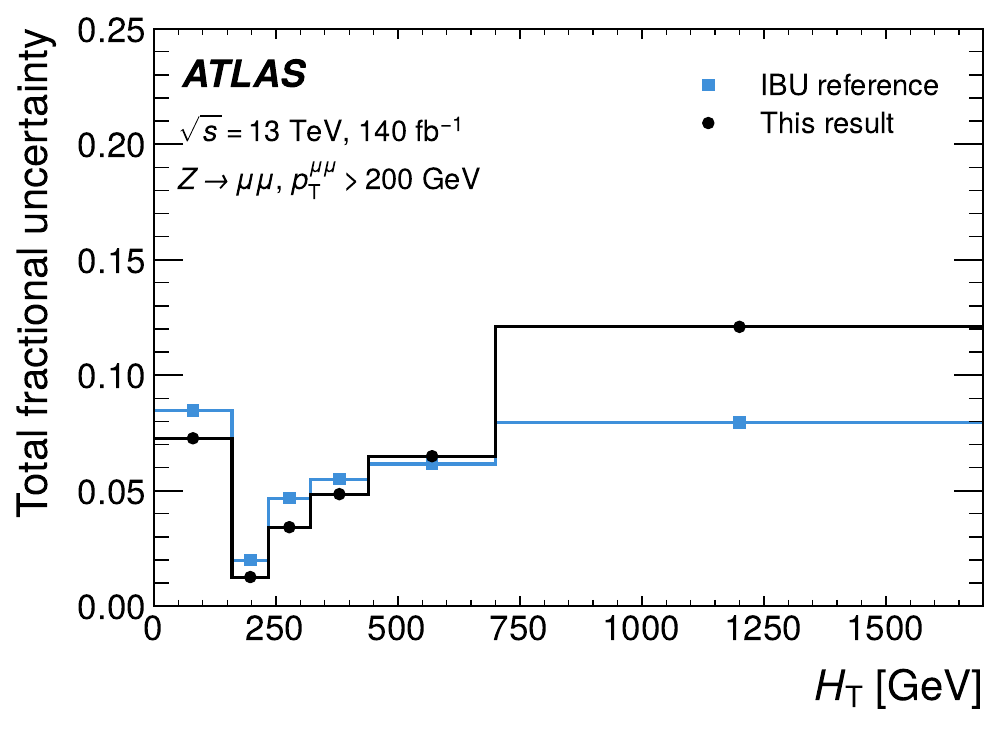}
}
\caption{Comparison of the total uncertainty estimates obtained with IBU (labeled `IBU') and \omnifold{} (labeled `This result') for the (a) \nch{} and (b) \HT{} observables. These observables were not measured using \multifold{}~\cite{STDM-2020-17}.}
\label{fig:comparison_uncertainty_5}
\end{figure}

Figures~\ref{fig:comparison_uncertainty_1}--\ref{fig:comparison_uncertainty_5} compare the total uncertainties obtained with the \omnifold{}, \multifold{}, and IBU measurements for each of the 26 observables. Overall, the projected uncertainties are remarkably similar across the three approaches. Despite deriving a measurement that is differential in the full charged-particle phase space, the \omnifold{} measurement incurs little or no uncertainty penalty relative to dedicated one-dimensional IBU measurements for most observables. For some jet observables, most notably the leading and subleading jet masses, the \omnifold{} uncertainties are even smaller than those obtained with IBU\@. The largest improvements relative to the previous \multifold{} measurement are observed for observables dominated by the muon kinematics, reflecting the improved stability of the neural-network training used in the present work.

\begin{figure}[p]
\centering
\subfloat[]{
\includegraphics[width=0.47\textwidth]{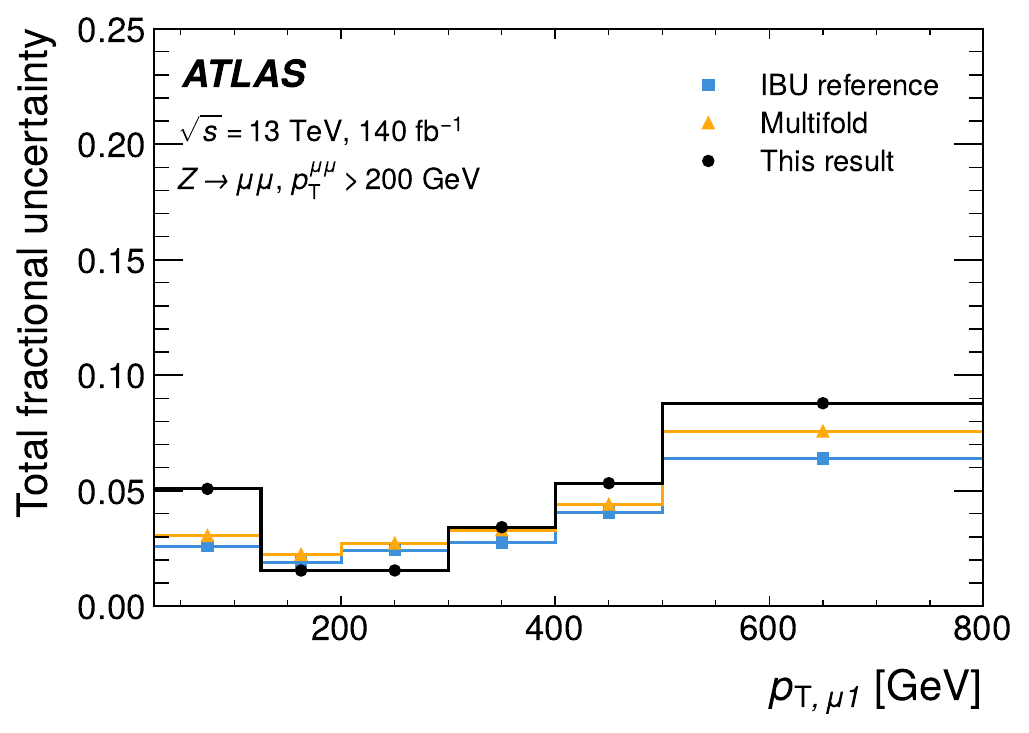}
}
\subfloat[]{
\includegraphics[width=0.47\textwidth]{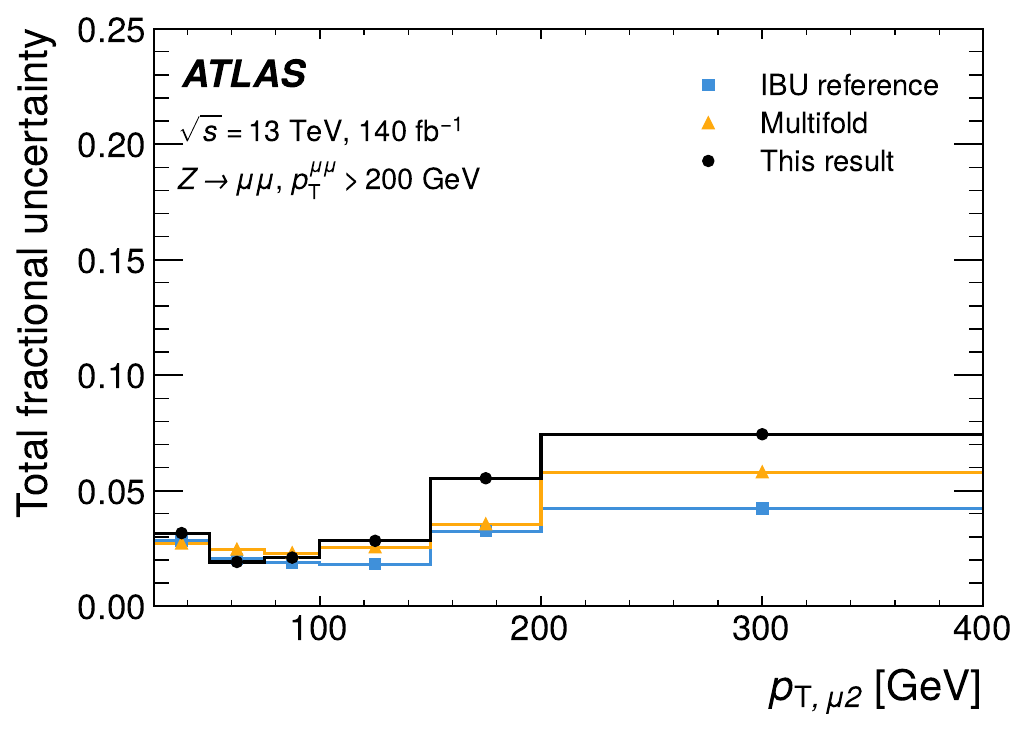}
}\\
\subfloat[]{
\includegraphics[width=0.47\textwidth]{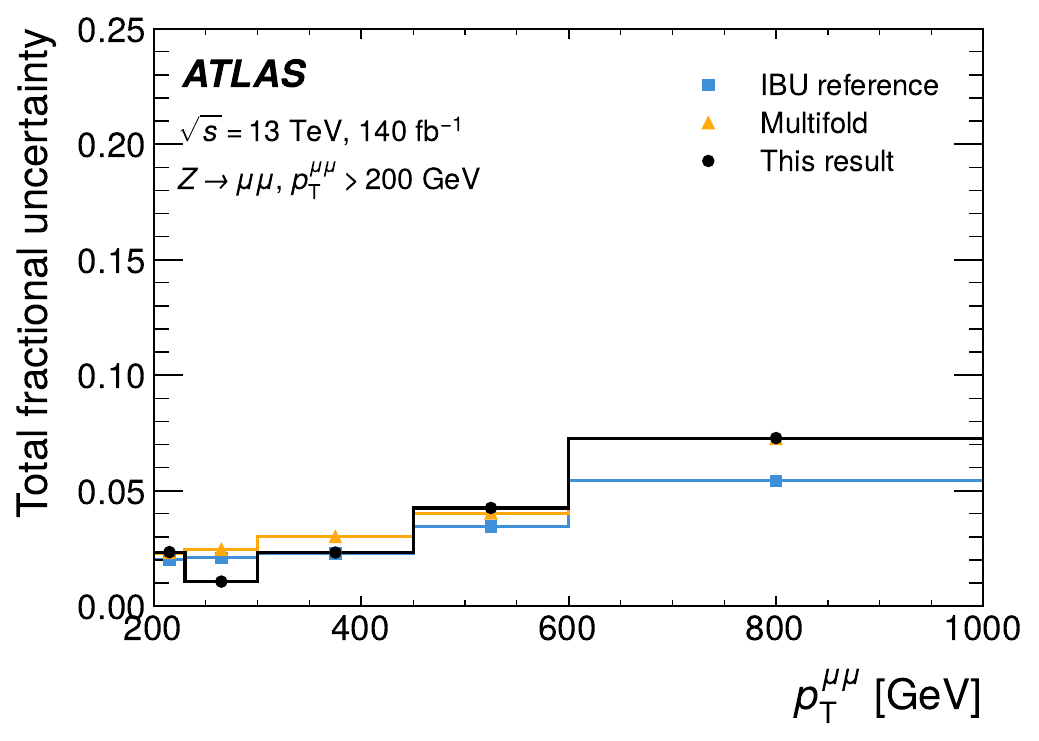}
}
\subfloat[]{
\includegraphics[width=0.47\textwidth]{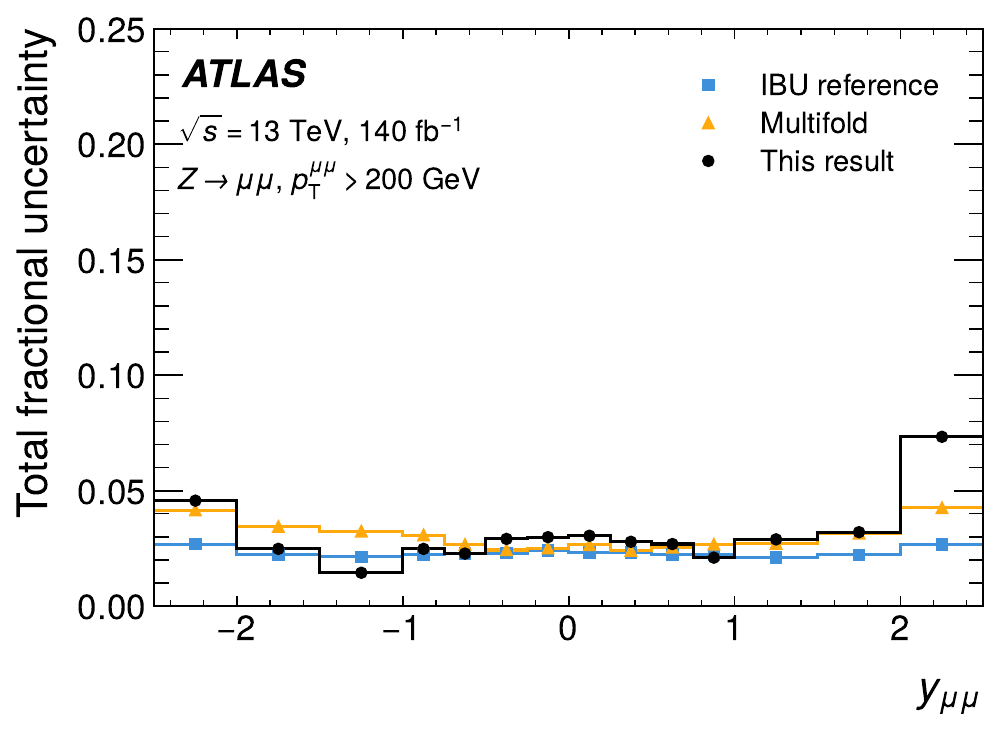}
}\\
\subfloat[]{
\includegraphics[width=0.47\textwidth]{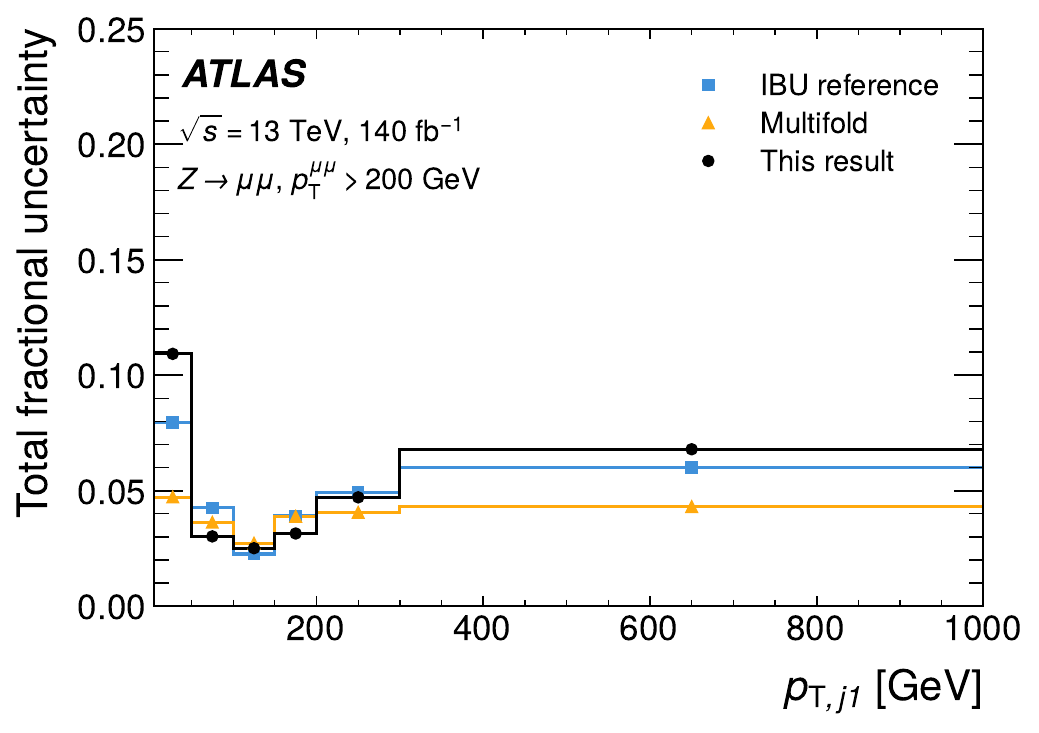}
}
\subfloat[]{
\includegraphics[width=0.47\textwidth]{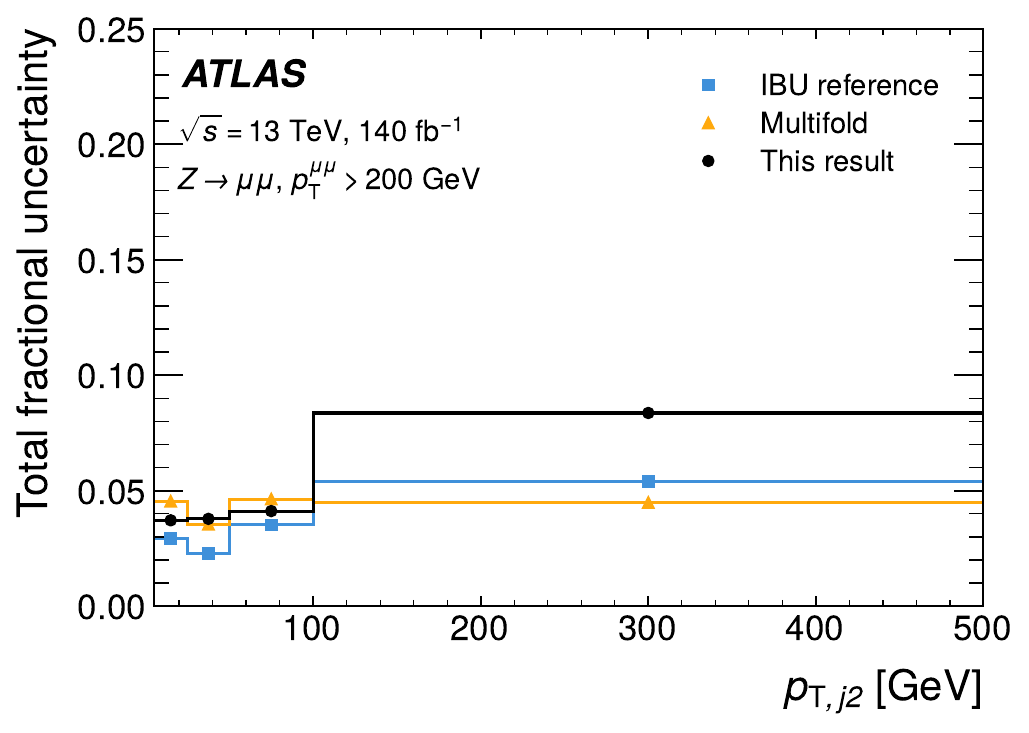}
}
\caption{Comparison of the total uncertainty estimates obtained with IBU, \multifold{}~\cite{STDM-2020-17} and \omnifold{} (labeled `This result') for the \pt of the (a) leading and (b) subleading muon, the (c) transverse momentum and the (d) rapidity of the dimuon system, and the transverse momentum of the (e) leading and (f) subleading charge-hadron jets.}
\end{figure}

\begin{figure}[p]
\centering
\subfloat[]{
\includegraphics[width=0.47\textwidth]{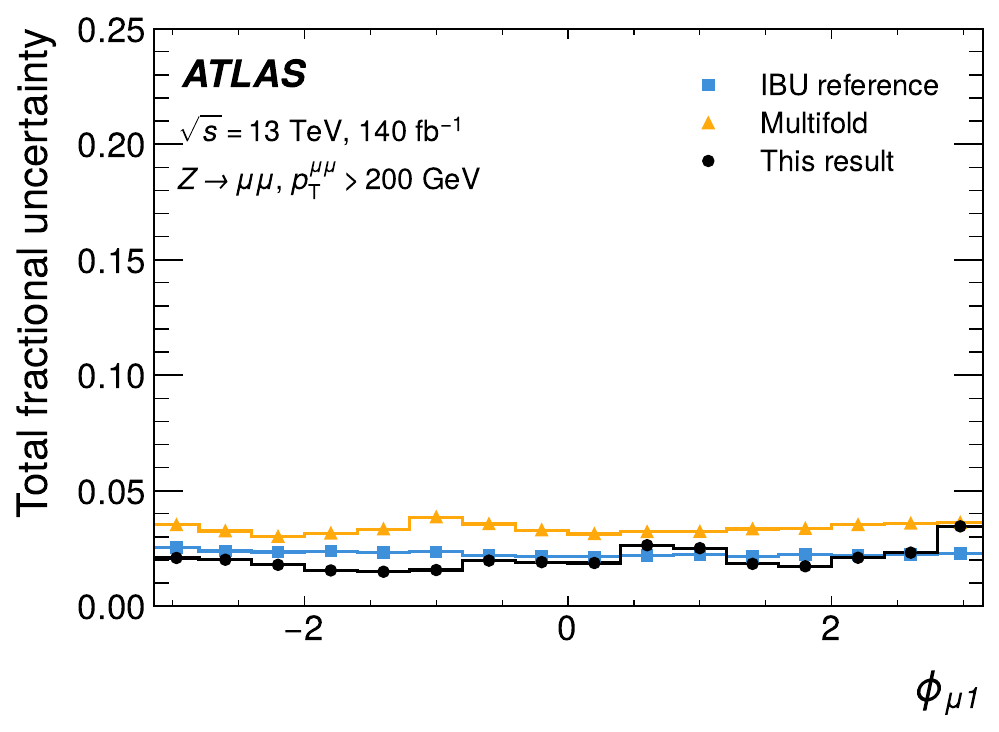}
}
\subfloat[]{
\includegraphics[width=0.47\textwidth]{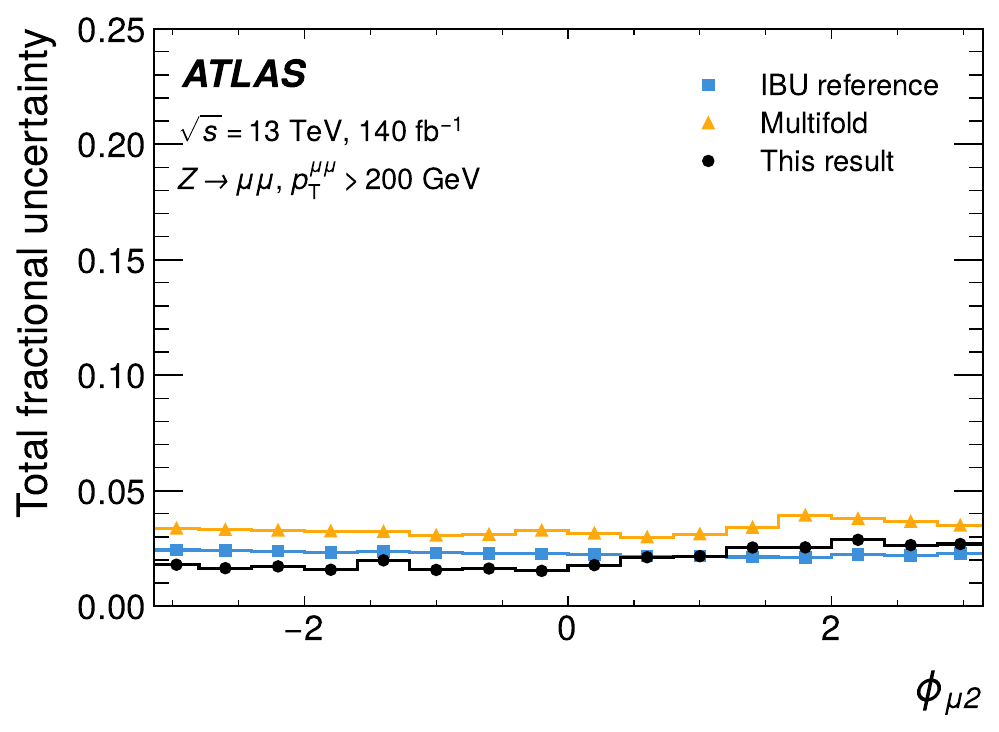}
}\\
\subfloat[]{
\includegraphics[width=0.47\textwidth]{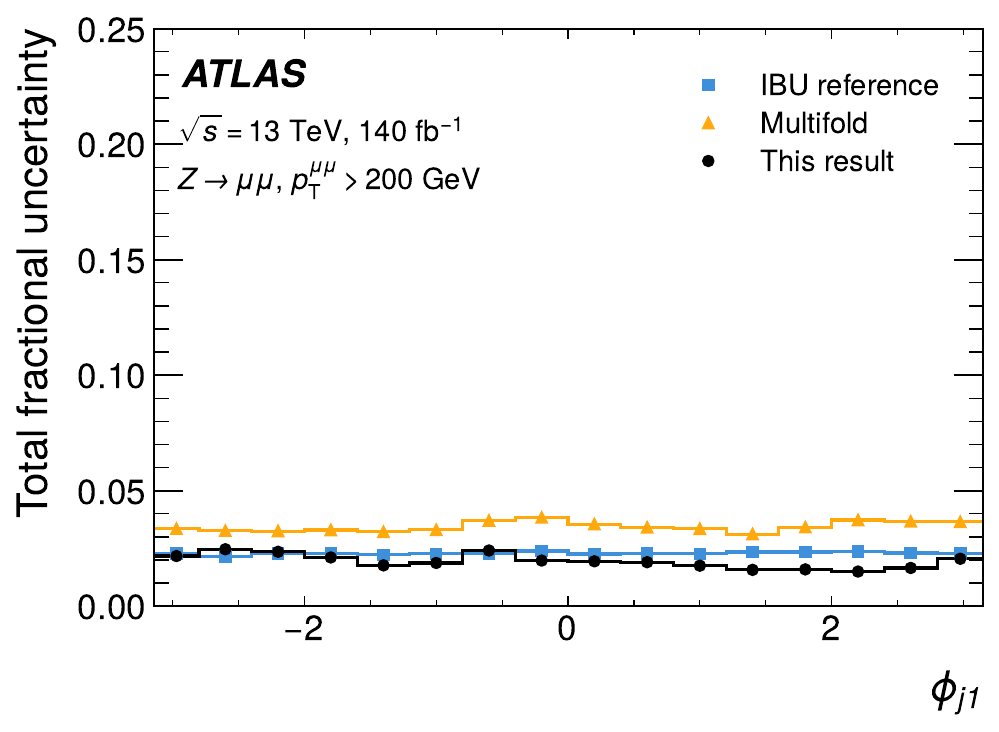}
}
\subfloat[]{
\includegraphics[width=0.47\textwidth]{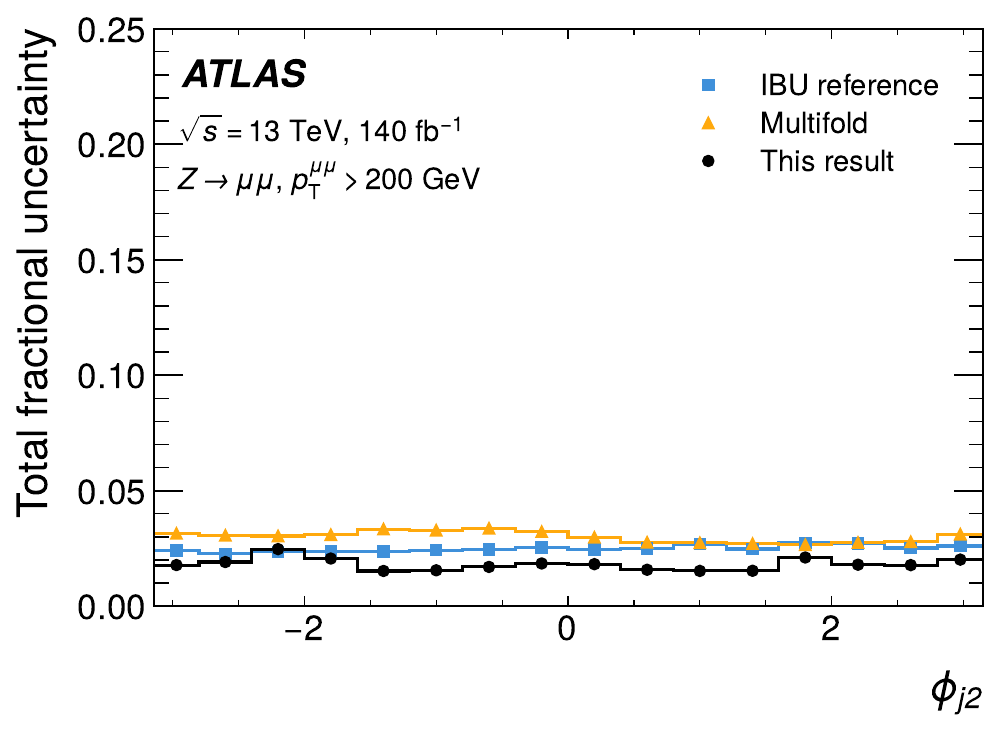}
}\\
\subfloat[]{
\includegraphics[width=0.47\textwidth]{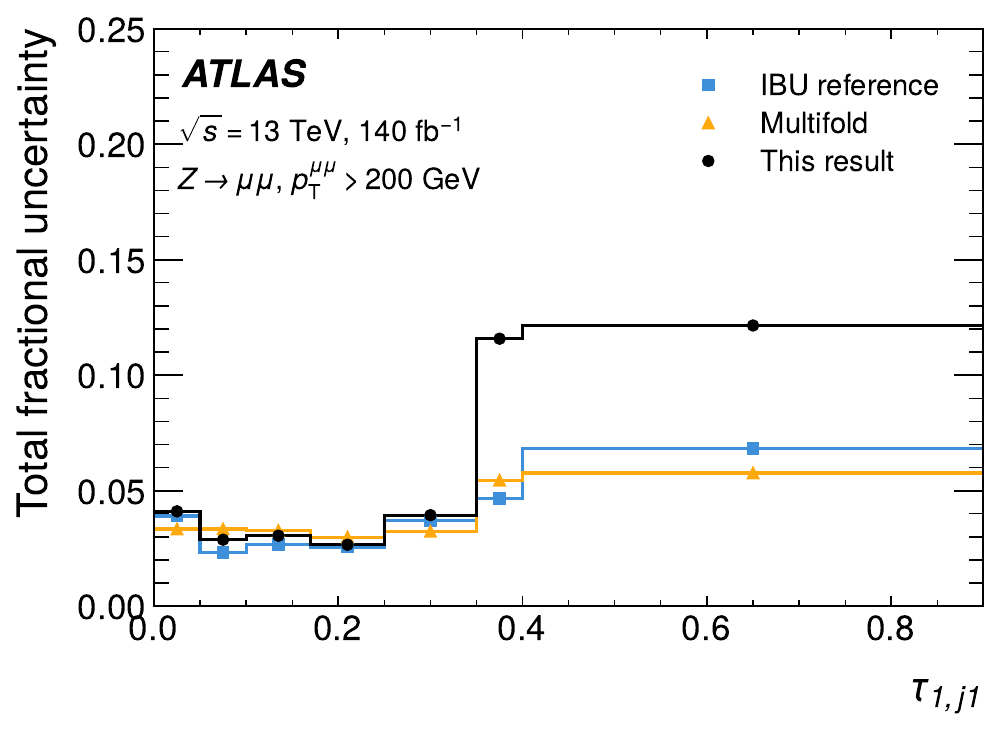}
}
\subfloat[]{
\includegraphics[width=0.47\textwidth]{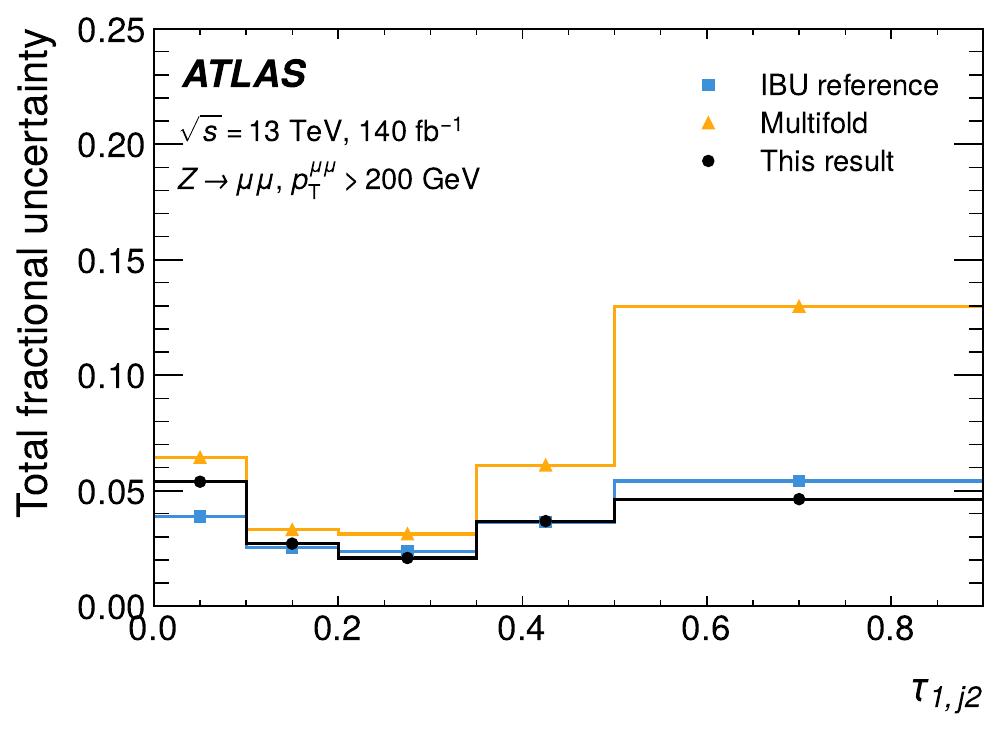}
}
\caption{Comparison of the total uncertainty estimates obtained with IBU, \multifold{}~\cite{STDM-2020-17} and \omnifold{} (labeled `This result') for the azimuthal angle of the (a) leading and (b) subleading muon, and (c) leading and (d) subleading charged-hadron jets, and the $1$-subjettiness~\cite{Thaler:2010tr} $\tau_1$ of the (e) leading and (f) subleading charged-hadron jets.}
\end{figure}

\begin{figure}[p]
\centering
\subfloat[]{
\includegraphics[width=0.47\textwidth]{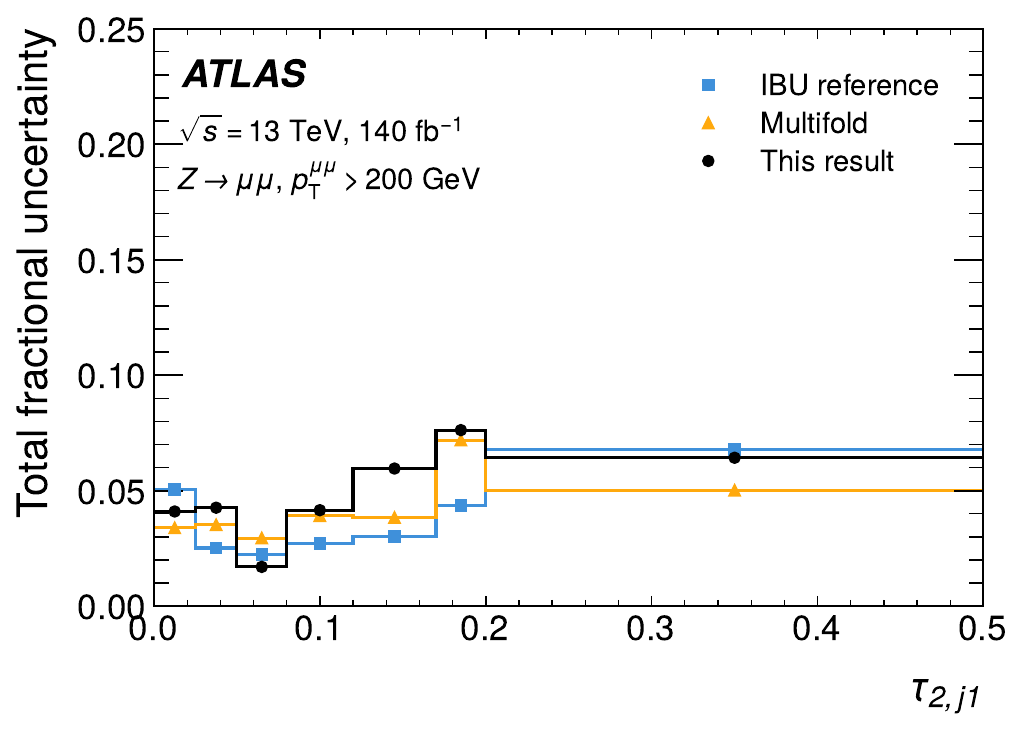}
}
\subfloat[]{
\includegraphics[width=0.47\textwidth]{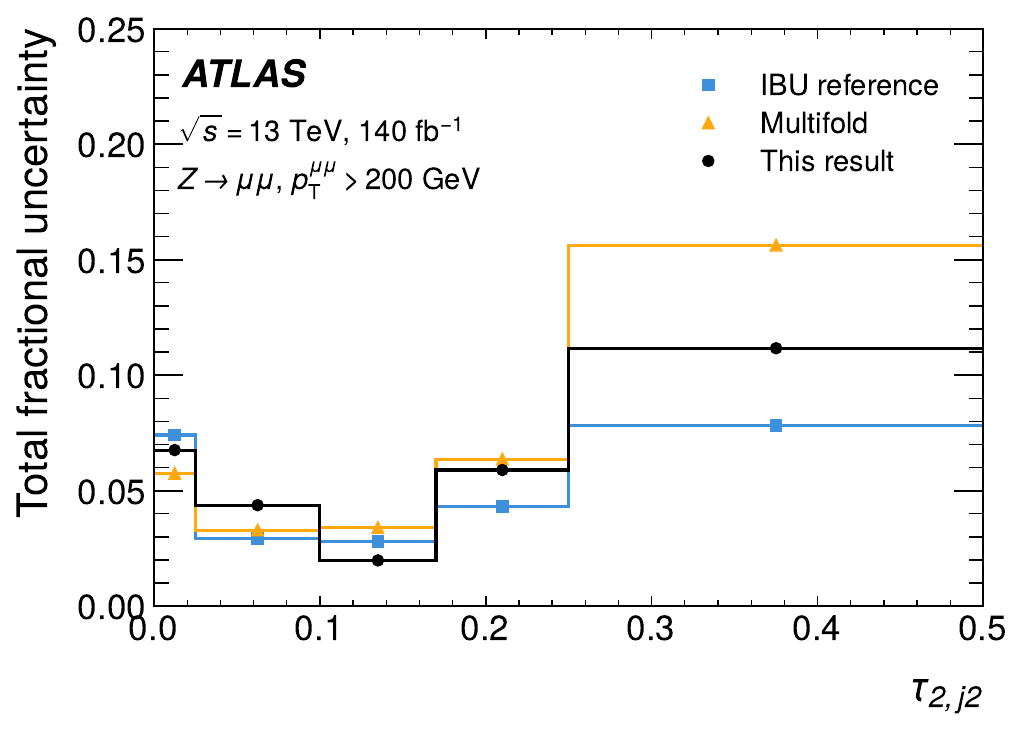}
}\\
\subfloat[]{
\includegraphics[width=0.47\textwidth]{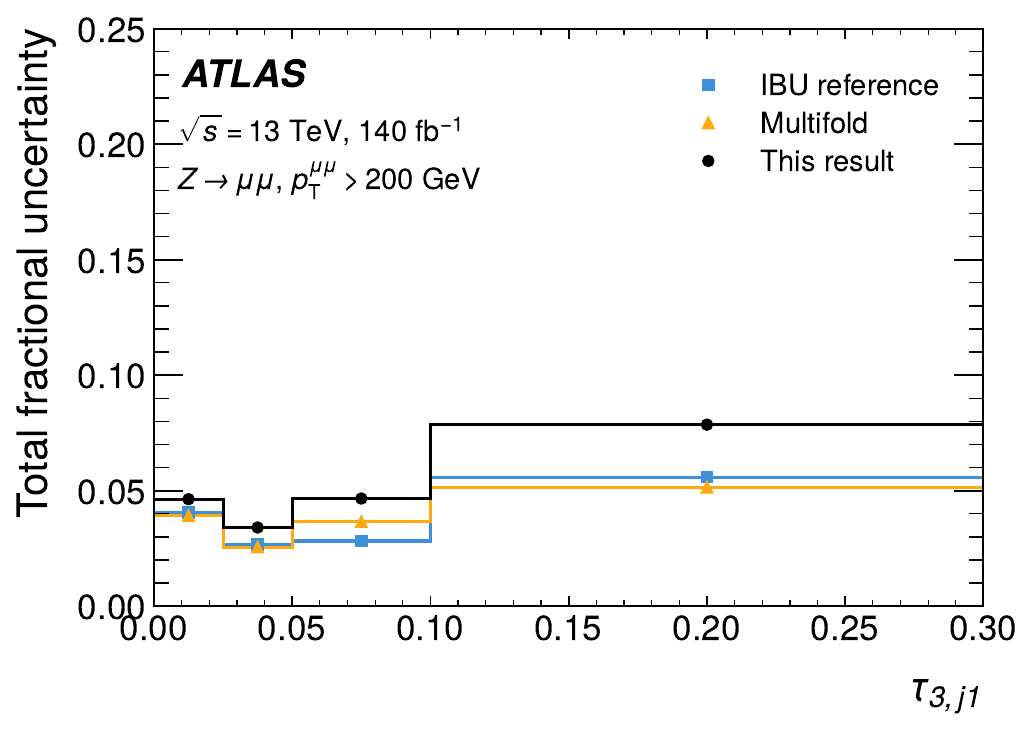}
}
\subfloat[]{
\includegraphics[width=0.47\textwidth]{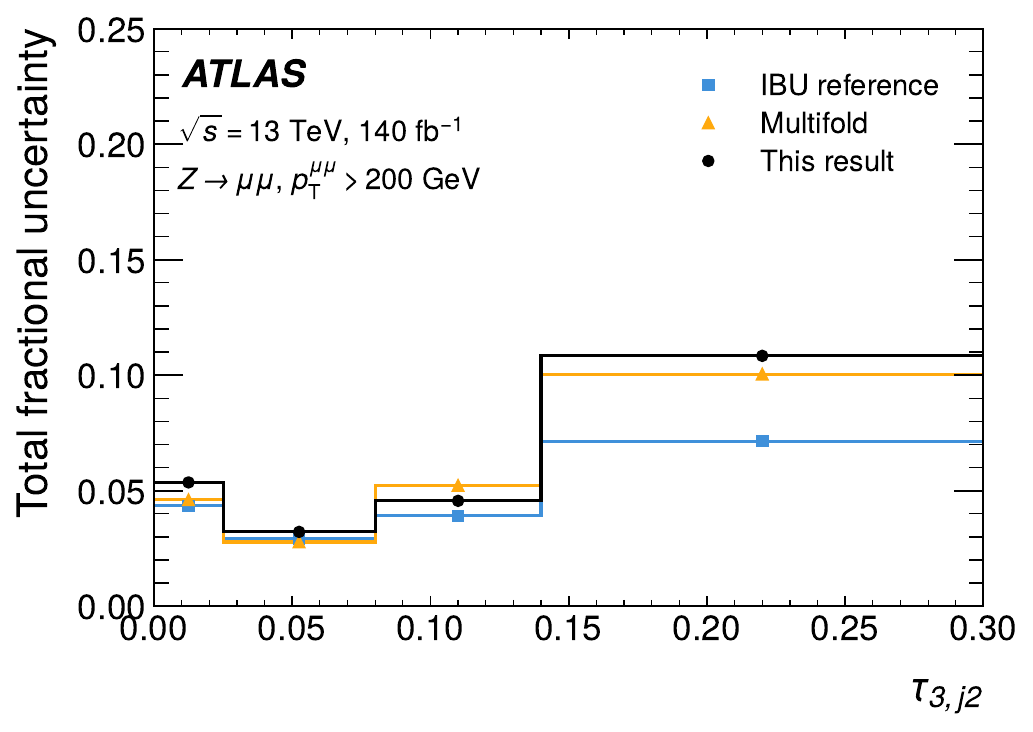}
}\\
\subfloat[]{
\includegraphics[width=0.47\textwidth]{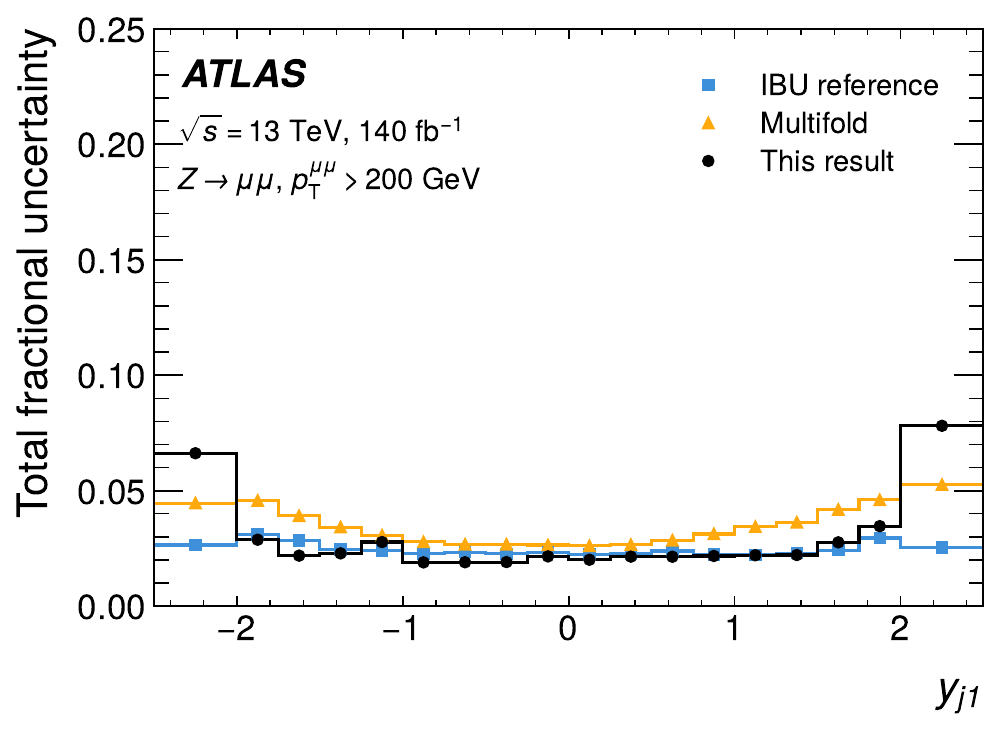}
}
\subfloat[]{
\includegraphics[width=0.47\textwidth]{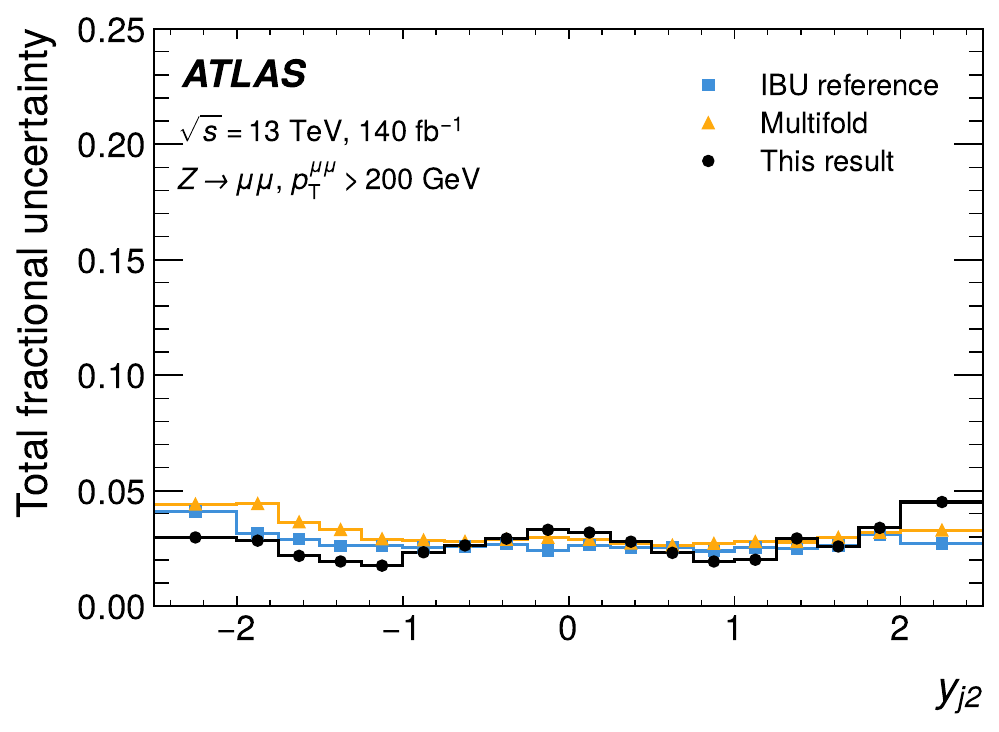}
}
\caption{Comparison of the total uncertainty estimates obtained with IBU, \multifold{}~\cite{STDM-2020-17} and \omnifold{} (labeled `This result') for the $2$-subjettiness $\tau_2$ of the (a) leading and (b) subleading charged-hadron jets, the $3$-subjettiness $\tau_3$ of the (c) leading and (d) subleading charged-hadron jets, and the rapidity of the (e) leading and (f) subleading charged-hadron jets.}
\end{figure}

\FloatBarrier

\subsection{Hidden variable uncertainty comparison}
\label{app:hv_uncert_comp}

\begin{figure}[p]
\centering
\subfloat[]{
\includegraphics[width=0.47\textwidth]{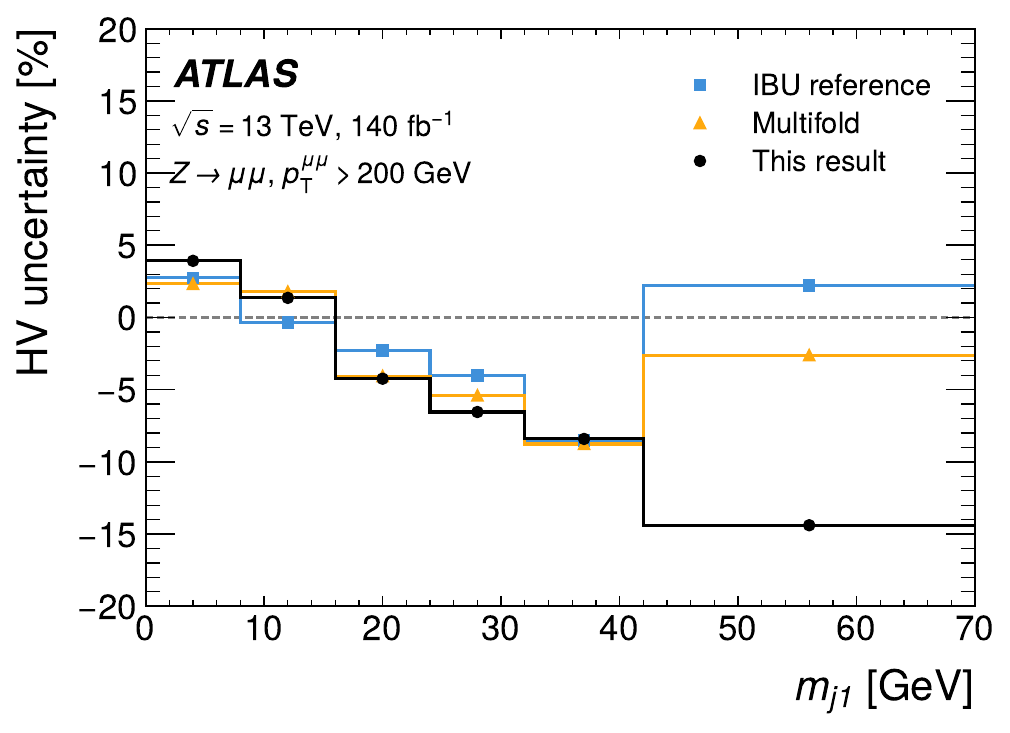}
}
\subfloat[]{
\includegraphics[width=0.47\textwidth]{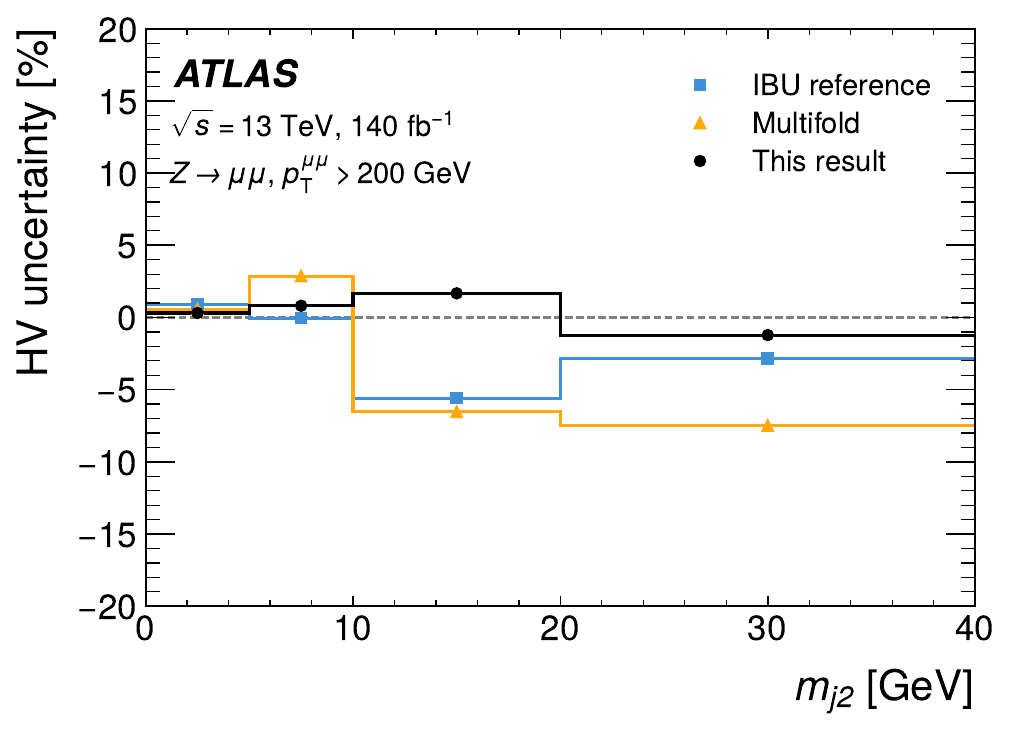}
\label{fig:comp_hv_mj2}
}\\
\subfloat[]{
\includegraphics[width=0.47\textwidth]{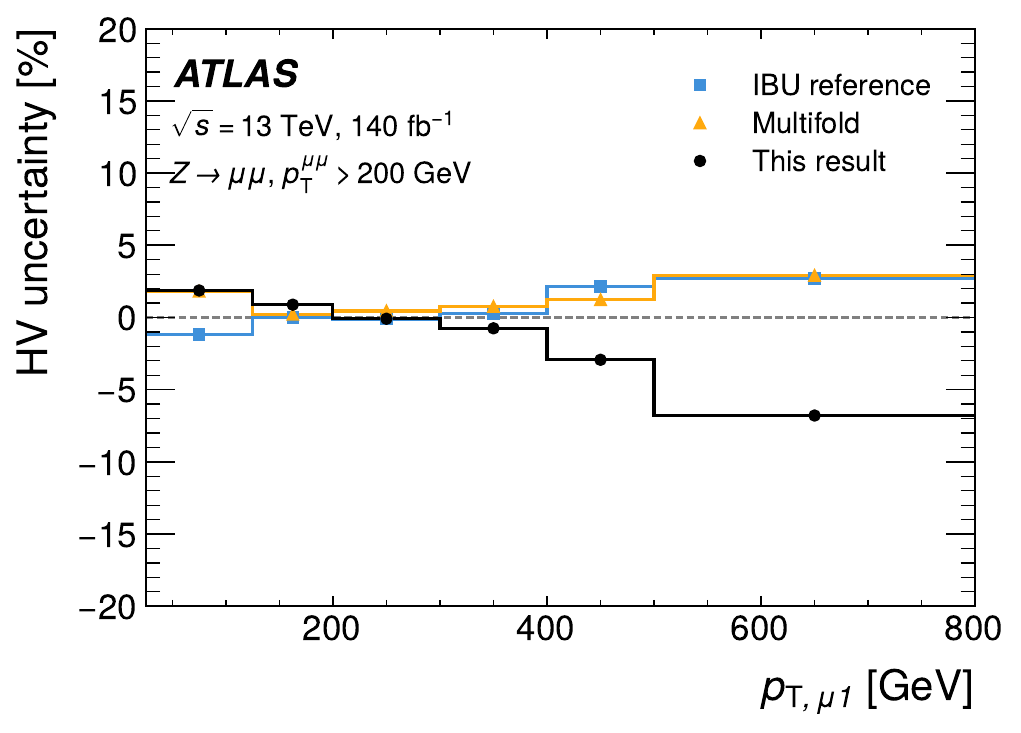}
}
\subfloat[]{
\includegraphics[width=0.47\textwidth]{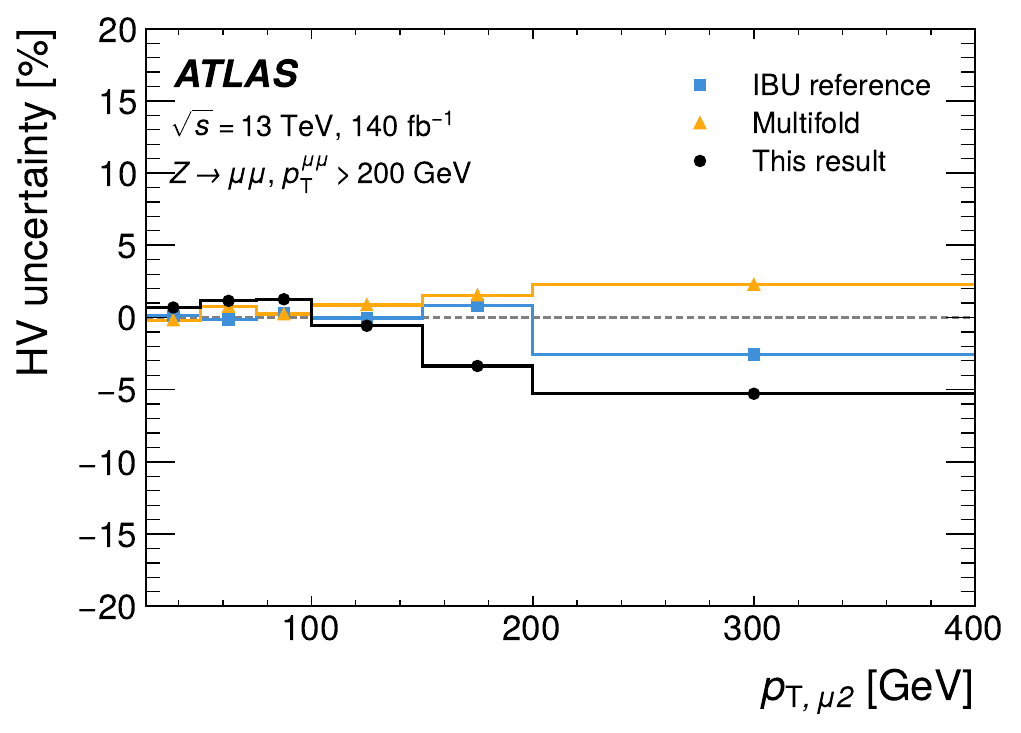}
}\\
\subfloat[]{
\includegraphics[width=0.47\textwidth]{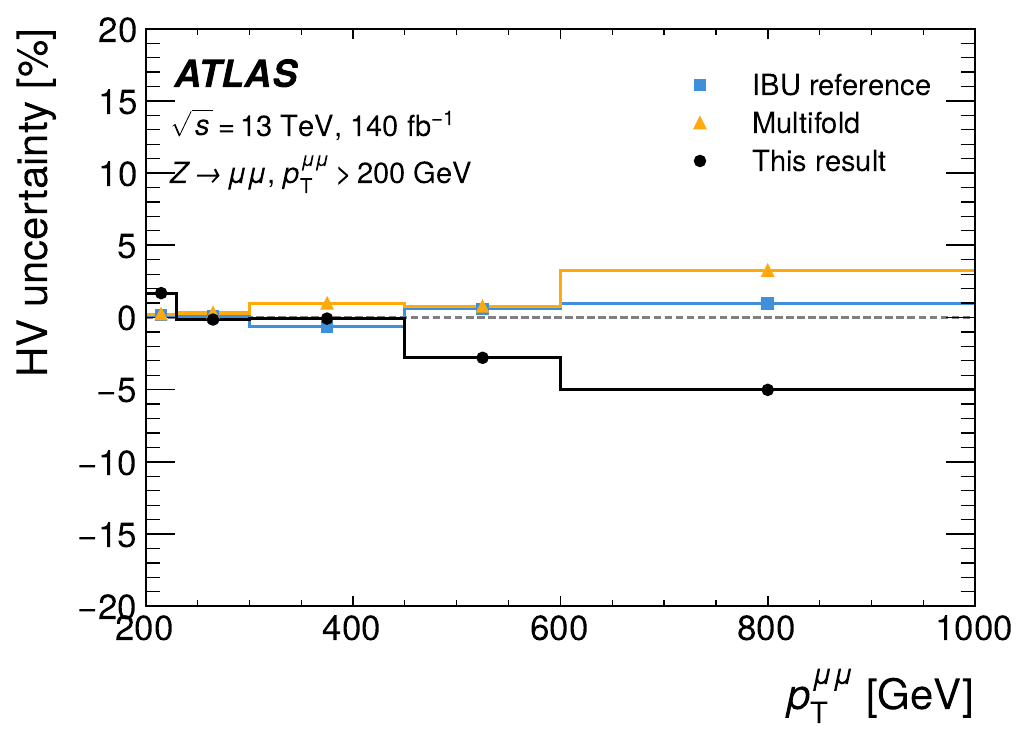}
\label{fig:comp_hv_ptll}
}
\subfloat[]{
\includegraphics[width=0.47\textwidth]{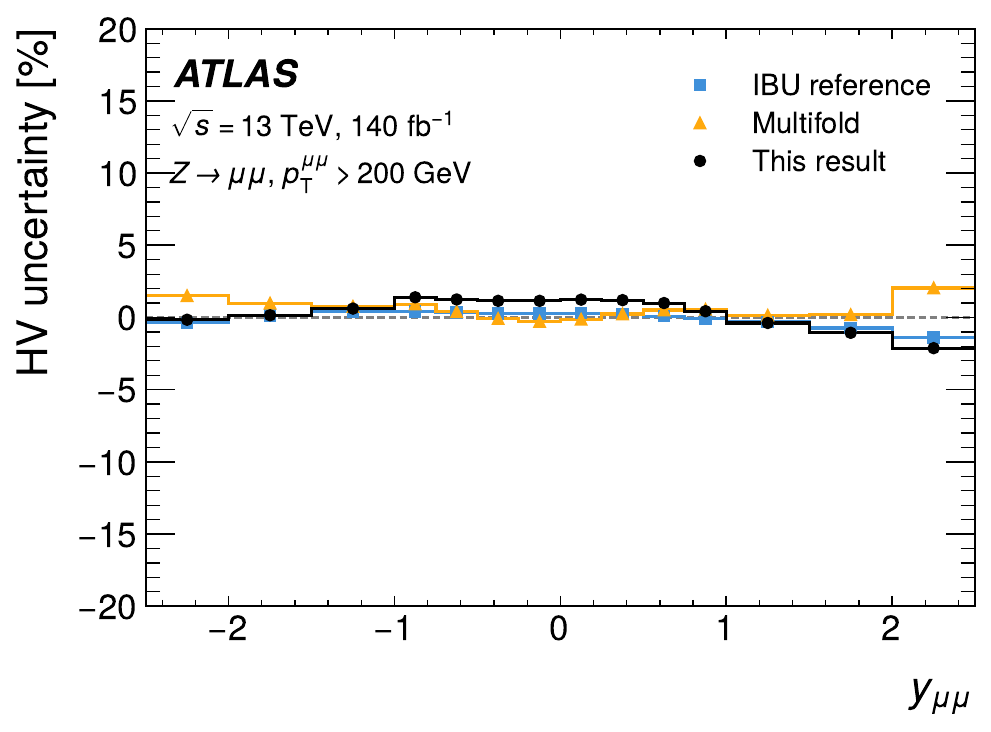}
}
\caption{Comparison of the hidden variable uncertainty obtained with IBU, \multifold{}~\cite{STDM-2020-17} and \omnifold{} (labeled `This result') for the masses of the (a) leading and (b) subleading charged-hadron jet, the transverse momentum of the (c) leading and (d) subleading muon, and the (e) transverse momentum and the (f) rapidity of the dimuon system.}
\label{fig:comparison_hv_uncertainty_1}
\end{figure}

\begin{figure}[p]
\centering
\subfloat[]{
\includegraphics[width=0.47\textwidth]{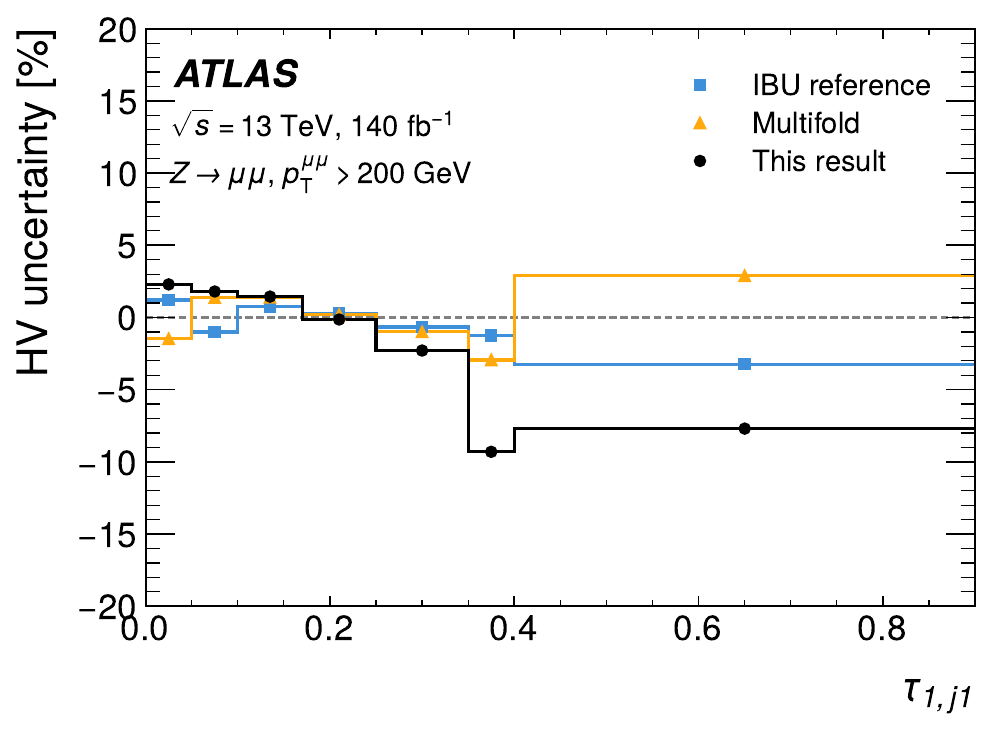}
}
\subfloat[]{
\includegraphics[width=0.47\textwidth]{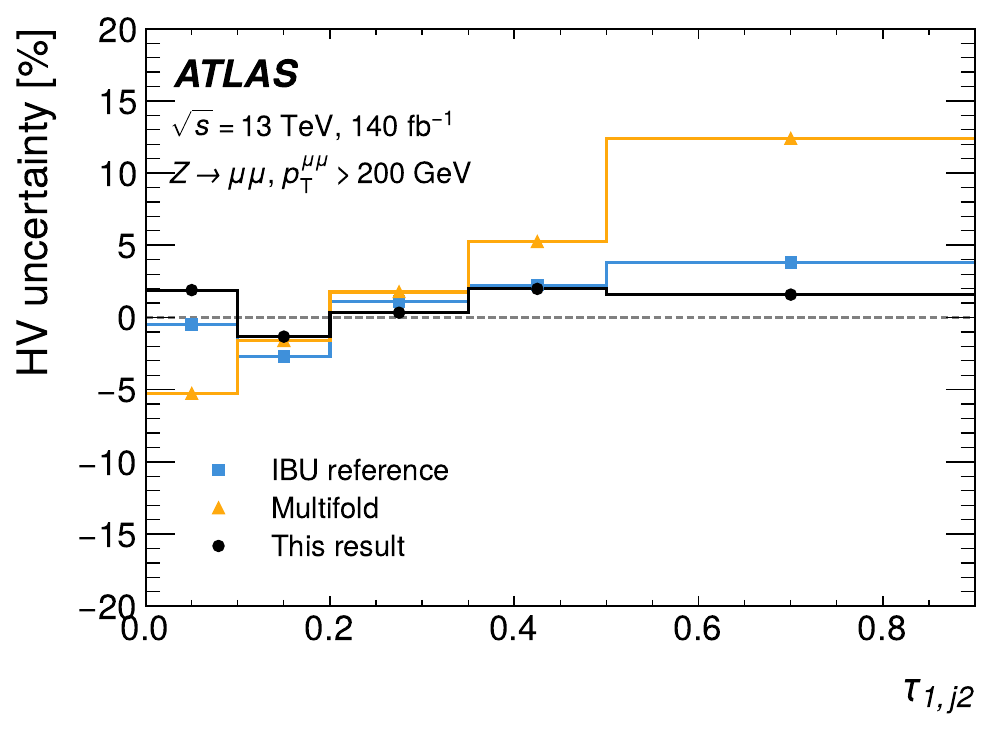}
}\\
\subfloat[]{
\includegraphics[width=0.47\textwidth]{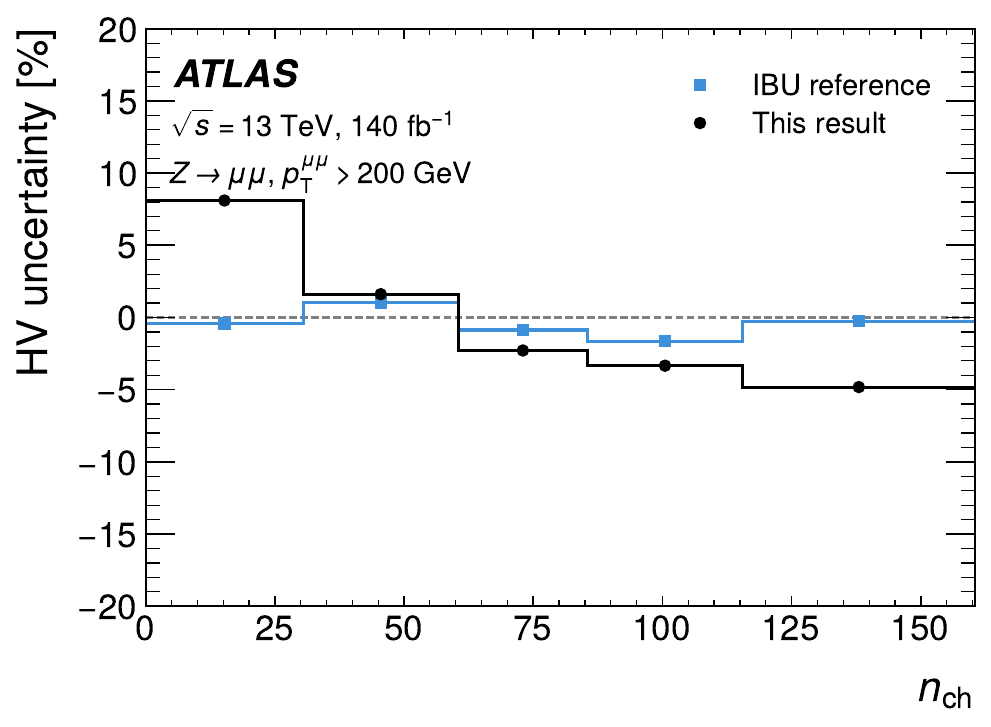}
}
\subfloat[]{
\includegraphics[width=0.47\textwidth]{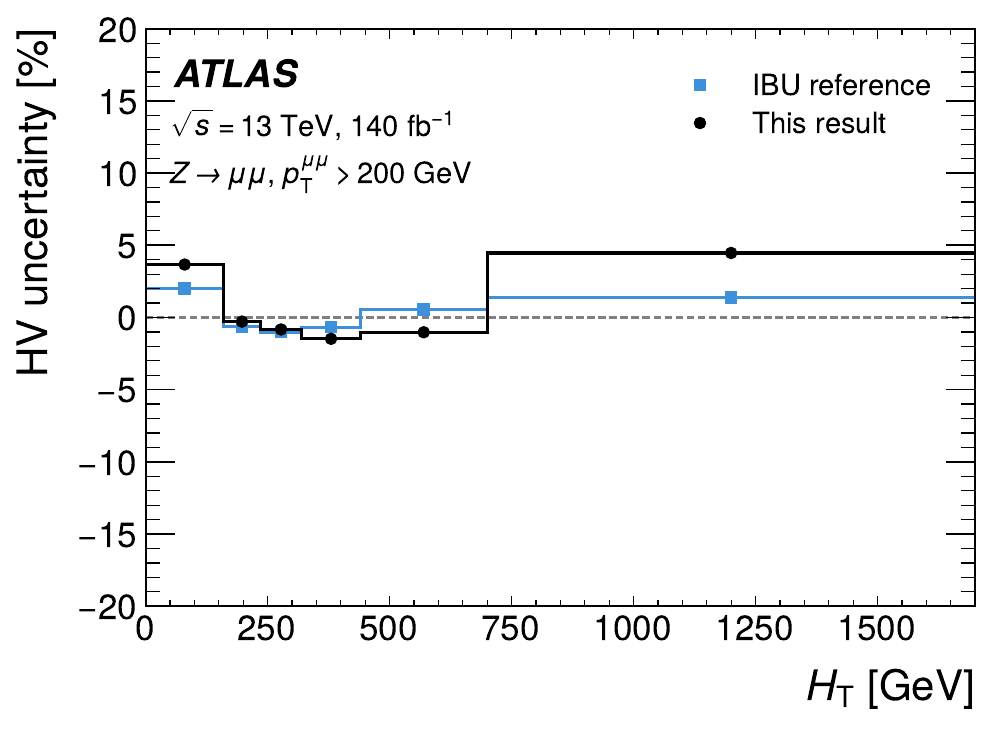}
}
\caption{Comparison of the hidden variable uncertainty obtained with IBU, \multifold{}~\cite{STDM-2020-17} and \omnifold{} (labeled `This result') for the $1$ subjettiness $\tau_1$ of the (a) leading and (b) subleading charged-hadron jet, the (c) number of charged hadrons \nch{}, and the (d) \HT{}.}
\label{fig:comparison_hv_uncertainty_2}
\end{figure}

Figures~\ref{fig:comparison_hv_uncertainty_1} and~\ref{fig:comparison_hv_uncertainty_2} compare the hidden-variable uncertainty estimated for the IBU, \multifold{}~\cite{STDM-2020-17}, and \omnifold{} measurements.
The figures show the signed uncertainty as opposed to the magnitude as in Appendix~\ref{app:uncert_comp}, since a single uncertainty is shown rather than many uncertainties added in quadrature.
The dominant hidden variables in this analysis, most notably the particle-level hadron composition, are only weakly constrained by the detector-level information. Consequently, even a highly differential unfolding cannot substantially reduce this uncertainty source, and comparable hidden-variable uncertainties are expected for all three approaches, as observed in Figs.~\ref{fig:comparison_hv_uncertainty_1} and~\ref{fig:comparison_hv_uncertainty_2}.
In some observables, such as the subleading charged-hadron jet mass (Figure~\ref{fig:comp_hv_mj2}), the \omnifold{} measurement yields a smaller hidden-variable uncertainty than either \multifold{} or IBU, whereas in others, such as \ptll{} (Figure~\ref{fig:comp_hv_ptll}), the uncertainty is slightly larger.
These larger uncertainties are primarily observed in low-statistics tails of distributions, where limited training data makes accurate estimation of the likelihood ratios used by the neural-network classifier more challenging. Nevertheless, these differences are small compared with the total uncertainty budget and have little impact on the overall precision of the projected measurements, as shown in Appendix~\ref{app:uncert_comp}.

\FloatBarrier

\section{Unbinned background subtraction}
\label{app:bkg_sub}
%


This appendix describes the procedure used to subtract the small \ttbar{} background from the pseudodata and data samples prior to unfolding with \omnifold{}. Since the unfolding is performed without binning, the background subtraction must also be performed in an unbinned manner. Following Ref.~\cite{Nachman:2020fff}, the subtraction is achieved by training a neural network classifier to reweight between the data distribution, and the background-subtracted data distribution, utilizing a background model provided by a Monte Carlo generator.
In the context of this analysis the unbinned subtraction is performed by training a neural network classifier to distinguish two sets of simulated events: the detector-level pseudodata sample described in Section~\ref{sec:pseudodata}, and the same set of simulated events concatenated with a negatively weighted detector-level MC sample that describes the \ttbar background contribution.
These negatively weighted events statistically remove the existing \ttbar events in pseudodata and leave only the signal contribution.
Classifiers are trained to distinguish the second set from the first set, and the neural network output can then be used to reweight the pseudodata sample to match the background removed sample.

The classifiers described above are trained with simulation.
Once trained the networks are evaluated over the data, and their outputs are used to remove the \ttbar background from the data.
A dedicated uncertainty, in which the background subtraction procedure is repeated with an alternative description of the \ttbar background furnished by an alternative generator, accounts for the generator dependence.
Both the nominal and alternative background subtractions are then propagated through the unfolding procedure to produce the final background subtraction uncertainty.

As described in Section~\ref{sec:bkg_sub}, the \ttbar background is small, accounting for only 0.25\% of events in the selected event yield.
This implies that differences between the detector-level pseudodata and the \ttbar subtracted pseudodata are small, making the classification task described above difficult.
Initial efforts to learn the required likelihood ratio directly from the full charged-particle phase-space representation were not successful.
In particular, Particle Transformer neural networks trained only on the low-level event representation described in Section~\ref{sec:nn_arch} were unable to isolate the \ttbar background contribution and remove it reliably.
Instead, the background subtraction procedure is performed differentially in five event-level summary observables, selected because the background contribution is localized in the distributions of these observables.
The most important of the five observables is the output of a Particle Transformer neural network, with the same architecture and optimized from scratch using the same methods as described in Sections~\ref{sec:nn_arch} and~\ref{sec:training}, that is trained to discriminate background \ttbar events from signal events.
The background contribution to the detector-level pseudodata is localized in the tail of this highly expressive observable, making it very useful for the training of the classifiers which perform the background subtraction as described above.
The four other observables used in the training are: \nch, \HT, $p_{\textrm{T}}^{j1}$, and $m^{j1}$.
Since the event representation used for background subtraction is fixed-length, MLP neural networks with two hidden layers of 512 nodes each and GeLU activation functions are used.
As with the neural networks used in the \omnifold{} procedure, the MLPs were implemented and trained in PyTorch.
Ensembles of 100 neural networks were trained for both the nominal and alternative background subtractions.
Given that the background subtraction should only remove density, the weights produced by the networks were clipped to a maximum of 1.0, and then averaged across the ensemble to produce the final background subtraction weights.

\begin{figure}[htb]
\centering
\includegraphics[width=0.6\textwidth]{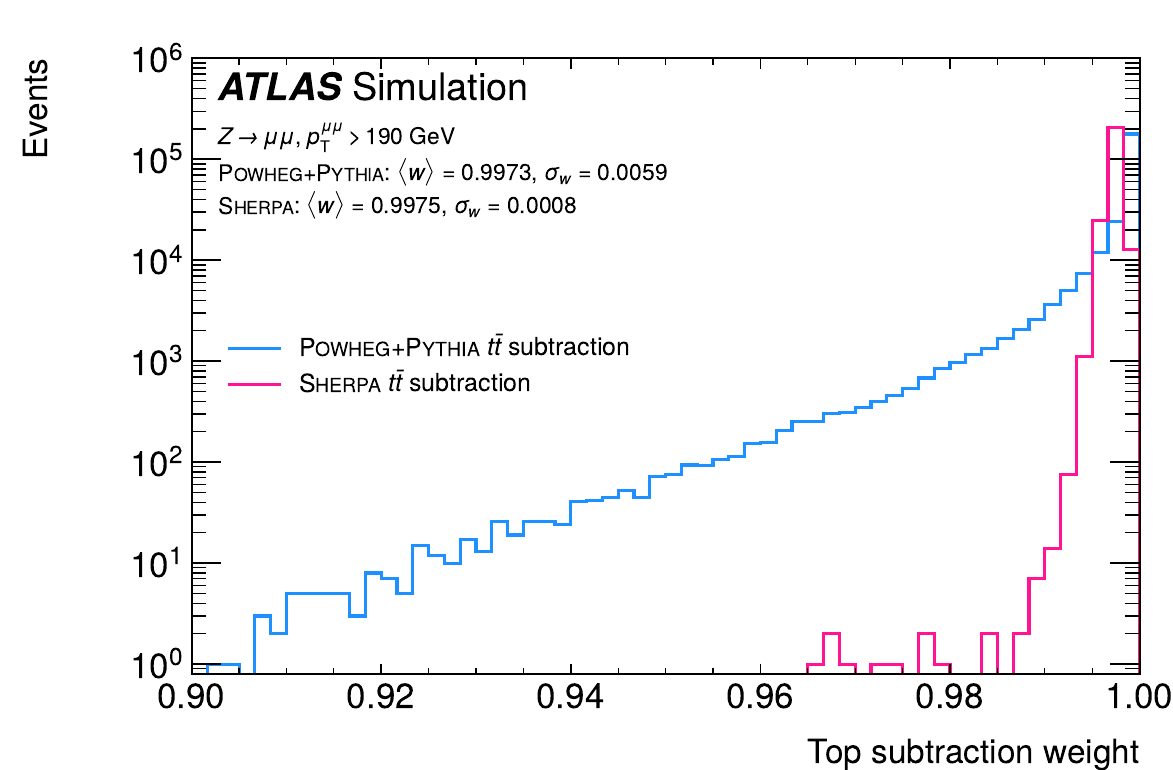}
\caption{Histograms of the \ttbar{} background subtraction weights produced using the nominal \POWHEG[v2]+\PYTHIA[8.230] sample (triangles) and the alternative \SHERPA[2.2.1] sample (squares) for the background estimate. Both methods remove approximately the same overall \ttbar{} contribution ($\approx0.25\%$ of the selected event yield), but assign different weights to individual events. The differences between the two sets of subtraction weights define the $t\bar{t}$ background subtraction uncertainty.}
\label{fig:top_subtraction_weights}
\end{figure}

\begin{figure}[htb]
\centering
\subfloat[]{
\includegraphics[width=0.45\textwidth]{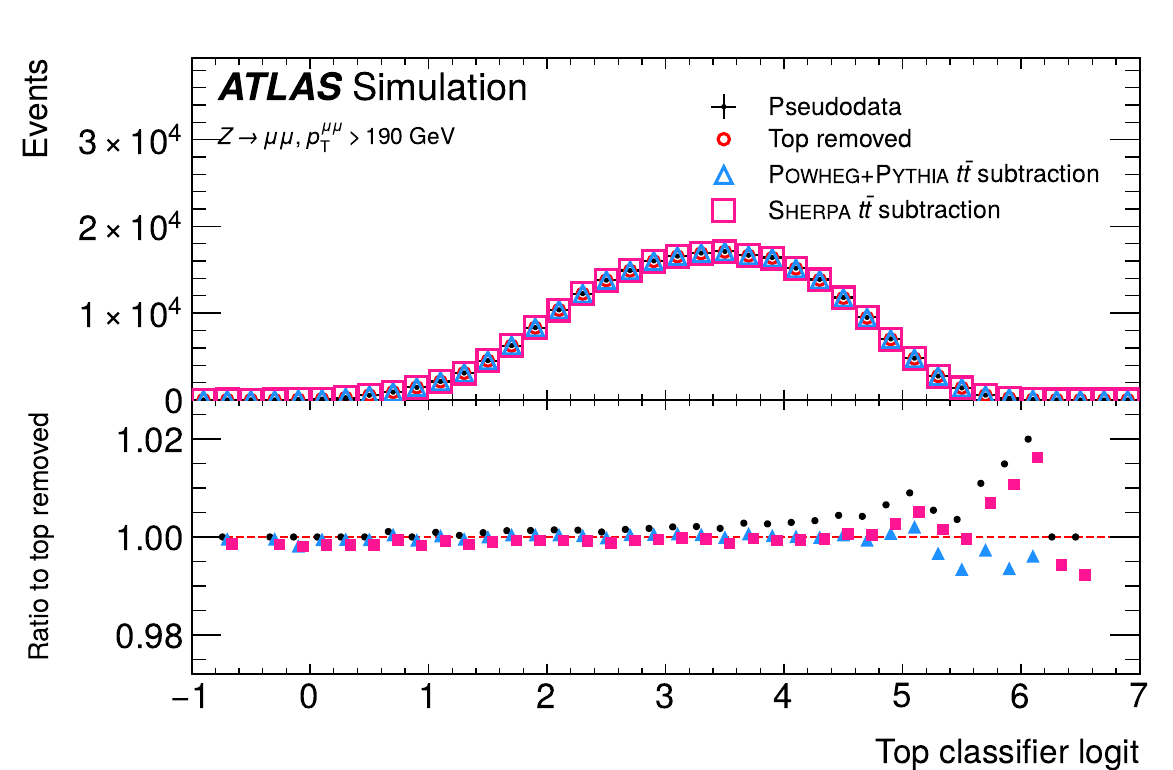}
\label{fig:top_subtraction_logit}
}
\subfloat[]{
\includegraphics[width=0.45\textwidth]{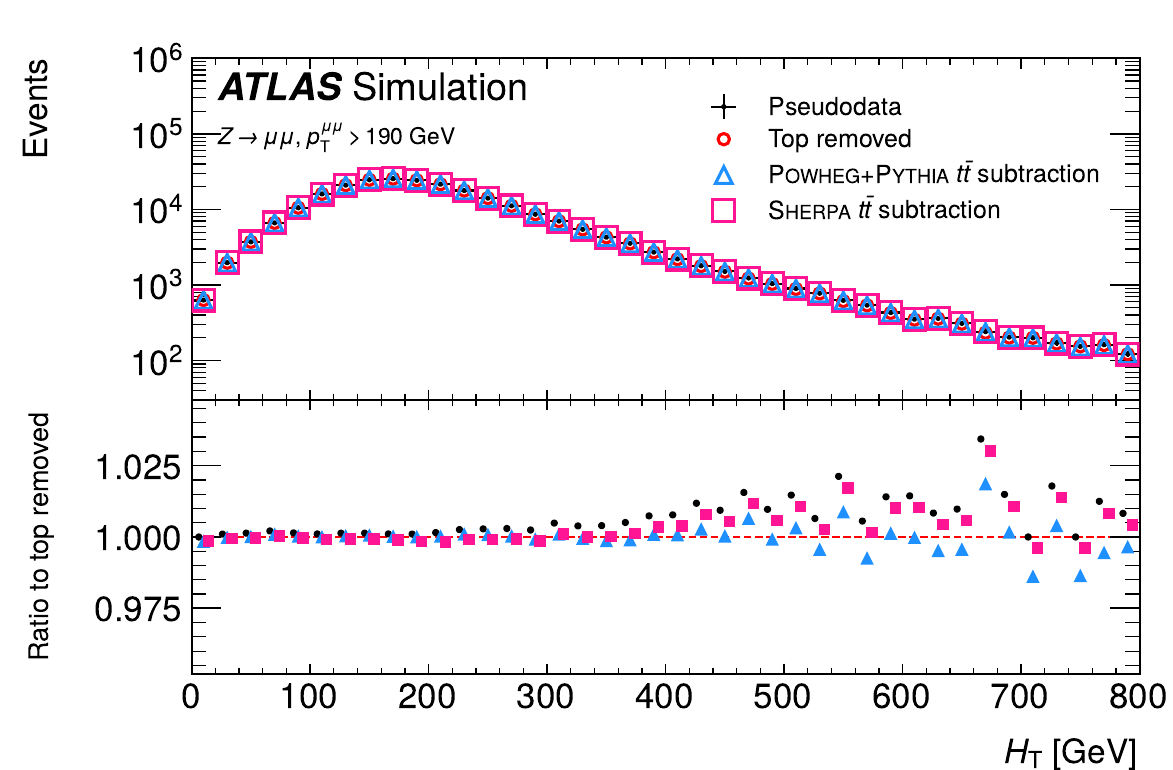}
\label{fig:top_subtraction_ht}
}
\caption{The \ttbar background subtraction for the detector-level pseudodata sample. The pre-subtraction pseudodata is plotted as points, the signal-only pseudodata with the \ttbar contribution removed is plotted as circles, and the pseudodata with the nominal and alternative subtraction weights applied are shown as triangles and squares respectively. Open markers are used in the upper pad for visibility but denote the same quantity. Ratio pads show the ratio to the signal-only pseudodata, which is the target of the subtraction procedure. The subtraction is shown for the (a) \ttbar classifier output (logit) and (b) \HT{} observables.}
\label{fig:top_subtraction_observables}
\end{figure}

Histograms of the background subtraction weights for the pseudodata sample are shown in Figure~\ref{fig:top_subtraction_weights}.
Similar weight distributions were obtained for the data sample.
The weight distributions produced using the nominal \POWHEG[v2]+\PYTHIA[8.230] sample and the alternative \SHERPA[2.2.1] sample both
have mean values close to 0.998, consistent with the expected $\approx0.25\%$ background fraction.
The weight distributions however have substantially different variances, with the distribution of the \POWHEG[v2]+\PYTHIA[8.230] weights having a substantially larger tail.
The difference between these weight distributions corresponds to the top background uncertainty which is propagated through the unfolding procedure.
Illustrations of the top subtraction in two observables are shown in Figure~\ref{fig:top_subtraction_observables}.
Figure~\ref{fig:top_subtraction_logit} shows the \ttbar classifier output, where the \ttbar contribution is localized in the tail of the distribution.
Both the nominal and alternative subtractions are able to remove the background, except in the very low-statistics tail above a classifier logit of $5.0$.
Figure~\ref{fig:top_subtraction_ht} shows the \HT observable, where the background contribution is less localized but both subtractions generally succeed in removing the background.


\FloatBarrier

\section{\omnifold{} hyperparameter optimization methods and results}
\label{app:hpo}

This appendix summarizes the hyperparameter optimization strategy used for the \omnifold{} procedure. Unlike conventional classification tasks, the objective of the neural networks within \omnifold{} is not to maximize classification accuracy but to estimate likelihood ratios as accurately as possible. Although these objectives become equivalent in the limit of infinite data and infinitely expressive models, finite training samples and limited model capacity mean that standard classification metrics, such as the area under the ROC curve, are not necessarily optimal proxies for unfolding performance. Furthermore, because \omnifold{} is an iterative algorithm, stable likelihood-ratio estimates that produce moderate corrections at each iteration are generally preferred over highly variable estimates that maximize classification performance in a single step.

To directly optimize the quality of the unfolded measurement, the objective function for the hyperparameter optimization was defined as the sum of the one-dimensional Wasserstein distances evaluated for the 24 observables measured in Ref.~\cite{STDM-2020-17}, together with the \nch{} and \HT{} observables introduced in this work. This metric was found to correlate significantly better with the quality of the resulting reweighting than traditional classification metrics.

A complete hyperparameter optimization across every neural-network training within the \omnifold{} procedure would be computationally prohibitive.
Hyperparameter optimization was therefore restricted to the pretraining stage and the step-1 and step-2 classifiers of the first \omnifold{} iteration.
The chosen hyperparameters for steps one and two were then used for all subsequent \omnifold{} iterations.
The hyperparameters that were optimized in each of the three trainings are marked in Table~\ref{tab:hparams}.
The ranges used and final values are given in Tables~\ref{tab:hparams-pre}, \ref{tab:hparams-step1}, and \ref{tab:hparams-step2}.
Architectural hyperparameters (e.g., the number of attention blocks and class-attention blocks) were optimized only during pretraining, since the same network architecture is subsequently reused throughout the unfolding. Training hyperparameters, including the learning-rate schedule and regularization parameters, were optimized independently for each step.
All sweeps were performed using a random search method and the configuration which produced the best sum of Wasserstein distances was selected.
This process was observed to be important for the overall performance of the \omnifold{} procedure.
In particular, the learning rates, including the length of the annealing periods in the learning rate schedule, were very important to tune properly.
Although the optimization required training hundreds of neural networks, the associated computational cost remained a small fraction of the total resources required for the full unfolding analysis.

\begin{table}
\centering
\begin{tabular}{c|c|c|c}
\hline \hline
Hyperparameter & pretraining & Step 1 & Step 2 \\
\hline
Number of attention blocks & \checkmark &  &  \\
Number of class attention blocks & \checkmark &  &  \\
Attention block dropout & \checkmark & \checkmark & \checkmark \\
Class attention block dropout & \checkmark &  &  \\
MLP dropout & \checkmark &  &  \\
Minimum learning rate & \checkmark & \checkmark & \checkmark \\
Optimizer & \checkmark & \checkmark & \checkmark \\
Weight decay &  & \checkmark & \checkmark \\
Cosine annealing period &  & \checkmark & \checkmark \\
\hline \hline
\end{tabular}
\vspace{8pt}
\caption{Hyperparameters optimized in the pretraining, step 1, and step 2 trainings. A \checkmark means the hyperparameter was optimized in that training.}
\label{tab:hparams}
\end{table}

\begin{table}
\centering
\begin{tabular}{c|c|c|c}
\hline \hline
Hyperparameter & Lower bound & Upper bound & Final value \\
\hline
Number of attention blocks & 3 & 8 & 6 \\
Number of class attention blocks & 1 & 3 & 1 \\
Attention block dropout & 0.0 & 0.3 & 0.02 \\
Class attention block dropout & 0.0 & 0.5 & 0.02 \\
MLP dropout & 0.0 & 0.5 & 0.0 \\
Minimum learning rate & 1e-7 & 1e-3 & 1e-5 \\
Optimizer & -- & -- & Lion \\
\hline \hline
\end{tabular}
\caption{Hyperparameter ranges and final values for the pretraining task. A random search with 150 runs was performed. The AdamW and Lion optimizers were considered for the optimizer choice.}
\label{tab:hparams-pre}
\end{table}

\begin{table}
\centering
\begin{tabular}{c|c|c|c}
\hline \hline
Hyperparameter & Lower bound & Upper bound & Final value \\
\hline
Block dropout & 0.0 & 0.3 & 0.02 \\
Minimum learning rate & 1e-7 & 1e-3 & 8e-6 \\
Weight decay & 0.0 & 0.2 & 0.01 \\
Cosine annealing period & 800 & 4000 & 3000\\
Optimizer & -- & -- & Lion \\
\hline \hline
\end{tabular}
\caption{Hyperparameter ranges and final values for the step 1 task. A random search with 100 runs was performed. The AdamW and Lion optimizers were considered for the optimizer choice.}
\label{tab:hparams-step1}
\end{table}

\begin{table}
\centering
\begin{tabular}{c|c|c|c}
\hline \hline
Hyperparameter & Lower bound & Upper bound & Final value \\
\hline
Block dropout & 0.0 & 0.3 & 0.02 \\
Minimum learning rate & 1e-7 & 1e-3 & 1e-5 \\
Weight decay & 0.0 & 0.2 & 0.01 \\
Cosine annealing period & 1000 & 20000 & 15000 \\
Optimizer & -- & -- & Lion \\
\hline \hline
\end{tabular}
\caption{Hyperparameter ranges and final values for the step 2 task. A random search with 50 runs was performed. The AdamW and Lion optimizers were considered for the optimizer choice.}
\label{tab:hparams-step2}
\end{table}

\FloatBarrier


%
%
\printbibliography
\clearpage
 
\begin{flushleft}
\hypersetup{urlcolor=black}
{\Large The ATLAS Collaboration}

\bigskip

\AtlasOrcid[0000-0002-6665-4934]{G.~Aad}$^\textrm{\scriptsize 102}$,
\AtlasOrcid[0000-0001-7616-1554]{E.~Aakvaag}$^\textrm{\scriptsize 17}$,
\AtlasOrcid[0000-0002-5888-2734]{B.~Abbott}$^\textrm{\scriptsize 121}$,
\AtlasOrcid[0000-0002-0287-5869]{S.~Abdelhameed}$^\textrm{\scriptsize 83b}$,
\AtlasOrcid[0000-0002-1002-1652]{K.~Abeling}$^\textrm{\scriptsize 54}$,
\AtlasOrcid[0000-0001-5763-2760]{N.J.~Abicht}$^\textrm{\scriptsize 48}$,
\AtlasOrcid[0000-0002-8496-9294]{S.H.~Abidi}$^\textrm{\scriptsize 30}$,
\AtlasOrcid[0009-0003-6578-220X]{M.~Aboelela}$^\textrm{\scriptsize 44}$,
\AtlasOrcid[0000-0002-9987-2292]{A.~Aboulhorma}$^\textrm{\scriptsize 36e}$,
\AtlasOrcid[0000-0001-5329-6640]{H.~Abramowicz}$^\textrm{\scriptsize 154}$
% ...
\bigskip
\\

% ...
$^{17}$Department for Physics and Technology, University of Bergen, Bergen; Norway.\\
% ...
$^{30}$Physics Department, Brookhaven National Laboratory, Upton NY; United States of America.\\
% ...
$^{36}$$^{(a)}$Facult\'e des Sciences Ain Chock, Universit\'e Hassan II de Casablanca;$^{(b)}$Facult\'{e} des Sciences, Universit\'{e} Ibn-Tofail, K\'{e}nitra;$^{(c)}$Facult\'e des Sciences Semlalia, Universit\'e Cadi Ayyad, LPHEA-Marrakech;$^{(d)}$LPMR, Facult\'e des Sciences, Universit\'e Mohamed Premier, Oujda;$^{(e)}$Facult\'e des sciences, Universit\'e Mohammed V, Rabat;$^{(f)}$Institute of Applied Physics, Mohammed VI Polytechnic University, Ben Guerir; Morocco.\\
% ...
$^{44}$Physics Department, Southern Methodist University, Dallas TX; United States of America.\\
% ...
$^{48}$Fakult\"{a}t Physik, Technische Universit{\"a}t Dortmund, Dortmund; Germany.\\
% ...
$^{54}$II. Physikalisches Institut, Georg-August-Universit\"{a}t G\"ottingen, G\"ottingen; Germany.\\
% ...
$^{83}$$^{(a)}$Khalifa University of Science and Technology, Abu Dhabi;$^{(b)}$New York University Abu Dhabi, Abu Dhabi;$^{(c)}$United Arab Emirates University, Al Ain;$^{(d)}$University of Sharjah, Sharjah; United Arab Emirates.\\
% ...
$^{102}$CPPM, Aix-Marseille Universit\'e, CNRS/IN2P3, Marseille; France.\\
% ...
$^{121}$Homer L. Dodge Department of Physics and Astronomy, University of Oklahoma, Norman OK; United States of America.\\
% ...
$^{154}$Raymond and Beverly Sackler School of Physics and Astronomy, Tel Aviv University, Tel Aviv; Israel.\\
% ...
$^{a}$ Also at Affiliated with an institute formerly covered by a cooperation agreement with CERN.\\
$^{b}$ Also at An-Najah National University, Nablus; Palestine.\\
$^{c}$ Also at Borough of Manhattan Community College, City University of New York, New York NY; United States of America.\\
$^{d}$ Also at Center for Interdisciplinary Research and Innovation (CIRI-AUTH), Thessaloniki; Greece.\\
$^{e}$ Also at Centre of Physics of the Universities of Minho and Porto (CF-UM-UP); Portugal.\\
$^{f}$ Also at CERN, Geneva; Switzerland.\\
% ...

\end{flushleft}


%

%
%
%
%

%
%
%
%

\end{document}